\documentclass[onecolumn, numberedappendix]{aastex631}
\usepackage{graphicx}
\usepackage{color}

\newcommand{\av}{$A_V$}

\newcommand{\eg}{{\it e.g.}}
\newcommand{\etal}{et~al.}

\newcommand{\ks}{$K_{\rm s}$}

\newcommand{\mum}{$\mu$m}

\begin{document}

\title{Young Stellar Object Candidates in IC~417 }

\reportnum{Version from \today}

\author[0000-0001-6381-515X]{L.~M.~Rebull}
\affiliation{Infrared Science Archive (IRSA) and 
NASA/IPAC Teacher Archive Research Project (NITARP), IPAC, MS 100-22, 1200 E.\
California Blvd., California Institute of Technology, Pasadena, CA
91125, USA; rebull@ipac.caltech.edu}
\author{R.~L.~Anderson III}
\affiliation{J.L. Mann High School, Academy of Mathematics, Science, and Technology,160 Fairforest Way, Greenville, SC 29607 USA}
\affiliation{NASA/IPAC Teacher Archive Research Project (NITARP), c/o IPAC, MS 100-22, 1200 E.\
California Blvd., California Institute of Technology, Pasadena, CA 91125, USA}
\author{G.~Hall}
\affiliation{Roper Mountain Science Center Planetarium, 402 Roper Mountain Road, Greenville, SC 29615  USA}
\affiliation{NASA/IPAC Teacher Archive Research Project (NITARP), c/o IPAC, MS 100-22, 1200 E.\
California Blvd., California Institute of Technology, Pasadena, CA 91125, USA}
\author[0000-0003-4269-260X]{J.~D.~Kirkpatrick}
\affiliation{IPAC, Mail Code 100-22, Caltech, 1200 E. California Boulevard, Pasadena, CA 91125, USA}
\author[0000-0002-9478-4170]{X.~Koenig}
\affiliation{NASA/IPAC Teacher Archive Research Project (NITARP), c/o IPAC, MS 100-22, 1200 E.\
California Blvd., California Institute of Technology, Pasadena, CA 91125, USA}
\author{C.~E.~Odden}
\affiliation{Phillips Academy, 180 Main Street, Andover, MA 01810 USA}
\affiliation{NASA/IPAC Teacher Archive Research Project (NITARP), c/o IPAC, MS 100-22, 1200 E.\
California Blvd., California Institute of Technology, Pasadena, CA 91125, USA}
\author{B.~Rodriguez}
\affiliation{NASA Jet Propulsion Laboratory 4800 Oak Grove Dr, Pasadena, CA 91109 USA}
\affiliation{Crescenta Valley High School, 2900 Community Ave, La Crescenta, CA 91214 USA}
\affiliation{NASA/IPAC Teacher Archive Research Project (NITARP), c/o IPAC, MS 100-22, 1200 E.\
California Blvd., California Institute of Technology, Pasadena, CA 91125, USA}
\author{R.~Sanchez}
\affiliation{Clear Creek Middle School, 361 W.\ Gatchell, Buffalo, WY 82834 USA}
\affiliation{NASA/IPAC Teacher Archive Research Project (NITARP), c/o IPAC, MS 100-22, 1200 E.\
California Blvd., California Institute of Technology, Pasadena, CA 91125, USA}
\author{B. Senson}
\affiliation{Physical Sciences Department, Madison College, 1701 Wright Street, Madison, WI 53704 USA}
\affiliation{NASA/IPAC Teacher Archive Research Project (NITARP), c/o IPAC, MS 100-22, 1200 E.\
California Blvd., California Institute of Technology, Pasadena, CA
91125, USA}
\author{V.~Urbanowski}
\affiliation{Academy of Information Technology \& Engineering, 411 High Ridge Road, Stamford, CT 06905 USA}
\affiliation{NASA/IPAC Teacher Archive Research Project (NITARP), c/o IPAC, MS 100-22, 1200 E.\
California Blvd., California Institute of Technology, Pasadena, CA 91125, USA}
\author{M.~Austin}
\affiliation{J.L. Mann High School, Academy of Mathematics, Science, and Technology,160 Fairforest Way, Greenville, SC 29607 USA}
\affiliation{NASA/IPAC Teacher Archive Research Project (NITARP), c/o IPAC, MS 100-22, 1200 E.\
California Blvd., California Institute of Technology, Pasadena, CA 91125, USA}
\author{K.~Blood}
\affiliation{Crescenta Valley High School, 2900 Community Ave, La Crescenta, CA 91214 USA}
\affiliation{NASA/IPAC Teacher Archive Research Project (NITARP), c/o IPAC, MS 100-22, 1200 E.\
California Blvd., California Institute of Technology, Pasadena, CA 91125, USA}
\author{E.~Kerman}
\affiliation{Crescenta Valley High School, 2900 Community Ave, La Crescenta, CA 91214 USA}
\affiliation{NASA/IPAC Teacher Archive Research Project (NITARP), c/o IPAC, MS 100-22, 1200 E.\
California Blvd., California Institute of Technology, Pasadena, CA 91125, USA}
\author{J.~Long}
\affiliation{J.L. Mann High School, Academy of Mathematics, Science, and Technology,160 Fairforest Way, Greenville, SC 29607 USA}
\affiliation{NASA/IPAC Teacher Archive Research Project (NITARP), c/o IPAC, MS 100-22, 1200 E.\
California Blvd., California Institute of Technology, Pasadena, CA 91125, USA}
\author{N.~Roosa}
\affiliation{Crescenta Valley High School, 2900 Community Ave, La Crescenta, CA 91214 USA}
\affiliation{NASA/IPAC Teacher Archive Research Project (NITARP), c/o IPAC, MS 100-22, 1200 E.\
California Blvd., California Institute of Technology, Pasadena, CA 91125, USA}

\begin{abstract}

IC~417 is in the Galactic Plane, and likely part of the Aur OB2
association; it is $\sim$2  kpc away. Stock~8 is one of the densest
cluster constituents; off of it to the East, there is a `Nebulous
Stream' (NS) that is dramatic in the infrared (IR). We have assembled
a  list of literature-identified young stellar objects (YSOs), new
candidate YSOs from the  NS, and new candidate YSOs from IR excesses.
We vetted this list via inspection of the images, spectral energy
distributions (SEDs), and color-color/color-magnitude diagrams. We
placed the 710 surviving YSOs and candidate YSOs in ranked bins,
nearly two-thirds of which have more than 20 points defining their
SEDs. The lowest-ranked bins include stars that are confused, or
likely carbon stars. There are 503 in the higher-ranked bins; half are
SED Class III, and $\sim$40\% are SED Class II. Our results agree with
the literature in that we find that the NS and Stock~8 are at about
the same distance as each other (and as the rest of the YSOs), and
that the NS is the youngest region, with Stock~8 a little older. We do
not find any evidence for an age spread within the NS, consistent with
the idea that the star formation trigger came from the north. We do
not find that the other literature-identified clusters here are as
young as either the NS or Stock~8; at best they are older than
Stock~8, and they may not all be legitimate clusters.

\end{abstract}

\section{Introduction}

\label{sec:intro}

IC~417 (also LBN 173.46-00.16 and SH 2-234) is a young cluster in the
Galactic Plane, essentially in the direction of the Galactic
anti-center ($l$=173.38\arcdeg, $b=-$00.20\arcdeg), and $\sim$2 kpc
away. It has been thought to be part of the Aur OB2 association,
though evidence is mixed (see, e.g., Marco \& Negueruela 2016 and
references therein).  

\begin{figure}[htb!]
\epsscale{1.0}
\plotone{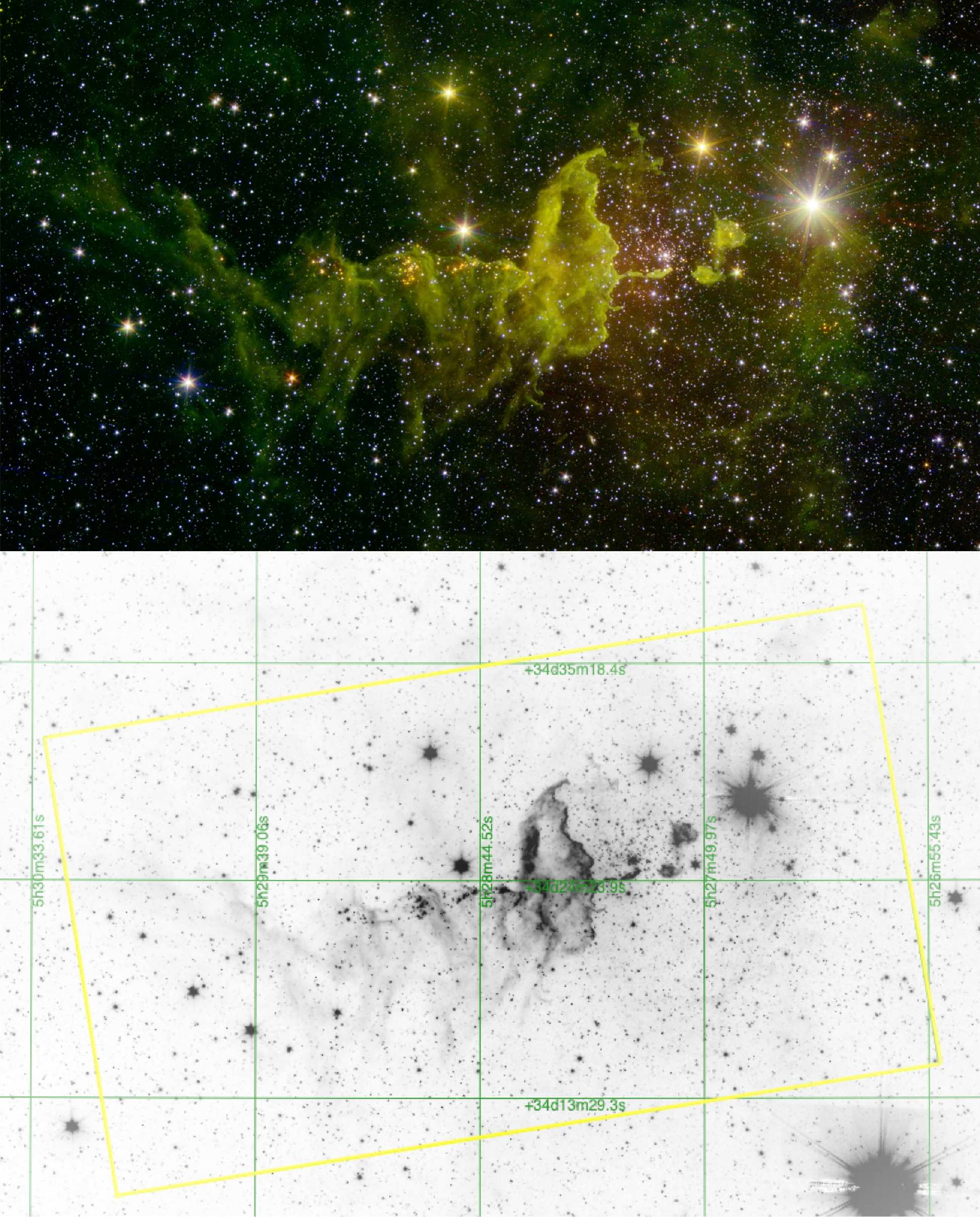}
\caption{ Top: three-color image of the heart of IC~417. Red=4.5 \mum\ (from
Spitzer/IRAC), green=3.6 \mum\ (from Spitzer/IRAC), and blue=1.3 \mum\
($J$ band, from 2MASS).  This image is a few degrees counter-clockwise
of North-up, and it is about 0.5$\arcdeg$ across.  Stock~8 is the
cluster at the center-right, creating a `bowl' in the green
nebulosity. Clusters of red objects can be seen in the `nebulous
stream' in the center-left of this figure. (Image credit:
NASA/JPL-Caltech, sig16-008.) Bottom: IRAC-2 (4.5 \mum) image from
GLIMPSE360 in reverse greyscale, north-up, showing sexigesimal 
coordinates (green) and the footprint of the media image (yellow).
Also see next Figure. }
\label{fig:where1}
\end{figure}

IC~417 has gained some notoriety in astrophotography circles; when
combined with NGC 1931 (to its southeast; not shown or considered
here), it makes for a dramatic image\footnote{See, \eg,
http://apod.nasa.gov/apod/ap061027.html or\\
https://slate.com/technology/2013/12/ic-417-star-forming-nebula-astrophoto.html},
and is sometimes called ``the Spider and the Fly."  There is a
relatively dense cluster, called Stock~8 (Stock 1956), apparent in
optical images, which appears to be nestled in a bubble of nebulosity 
particularly obviously in the infrared (IR); there is nebulosity
extending from it off to the East in a region called the ``Nebulous
Stream'' (Jose \etal\ 2008; hereafter J08), which we abbreviate
``NS.'' In the mid-infrared, it is particularly impressive; see
Figure~\ref{fig:where1}, where there are clusters of red objects
apparently embedded in the nebulosity. It is not entirely clear what
the sequence of star formation is in this region, for example, how (or
even whether) the star formation has been triggered (see, e.g., 
J08; Camargo \etal\ 2012; Dewangan \etal\ 2018). Towards that end,
it is useful to identify the cluster members.

Young stellar objects (YSOs)\footnote{We use the term YSO here to
encompass young objects all the way from SED Class I, where the
forming star is still surrounded by a substantial cocoon of matter,
through H ignition, where the star is still young although on
the main sequence; also see Appendix~\ref{app:textbook}.} 
can be identified from infrared (IR) excess from a
circumstellar disk; ultraviolet (UV) or even blue excess from
accretion; H$\alpha$ excess from accretion and/or coronal emission;
variability at nearly any wavelength; and/or from clustering on the
sky. All of these methods have been used to identify YSO candidates in
this region, but relatively few studies have focused narrowly just on
IC~417 and the region immediately surrounding it.  Since IC~417 is in
the Galactic Plane, it has been serendipitously observed by many
surveys, but few articles have pulled together data from a variety of
optical and IR instruments and focused on the stellar population as we
do here.  We now summarize what work has been done to date in this
region.

The distance to IC~417 has fluctuated near $\sim$2 kpc. Mayer \& Macak
(1971) estimated 2.97 kpc;  Fich \& Blitz (1984) deduced a kinematic
distance of 2.3 $\pm$0.7 kpc. Malysheva (1990) estimated 1.897 kpc,
the closest distance estimate available. Mel'Nik \& Efremov (1995)
derived  2.68 kpc. J08 obtained 2.05$\pm$0.1 kpc, with the most
detailed analysis to that point, based on optical and infrared data.
Camargo \etal\ (2012) obtained 2.7 kpc, and placed it in the near side
of the Perseus arm, with stellar ages $<$10 Myr. Marco \& Negueruela
(2016; hereafter MN16) estimated that the stars were 4-6 Myr and at a
distance of 2.80$^{+0.27}_{-0.24}$ kpc. Finally, Dewangan \etal\
(2018) estimate 2.8 kpc. Several authors (including J08; Camargo
\etal\ 2012; Dewangan \etal\ 2018) have attempted to assemble a story
of star formation across the region, with most concluding that there
is some sequential and/or triggered star formation here, but the
details of this have been difficult to pin down due to varying
distance estimates for subsets of this region (including the larger
population of structures thought to be part of Aur OB2), and the
likelihood that particularly in and around IC~417, the age uncertainty
is likely comparable to any age gradient (Camargo \etal\ 2012).  Most
of the stars we consider in this paper are thought to be at about the
same distance. We have adopted a distance estimate of $\sim$2 kpc, and
accept as likely members anything between 1 and 3 kpc, but
acknowledge that there is still uncertainty in this distance.
(Also see App.~\ref{app:distances}.)

\begin{figure}[htb!]
\epsscale{1.0}
\plotone{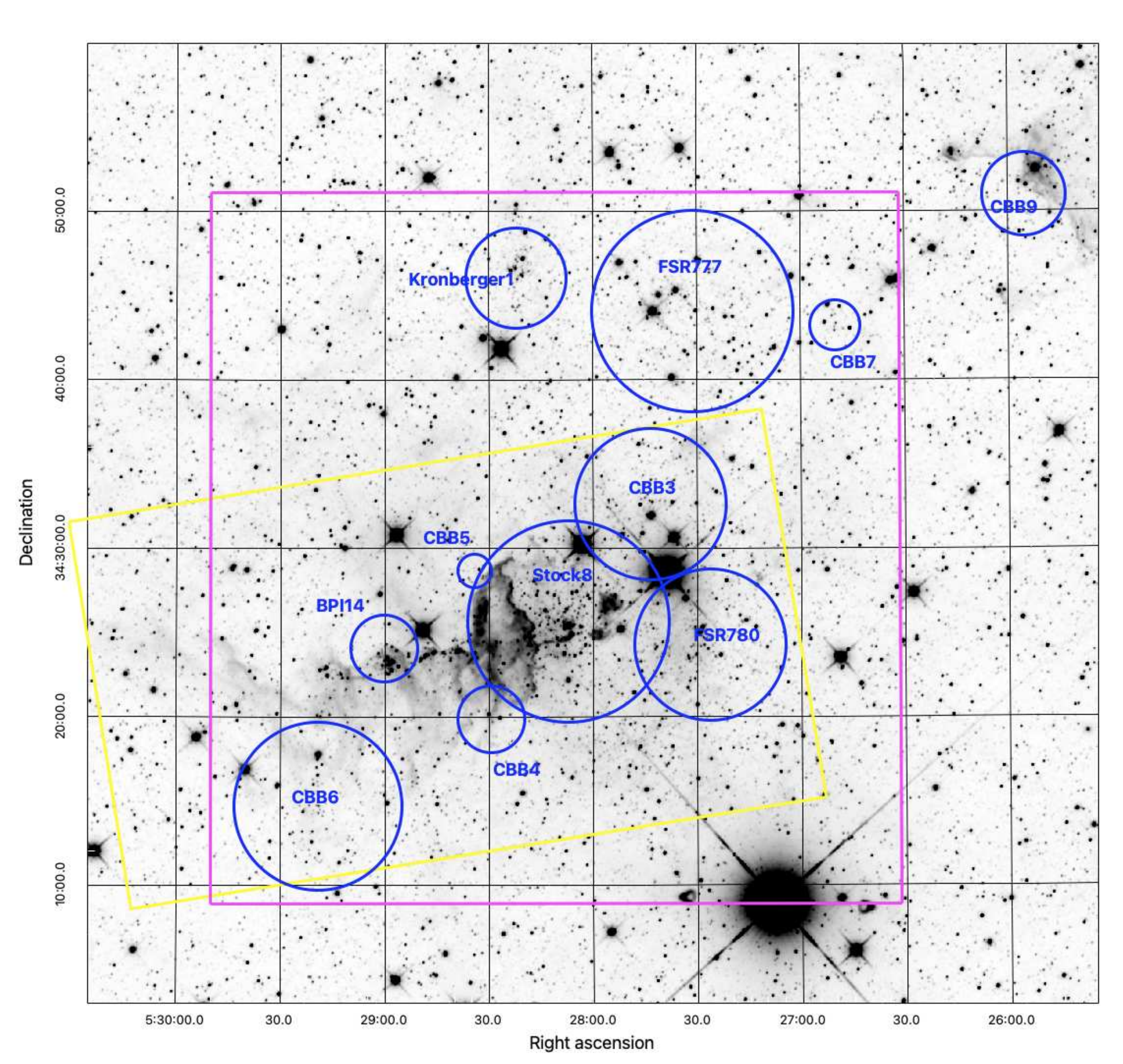}
\caption{WISE-2 (4.5 \mum; from unWISE) reverse greyscale image of
our region. This image is $\sim$1$\arcdeg$ on a side, centered on
5:28:00 34:30:00 (J2000).  The clusters from Figure 19 in C12 are
included as indicated; see the text for more information. 
The clusters are listed in Table~\ref{tab:clusters}. 
The yellow polygon indicating the footprint from 
Fig.~\ref{fig:where1} is also shown for reference.
The field we have focused upon is the magenta box, running from 
$\alpha,\delta$=05:26:31.5,+34:08:50.6 (the lower right
corner of the magenta box)  to 05:29:50,+34:51:05 
 (the upper left corner of the magenta box). 
(All coordinates are J2000).  }
\label{fig:where2}
\end{figure}

\begin{deluxetable}{lccp{6.5cm}}
\tabletypesize{\scriptsize}
\tablecaption{Literature Clusters in or near IC~417\label{tab:clusters} (also see Fig.~\ref{fig:where2})}
\tablewidth{0pt}
\tablehead{\colhead{Name} & \colhead{Center (J2000)} & \colhead{Approx.~Radius ($\arcsec$)}& 
\colhead{Notes} }
\startdata
CBB 9   & 05:25:55 +34:50:54 & 150 & Camargo \etal\ (2012); outside of the region considered in the rest of the paper \\
CBB 7   & 05:26:50 +34:43:10 & 90 & Camargo \etal\ (2012) \\
FSR 777 & 05:27:31 +34:44:01 & 360 & Froebrich \etal\ (2007); also Alicante 11 (MN16)\\
FSR 780 & 05:27:26 +34:24:12 & 270& Froebrich \etal\ (2007)\\
CBB 3   & 05:27:43.3 +34:32:36 & 270& Camargo \etal\ (2012)\\
Stock~8 & 05:28:07 +34:25:38 & 360& Stock (1956) \\
Kronberger 1 & 05:28:22 +34:46:01 & 180& Kronberger \etal\ (2006); also Alicante 12 (MN16)\\
CBB 4   & 05:28:29.3 +34:19:50 & 120& Camargo \etal\ (2012)\\
CBB 5   & 05:28:33.9 +34:28:37 & 60& Camargo \etal\ (2012)\\
BPI 14  & 05:29:00 +34:24:00 & 120 &  Borissova \etal\ (2003); also CC 14 (Ivanov \etal\ 2005); within the NS\\
CBB 6   & 05:29:19 +34:14:41.4 & 300& Camargo \etal\ (2012)\\
\enddata
\end{deluxetable}

\begin{figure}[htb!]
\epsscale{1.0}
\plotone{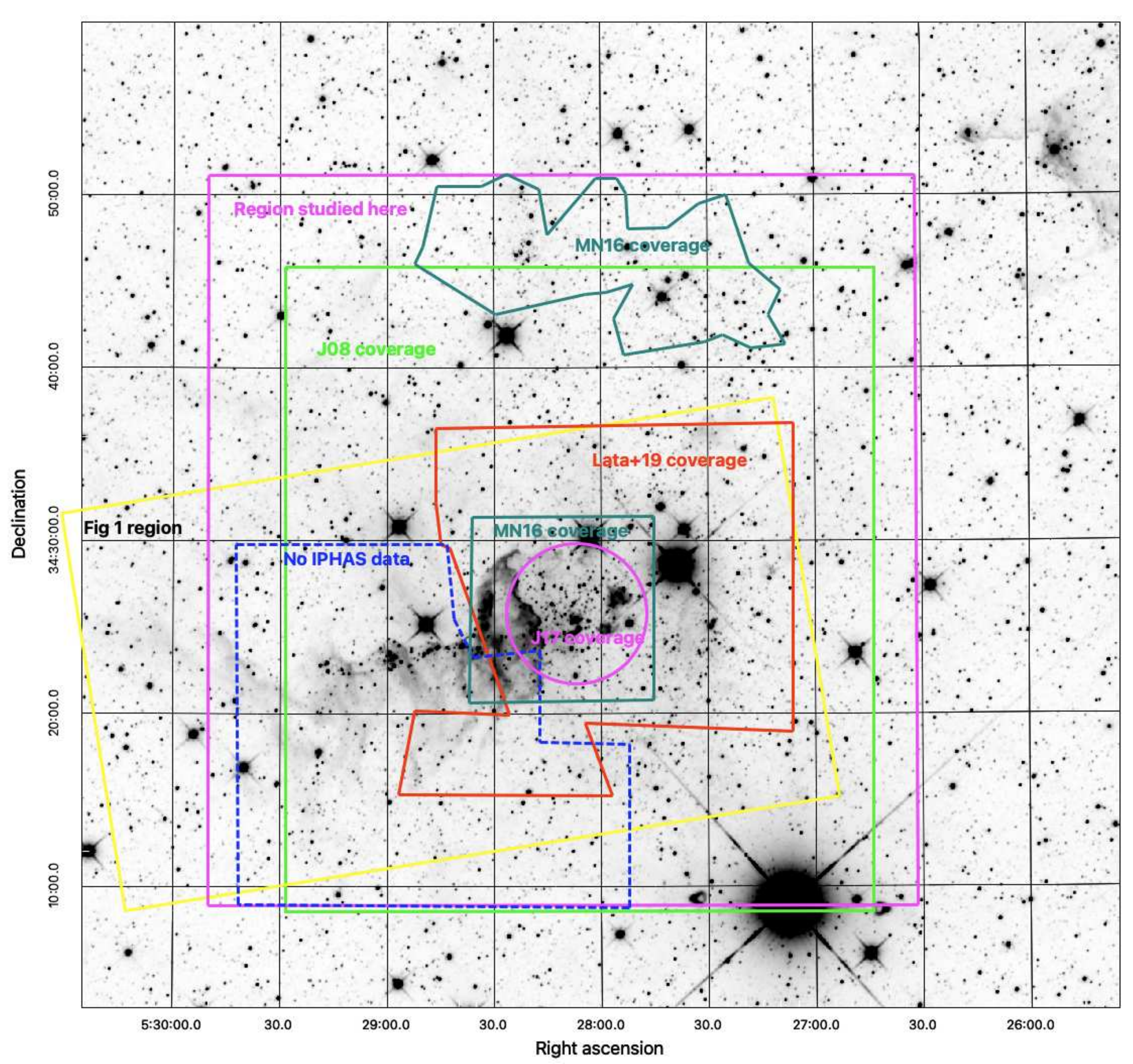}
\caption{WISE-2 (4.5 \mum) reverse greyscale image 
(as in Fig.~\ref{fig:where2}) of our region, with yellow polygon
(coverage of press image from Fig.~\ref{fig:where1}) and 
large magenta square (our region of study) included for reference. 
The dark blue irregular polygon with dashed lines is the region that is
NOT covered by IPHAS; note that it includes most of the NS.  Green is
J08 coverage; small teal square plus the irregular teal polygon to
north is MN16;  magenta circle is Jose \etal\ (2017); complex red
polygon is Lata \etal\ (2019); UKIDSS coverage extends down to
declination$\sim$34.25$\arcdeg$, which is about the lower edge of the
red Lata \etal\ (2019) polygon. Many studies have focused on Stock~8
(see also Fig.~\ref{fig:where2}); most of our work here is on the
larger area, with a particular focus on the NS (see next Figure). }
\label{fig:wherecoverage}
\end{figure}

\begin{figure}[htb!]
\epsscale{1.0}
\plotone{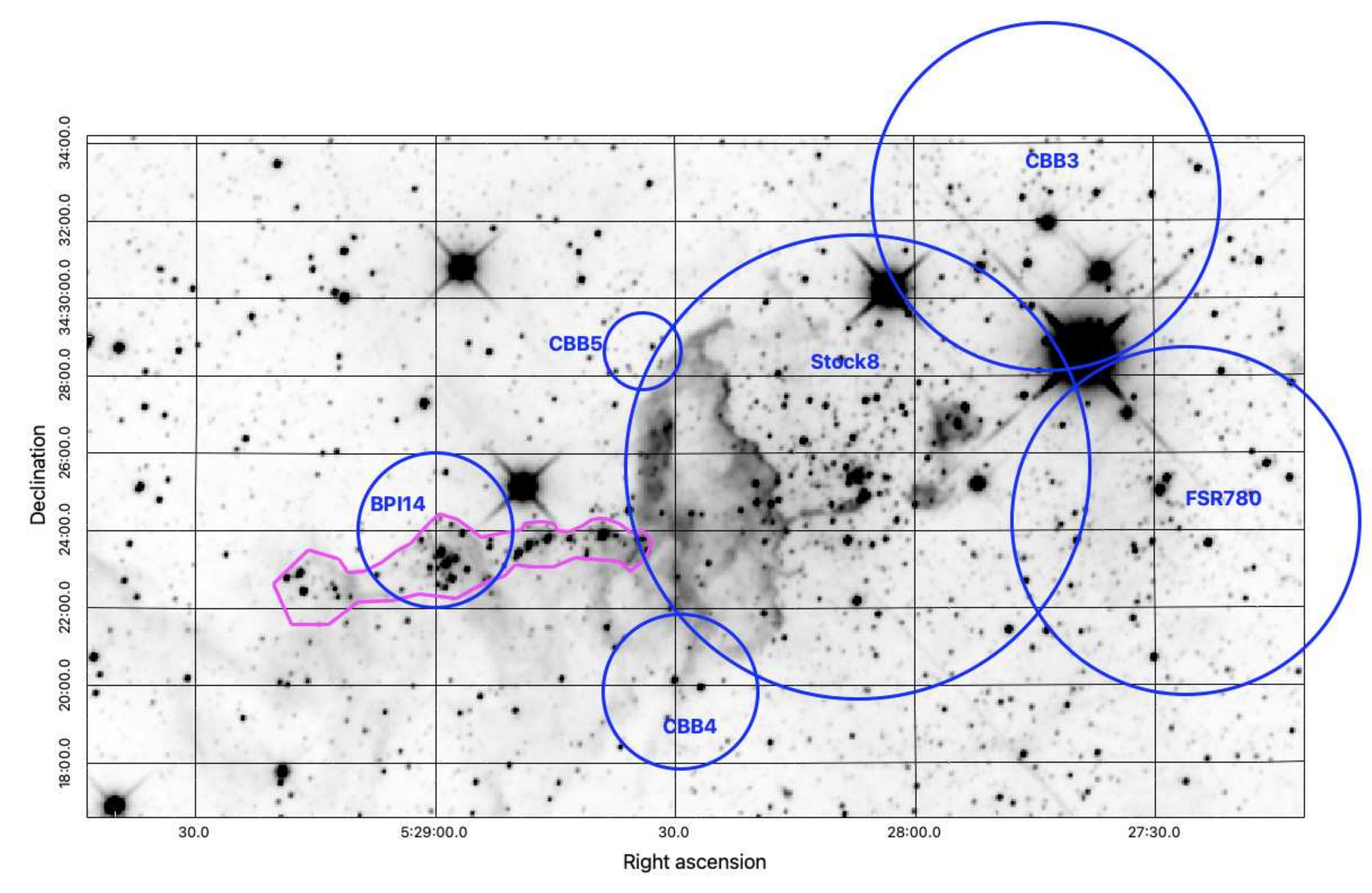}
\caption{WISE-2 (4.5 \mum) reverse greyscale image (from unWISE) of
our region, zoomed in on the Nebulous Stream
(NS). The blue clusters are as in  Fig.~\ref{fig:where2}, and only
include the clusters entirely or mostly within this image.  
The magenta polygon
is empirically derived here to include the regions of nebulosity and
obviously red stars in Fig.~\ref{fig:where1}.}
\label{fig:wherens}
\end{figure}

The earliest work explicitly on the stars in this region dates from
the 1970s-1980s and largely consists of identification of the
brightest stars (Georgelin \&  Georgelin et S.\ Roux 1973; Vetesnik
1978; Efremov \&  Sitnik 1988; Chargeishsvili 1988).  Malysheva (1990)
also identified the brightest stars, and concluded the stars here were
$\sim$12 Myr old. The brightest stars here are largely O and B stars,
but include OP~Aur, a carbon star.  Additionally, Kohoutek \& Wehmeyer
(1999) noted some H$\alpha$-bright stars here.

The next significant advance in this region was with the release of the Two
Micron All-Sky Survey (2MASS; Skrutskie \etal\ 2006). In the process
of identifying clusters in various regions of the Galactic Plane,
several authors used 2MASS-derived star count data to identify
clusters in the IC~417 region.  All the literature clusters in the IC~417
region are identified in Figure~\ref{fig:where2} and
Table~\ref{tab:clusters}.   Bica \etal\ (2003) called out the entire
region, listing it as identical to IC~417. Borissova \etal\ (2003)
identified BPI 14 (see Fig.~\ref{fig:where2} and
Table~\ref{tab:clusters}). Ivanov \etal\ (2005) also called attention
to BPI 14, labeling it CC 14.  Kronberger \etal\ (2006) contributed
the cluster marked Kronberger 1 in Fig.~\ref{fig:where2}. Froebrich
\etal\ (2007)  identified FSR 777 and 780. Camargo \etal\ (2012)
identified the clusters tagged ``CBB'' in Fig.~\ref{fig:where2} (and
Table~\ref{tab:clusters}), in addition to confirming clusters from the
literature, and providing the regional map on which our
Fig.~\ref{fig:where2} is primarily based (see their Fig.~19).  They 
find that all of these clusters are under 10 Myr old, all associated
with each other, and most likely in the Perseus arm. An independent
analysis by MN16 identified some of the same clusters from Camargo
\etal\ (2012), to which they added additional deep Str\"omgren optical
and new $JHK_s$ photometry, and some classification spectra (see
Fig.~\ref{fig:wherecoverage} and Table~\ref{tab:clusters}). They
confirmed clusters FSR 777 (which they call Alicante 11) and
Kronberger 1 (which they call Alicante 12).  They identify an O8 star
(HD 35633) as the source of the ionizing photons carving out the
`shell' or `bowl' around Stock~8 (Figs.~\ref{fig:where1} or
\ref{fig:where2}), and an SB2 with integrated type O8
(BD+34$\arcdeg$1058) as the ionizing source to the north of the NS.
They also reassessed the age and distance of the clusters in this
region, finding 4-6 Myr and  $\sim$2.80$^{+0.27}_{-0.24}$ kpc,
respectively. They suspect that it is in the Perseus arm, though
possibly on the near side of it, noting that the location of the arm
is not well-defined in this region.  Interestingly, they find that the
NS is not directly associated  with Stock~8. 

J08 was the first extensive survey of IC~417 itself (as opposed to as
an additional cluster in a set of many). They included 2MASS data and
deep optical ($UBVI_c$) imaging, and were primarily focused on the
cluster Stock~8, which appears to sit within the ``bowl'' of
nebulosity in Figs.~\ref{fig:where1} or \ref{fig:where2}.  They
identified near-infrared (NIR) excess and H$\alpha$ excess sources as
candidate YSOs. They derived a distance of 2.05$\pm$0.1 kpc, and ages
of 1-5 Myr. J08 is the first paper to identify the sinuous structure
that is so obvious in the IRAC bands  in Fig.~\ref{fig:where1}. They
found it in the NIR and dubbed it the ``Nebulous Stream.'' 
J08 conclude that the young clusters in the NS are at the same
distance as Stock~8, but not yet affected by the star formation
activity in Stock~8; instead, their formation was likely triggered by
an O8 star to the North.  The most prominent sub-cluster of the NS was
identified as BPI 14 = CC 14 (Fig.~\ref{fig:where2} \& 
Table~\ref{tab:clusters}).  

Jose \etal\ (2017; hereafter J17) returned to Stock~8, analyzing the
initial mass  function of Stock~8 with deep optical, near-IR, and
mid-IR data. They identified a ``large, irregular cavity'' at 350
\mum\ and at 12 \mum; IC~417 is at the southern edge of this cavity. 
They posit that the early-type stars identified in MN16 are creating
this cavity, and that they have triggered star formation here, though
not necessarily in Stock~8.  They find that the NS is likely to be
younger than Stock~8.

Dewangan \etal\ (2018) had an extensive discussion of the filamentary
structures in IC~417 (which they call S234) as well as other clusters
nearby in projected distance (which may or may not be part of Aur
OB2).  While largely focused on far-IR ($\geq$70 \mum) and radio
wavelengths  and the distribution of gas/dust, a section of the paper
includes figures with ``selected YSOs,'' however, no data table of the
YSOs was provided. These authors are attempting to deduce the sequence
of star formation in this region, and it is complicated at least in
part because of the variety of estimated distances to the
substructures of the complex. They also analyze the
NS; they break the NS into two pieces. In  Fig.~\ref{fig:where1},
their ``ns1'' is the portion we consider to be the entirety of the NS,
with the four subclusters of apparently red objects  and the `sinuous'
texture of the nebulosity, parallel to the image orientation; their
``ns2''  is the far less prominent (more diaphanous) structure at
about a  45$\arcdeg$ angle on  the left. Dewangan \etal\ (2018) find
far more interesting behavior in ns1; ns2 does not appear to be
forming stars, whereas ns1 is forming, by their estimate, $\sim$80
YSOs (substantially fewer than we estimate here). 

Lata \etal\ (2019) presented a variability analysis of
stars in Stock~8, finding more than 100 short-period variables. They
attribute many of the periodic signals they find to pulsation; they 
determine the age of their pre-main-sequence periodic variables  to be
$\lesssim$5 Myr. No analysis of the non-periodic variables is included
in that paper.  We chose to be more expansive and investigated all 130
periodic variables as possible YSO candidates, as opposed to just those
identified as YSO candidates in Lata \etal\ (2019). 

Pandey \etal\ (2020) explored star formation on a large scale in the
larger Auriga region, of which IC~417 was just a part. They identified
YSOs based on IR excess, finding two large bubble structures in the
nebulosity and the YSO distribution;  IC~417 is on the southern edge
of one of their bubbles. They find far fewer YSOs North of IC~417
(within the bubble) than in or around it.

If the stars in the IC~417 region (including Stock~8, the NS, and 
the clusters discovered by star counts) are really
$\sim$10 Myr old or as young as $\sim$3 Myr as some have claimed, 
at least $\gtrsim$10\% of the member stars here should
still have substantial circumstellar dust disks (e.g., Mamajek 2009).
Exploring the disk fraction on the whole and as a function of location
in this region may be able to provide constraints on the age (or age
spread) of the clusters here. It is therefore worth looking for new YSO
candidates based on IR excess. The stars in the NS are visibly red in
Fig.~\ref{fig:where1}. Given that there are numerous optical data sets
available in this region, we should be able to determine if these stars are
red primarily due to interstellar extinction or due to circumstellar
dust. It is also worth exploring the clusters within the NS; four are
evident by eye.  
 
In this work, we collect the YSO candidates identified in the
literature in this region, add objects apparently coincident
with the NS, and add to that list new objects selected based on
WISE+2MASS IR colors. We then investigate the optical+IR properties of this
unified list of YSO candidates. We focus on a region from
$\alpha,\delta$=05:26:31.5,+34:08:50.6 to 05:29:50,+34:51:05 (see
Fig.~\ref{fig:where2}; these coordinates are the lower right and 
the upper left corners of the magenta box in Fig.~\ref{fig:where2}) 
because it covers most of the clusters
identified here. We explore, where possible, the properties of
the literature-identified clusters.

In Section~\ref{sec:data}, we present all the archival data we used,
and how we merged the catalogs. Section~\ref{sec:ysocand} describes
the assembly of the list of YSO candidates from the literature,
by position in the NS, and via selection based on IR excess using
2MASS+WISE. Section~\ref{sec:vettingysos} goes into detail of how
we vetted the YSO candidates, using image inspection, SED inspection, 
color-magnitude and color-color diagram inspection, and our procedure
for final ranking of the YSO candidates. In 
Section~\ref{sec:ensembleysos}, we describe many properties of the
entire ensemble of YSO candidates, and Section~\ref{sec:obvclusters}
delves into more detail about Stock~8 and the NS. Section~\ref{sec:obc}
considers the OB and carbon stars in this region, and 
Section~\ref{sec:2massclusters} explores the clusters found via
2MASS star counts. Section~\ref{sec:summary} summarizes our main results. 

The Appendix has a variety of supporting information including
more information on the distance to IC~417 (App.~\ref{app:distances}), 
more on how we matched sources across catalogs (App.~\ref{app:srcmatch}),
background information on how and where to find young stars in various 
color-color and color-magnitude diagrams (App.~\ref{app:textbook}), and 
then detailed examples of how we ranked a dozen sources out of our final YSO 
candidate list (App.~\ref{app:briefexamples}).

\begin{rotatetable}
\begin{deluxetable}{p{2cm}p{2.5cm}cp{2.5cm}ccccp{5cm}}
\movetableright=1mm
\tabletypesize{\scriptsize}
\tablecaption{Archival Data in IC~417\label{tab:datalist}}
\tablewidth{0pt}
\tablehead{\colhead{Band} & \colhead{Wavelengths} & \colhead{Resolution}& \colhead{Limiting
mag\tablenotemark{a}} & \colhead{Origin}& \colhead{Catalog}& \colhead{YSO Candidate}& 
\colhead{Color/symbol}& 
\colhead{Notes} \\[-0.3cm]
& \colhead{(\mum)} & \colhead{($\arcsec$)}& \colhead{(mag)}& &\colhead{Fraction\tablenotemark{b}} &\colhead{Fraction\tablenotemark{c}}& \colhead{in SEDs\tablenotemark{d}}}
\startdata
$UBVI_c$ & 0.36, 0.44, 0.55, 0.79 & $\sim$3 & 17, 20, 19, 17 & J08 & 15\% & 52\% & black +& Published entire
catalog, not just cluster members. Partial coverage, 
green box in Fig.~\ref{fig:wherecoverage}.\\
$uvby\beta$  & 0.34, 0.41, 0.47, 0.55, 0.66 & $\sim$0.7 & 17, 16, 15, 15, 2.8&
MN16 & 0.5\%& 10\% & purple + & Primarily bright stars. Partial coverage, cyan polygons in Fig.~\ref{fig:wherecoverage}.\\
$grizy$ & 0.481, 0.617, 0.752, 0.866, 0.926 & $\sim$0.6 & 22, 22, 20, 19.5, 19  & Pan-STARRS &60\% & 
84\% & cyan $\Diamond$ & Covers whole region.\\
$VI_c$ & 0.44, 0.79 & $\sim$0.7 & 21, 19 & J17 & 3\% & 29\% & black + &  Partial coverage, magenta circle in Fig.~\ref{fig:wherecoverage}.\\
$r^{\prime},i^{\prime}$, H$\alpha$ & 0.624, 0.774, 0.656 & $\lesssim$0.9 &  18.7, 17.7,
15.3 & IPHAS & 32\% & 44\% & yellow $\Diamond$& Complex coverage, region that is NOT covered is the dark blue polygon in Fig.~\ref{fig:wherecoverage}.  \\
$G_{BP},G_p,G_{RP}$ & 0.511, 0.622, 0.777\tablenotemark{e} & $\sim$0.4 &  20, 20, 19 & Gaia DR2, DR3 & 45\%& 67\% & green $\Box$ & Covers whole region; 26\% have parallaxes from DR2 and 41\% from DR3; 32\% have
distances from Bailer-Jones \etal\ (2018), and 41\% from Bailer-Jones \etal\ (2021).\\
$JHK_s$ & 1.248, 1.631, 2.201 & $\sim$1 & 16.5, 16, 15.5 & UKIDSS & 48\% & 85\% & red $\Diamond$ & Covers most of 
the region, down to $\sim$34.25.\\
$JHK_s$ & 1.235, 1.662, 2.159  & $\sim$1.6& 16.7, 15.8, 15.4 & 2MASS & 24\% & 71\% & black $\Diamond$& Covers whole region.\\
$JHK_s$ & 1.248, 1.631, 2.201 & $\sim$2& 16, 15.5, 15 & MN16 & 1\% & 17\% & purple + & Partial coverage, cyan polygons in Fig.~\ref{fig:wherecoverage}.\\
IRAC-1,2 & 3.6,  4.5 & $\sim$1.6 & 15.2, 15.2 & GLIMPSE  & 54\% & 91\% &  black $\bigcirc$ & highest spatial
resolution and sensitivity available at these bands. Covers whole region. If source 
visible in image but not in GLIMPSE catalog, photometry taken from J17; 
yellow circles) 
or done anew here. \\
WISE-1,2,3,4 & 3.4, 4.6, 12, 22  & $\sim$6-12 & 14.6, 14.9, 12.3, 8.9 & WISE & 73\% & 58 \% &
black $\star$ & 
AllWISE release, plus CatWISE (blue $\Box$) and unWISE (green +). Covers whole region. \\
MSX B1,B2, A, C, D, E & 4.29, 4.35, 7.76, 11.99, 14.55, 20.68 & $\sim$9-15 & \ldots & MSX catalog & $<$0.08\% & 1.4\% & cyan $\Box$ & Too few points to assess limiting mag here ($<$40 sources). Covers whole region.\\
AKARI IRC, FIS & 9, 18; 65, 90, 140, 160 & $\sim$2 & \ldots & AKARI IRC, FIS & $<$0.05\% & 1\% &  yellow $\times$ &
Too few points to assess limiting mag here ($<$25 sources). Covers whole region.\\
PACS 70,160 & 70, 160 & 5.6, 10.7 & \ldots & PACS PSC & $<$0.05\% & 0.7\% & green $\Box$ & Too few points to assess limiting mag here ($<$25 sources). Covers whole region.
\enddata
\tablenotetext{a}{Empirical limiting magnitude, \eg, a histogram of the observed
magnitudes at this band in this region peaks at about this value.}
\tablenotetext{b}{Out of the entire catalog of $\sim$46,000 sources,
what fraction has a counterpart in the catalog given? Ex: 15\% of the entire
catalog has a $UBVRI$ counterpart from J08; 45\% have a counterpart from Gaia.}
\tablenotetext{c}{Out of the catalog of 710 YSO candidates,
what fraction has a counterpart in the catalog given? Ex: 52\% of the YSO candidates
have a $UBVRI$ counterpart from J08; 67\% from Gaia.}
\tablenotetext{d}{Color/symbol used for these data in SEDs later in this paper and in the IRSA delivery.}
\tablenotetext{e}{Gaia DR2 wavelengths for $G_{BP},G_p,G_{RP}$ are 0.532, 0.673, 0.797 \mum, respectively; the wavelengths given in the table are for E/DR3.}
\end{deluxetable}
\end{rotatetable}

\clearpage

\section{Data}
\label{sec:data}

\subsection{Overview}
\label{sec:dataoverview}

In order to look for candidate YSOs in the IC~417 region, we first
assembled data from a wide variety of places, summarized in 
Table~\ref{tab:datalist}, and discussed in this section.  Because IC
417 is in the Galactic Plane, it has been serendipitously observed by
several different surveys.  Several data sets are available over our
entire region (from  $\alpha,\delta$=05:26:31.5,+34:08:50.6 to
05:29:50,+34:51:05; see Fig.~\ref{fig:where2}), and some data are
available only over a portion of the field (see
Fig.~\ref{fig:wherecoverage}). We kept track of the sources
identified in the literature as YSOs or YSO candidates where relevant.
In practice, we started with 2MASS to establish a reliable coordinate
system and then found matches by position with sources from both
shorter and longer wavelengths (see Sec.~\ref{sec:datamerging}). 
Sources that were optical-only were
not often retained unless they were listed in the literature as
possible or confirmed YSOs.

All of these data were combined (bandmerged) initially by looking for
matches by position, typically within 1$\arcsec$. Most of these
catalogs have very good positions, so a larger radius is not
required, except where specified below. After merging the catalogs,
there are $\sim$46,000 objects, with data included from 0.34 to 160 \mum.
We used these catalogs to create spectral energy distributions (SEDs);
see Sec.~\ref{sec:vettingseds}. By checking the SEDs, we can identify
sources that are incorrectly bandmerged, because in those cases, 
the SEDs have obvious discontinuities. Those sources were given special 
attention and manually matched to the correct source as needed.

\subsection{Optical}
\label{sec:dataoptical}

\subsubsection{Pan-STARRS}

Data from the Panoramic Survey Telescope and Rapid Response System 
(Pan-STARRS) DR1 (Chambers \etal\ 2016) were pulled from the Mikulski 
Archive for Space Telescopes (MAST) for this region. This survey 
covers this region in five bands ($grizy$) with a spatial resolution 
of $\sim$0.6$\arcsec$. In this region, this survey reaches 
$\sim$20th mag in most bands; see Table~\ref{tab:datalist}. We have
Pan-STARRS counterparts for $\sim$60\% of the sources in our master
catalog of this region.

\subsubsection{Gaia}

Gaia DR2 (Gaia Collaboration 2018ab), 
EDR3 (Gaia Collaboration \etal\ 2021a,b), 
and DR3 (Gaia Collaboration 2022ab)  
data were obtained for this region via the Infrared Science
Archive (IRSA; https://irsa.ipac.caltech.edu). The bands for this are
$G, G_{BP},$ and $G_{RP}$; data go to $\sim$20th mag here.  
The effective
wavelengths are slightly different between the two data releases.  
Gaia DR2 wavelengths for $G_{BP},G_p,G_{RP}$ are 0.532, 0.673, 
\& 0.797 \mum, respectively, and E/DR3 wavelengths are 0.511, 0.622, 
\& 0.777 \mum.
We have Gaia DR2
parallaxes for 26\% of the master catalog, and 41\% from DR3. Since 
some of the measured parallaxes are negative, we used distances from 
Bailer-Jones \etal; $\sim$32\%
have distances from Bailer-Jones \etal\ (2018), and $\sim$41\% from 
Bailer-Jones \etal\ (2021). (Also see Appendix~\ref{app:distances}
on distances.) 
Data were collected and merged (by
position) from DR2, EDR3, and DR3 because work for this project
extended over enough years that data from all releases were relevant,
and encompassed slightly different stars. In practice, the 
photometry and distances were matched by position and bookkept 
separately for each delivery, for each source.

\subsubsection{IPHAS}

The INT Photometric H$\alpha$ Survey (IPHAS; Barentsen \etal\ 2014)
covered the Galactic plane in the Northern hemisphere, including the
IC~417 region, in $r^{\prime}, i^{\prime}$, and H$\alpha$.  The
spatial resolution of this survey is $\sim1.1\arcsec$, and it goes to
$\sim$17-19 mags in this region in the broadband filters. These data
as served by VizieR (at least as of the time when we downloaded the
catalog) are missing in a relatively large polygon in the SE of our
field, over most of the NS (Fig.~\ref{fig:wherecoverage}). According
to J.\ Drew (2015, priv.~comm.), this region was not observed on a
photometric night and thus the photometry was not released as part of
DR2.  J.\ Drew kindly directly provided this lower-quality photometry
in 2015,
but for the stars that were detected in IPHAS, there are considerable
other optical data available now, and there is a relatively large
systematic offset (vividly apparent in the SEDs) between this
lower-quality IPHAS data and the rest of the data we have amassed.
Therefore, we did not use these lower-quality IPHAS data. There are
good IPHAS counterparts for $\sim$32\%  of the sources in our catalog.
Seventeen H$\alpha$-bright stars as identified in Witham \etal\ (2008)
from this region were tagged as YSO candidates in our database.

\subsubsection{J08}

J08 published their entire catalog of $UBVI_c$ measurements, not just
those for their candidate cluster members. This allows us to use their
data for investigation of objects other than their candidate cluster
members. Their spatial resolution is about $\sim$1.5$\arcsec$; their
data cover only the central region (see
Fig.~\ref{fig:wherecoverage}).  They include relatively faint objects;
histograms of the measured magnitudes peak at $\sim$17-20 mag.  In
order to match correctly these optical measurements to the 
established catalog, sources were first matched by position to 2MASS,
which revealed several duplicate sources and systematic offsets of 
reported positions of typically 0.5$\arcsec$, but with a long tail  to
larger separations that was strongly dependent on position on the sky.
Individual sources were matched by hand (using tools and procedures
developed in, e.g., Rebull 2015; see Appendix~\ref{app:srcmatch}) 
across surveys and across the field. 
In the end, 15\% of our catalog had data from J08. Candidate cluster
members from J08 were identified as candidate YSOs in our database.  

\subsubsection{MN16}

MN16 published Str\"omgren photometry, as well as spectroscopy
(specifically spectral types) in the central portion of the region and
just north of it (see Fig.~\ref{fig:wherecoverage}).  Their scientific
goals dictated that they focus on the O and B stars, so most of their
objects are bright. Their resolution was $\sim1\arcsec$, and we have
counterparts  from MN16 for just $\sim$6\% of the final catalog. We
incorporated all the reported photometry and spectral types from this
work into our database. Stars that were O and B stars were identified
as YSOs because stars that are that massive are at most a few million 
years old, and stars in the direction of and at the distance of IC~417 
that are a few million years old are YSOs.

\subsubsection{Spectral Types from the Literature}

Spectral types from the literature (Georgelin \& Georgelin et S. Roux
1973; Vetesnik 1978; Malysheva 1990; Efremov \& Sitnik 1988;
Chargeishsvili 1988) were included and all of the O and B stars were 
identified as YSOs in our database, matching by name rather 
than position.

\subsubsection{J17}

J17 returned to Stock~8 (alone; see Fig.~\ref{fig:wherecoverage}),
analysing the IMF of Stock~8  with deep optical data, which we
included here. J.\ Jose kindly provided (priv.\ comm., 2018) the entire
catalog, not just the YSOs. As for J08, these catalogs required a bit
of manipulation to consolidate internal duplicates and adjust the
astrometry to match 2MASS or Spitzer coordinates 
(see Appendix~\ref{app:srcmatch}). Just 3\% of our
catalog has a counterpart from J17. 

\subsubsection{Pandey et al.~(2020)}

Pandey \etal\ (2020) obtained observations in $VI$ over a very large
region in Auriga; these optical observations were largely superseded
by the optical data we already had amassed, so we retained solely
the identification of YSO candidates from Pandey \etal\ (2020).

\subsection{Near Infrared}
\label{sec:datanir}

\subsubsection{2MASS}

In 2MASS (Skrutskie \etal\ 2003, 2006), $JHK_s$ data in this 
region go to $\sim$15-17 mags, with a
spatial resolution of $\sim$1.5$\arcsec$. The infrared data more
easily penetrate the interstellar medium and can reveal stars not
easily detected in the optical bands here. As described above, 2MASS
provided the initial base catalog to which all other catalogs were
merged 
(see Sec.~\ref{sec:datamerging} for more on the merging process). 
However, only $\sim$24\% of our final, multi-wavelength merged
catalog has a 2MASS match. The detections from the 2MASS catalog
were only retained if the data quality flags were not D, E, F, or X. 
Upper limits were retained as such.

\subsubsection{UKIDSS}

The UKIRT Infrared Deep Sky Survey (UKIDSS) Galactic Plane Survey
(Lucas \etal\ 2008) covered this region in $JHK_s$ to slightly fainter
magnitudes than in 2MASS. At 0.8$\arcsec$, UKIDSS has higher spatial
resolution than 2MASS. UKIDSS data are available over most of our
field; its coverage includes the northern 90\% of our field, to
declination $\sim$34.25$\arcdeg$.  About half our sources have UKIDSS
counterparts.

\subsubsection{MN16}

MN16 published new $JHK_s$ photometry in the central portion of the
region and just north of it, focusing on bright objects. The
sensitivity of their $JHK_s$ data is comparable to, if not a little
shallower than, the $JHK_s$ data from other sources. Their resolution
was $\sim1\arcsec$. 

\subsubsection{J17}

J17, working in Stock~8 (alone) included near-IR  data we have already
included here (UKIDSS, 2MASS).

\subsection{Mid- and Far-Infrared}
\label{sec:datamfir}

\subsubsection{WISE}

The Widefield Infrared Survey Explorer (WISE; Wright \etal\ 2010a),
like 2MASS, is an all-sky survey, so the IC~417 region was entirely
included. The survey was conducted in four bands, 3.4, 4.6, 12, and 22
\mum. We primarily used the AllWISE release (Wright \etal\ 2010b), 
which sums up all available data
prior to 2011 February. The catalog reaches much fainter objects in
the 3.4 and 4.6 \mum\ than in 12 and 22 \mum. However, the spatial
resolution is relatively low, $\sim$6.1, 6.4, 6.5, and 12$\arcsec$ for
the four channels, respectively. In this crowded region, the WISE
sources often encompass more than one source seen at the shorter
bands. However, because the Spitzer data (see below) do not go past 
4.5 \mum\ here, WISE is the best available choice for IR data between 
5 and 25 \mum.  The detections from the AllWISE catalog
were retained if the data quality flags were A, B, or C; if the data
quality flag was Z, then the data were provisionally retained with a very
large error bar, 30\% larger than what appears in the catalog.
Upper limits were retained as such.

The AllWISE catalog includes data prior to 2011 Feb, but far more data
have been obtained from 2013 to date in WISE channels 1 and 2 (3.4
and  4.6 \mum). CatWISE (Marocco \etal\ 2021, Eisenhardt \etal\ 2020, 
CatWISE team 2020) and unWISE (Lang 2014;  Meisner \etal\ 2017a,b;
Schlafly \etal\ 2019, Meisner \etal\ 2019)  are both efforts to
include  these more recent data; both use images created by unWISE but
obtain independent photometry. We included catalogs from both CatWISE
and unWISE in our database; both cover the whole region. There is WISE
photometry from at least one origin (AllWISE, CatWISE, unWISE) for
73\% of the catalog.

Additionally, we had one more WISE data reduction.  When we
started this project, co-author Koenig had recently published  papers
with a new approach to identifying YSOs from WISE colors (Koenig \&
Leisawitz 2014 and Koenig \etal\ 2012). Koenig \etal\ (2012) describes
an approach for doing photometry on WISE images, PhotVis. We had the
output of PhotVis (and the YSO color selection) run on the  AllWISE
data. We only used the PhotVis photometry if there was not already
photometry for that source at that band  from AllWISE. 

\subsubsection{Spitzer}

The Spitzer Space Telescope (Werner \etal\ 2004) program called
Galactic Legacy
Infrared Mid-Plane Survey Extraordinare (GLIMPSE; Churchwell \etal\
2009) included the Galactic plane. The original GLIMPSE
survey did not include IC~417, but the 2-band post-cryogen 
continuation of GLIMPSE late in the Spitzer mission called GLIMPSE360
did include this region (GLIMPSE team 2014; Meade \etal\ 2014), such 
that only the two
shortest IRAC channels were used (3.6 and 4.5 \mum). Nonetheless, the
Spitzer data are more sensitive and are much higher spatial resolution
($\sim1.2\arcsec$) than the WISE data, so they are very useful in this
crowded region. GLIMPSE360 data are used in Fig.~\ref{fig:where1}. For
the catalogs, we used the `more complete, less reliable' catalog 
(the GLIMPSE360 Archive);
since we inspected each YSO candidate by hand
(Sec.~\ref{sec:vettingysos}), it is more important that we get
measurements of objects to as faint as possible, knowing that we can
reject the less reliable detections on a case-by-case basis. 

If the source could be seen in the IRAC images, but there was no 
corresponding row in the GLIMPSE360 catalog because it was 
just too faint, we performed standard
aperture photometry on the GLIMPSE360 mosaics, as needed.  (We used
aperture 3 px, annulus 3-7 px, and aperture corrections 1.124 and 1.127,
for IRAC-1 and -2, respectively; IRAC Instrument and 
Instrument Support Teams 2021.) J17 also used mid-IR (Spitzer/IRAC)
imaging data from GLIMPSE360, doing their own photometry, but just in
Stock~8. If no other photometry was available, we used the J17 IRAC
photometry.

\subsubsection{Winston et al.~(2020)}

Winston \etal\ (2020) used GLIMPSE360 data from Spitzer/IRAC combined
with WISE and 2MASS data to select YSO candidates along the Galactic
Plane (e.g., not just in this region). Since we already have the
Spitzer, WISE, and 2MASS data in our database, we simply tagged their
identified YSO candidates in our database.

\subsubsection{AKARI}

We also included AKARI (Murakami \etal\ 2007, AKARI team 2010a)  IRC
data at 9 and 18 \mum\ for stars in this region. AKARI was an all-sky
Japanese mission, but was not  as sensitive as Spitzer, so only a
handful of stars in our region have AKARI IRC counterparts (see
Table~\ref{tab:datalist}).  AKARI data required a 3$\arcsec$
matching radius to find counterparts.

\subsubsection{Other Long Wavelength Data}

There are long wavelength imaging data in this region, including AKARI
FIS (50-180 \mum; AKARI team 2010b), MSX (8-21 \mum; Egan \etal\
2003), and Herschel (Pilbratt \etal\ 2010) PACS (70-160 \mum;
Poglitsch \etal\ 2010;  Marton \etal\ 2017) and SPIRE (250-500 \mum;
Griffin \etal\ 2010)\footnote{The Herschel data in this region were taken as part of 
Hi-GAL (Molinari \etal\ 2010).}. Many of these datasets have been used in the
literature (e.g., J17,  Pandey \etal\ 2020, Dewangan \etal\ 2018).  However, the
long-wavelength data for point sources are prohibitively complicated
for us to use because of the relatively low spatial resolution,
relatively high source surface density, and relatively bright
nebulosity.  We matched our sources to most of these catalogs (all
except SPIRE), usually with large matching radii (AKARI:3$\arcsec$;
MSX:10$\arcsec$; Herschel:2$\arcsec$), but only retained the 
match if the SED made physical
sense, if the given measurements were consistent (e.g., WISE, AKARI,
and MSX all agreed), and/or if the source was not obviously confused. 
Correctly apportioning fractional long-wavelength flux among
nebulosity and individual constituent point sources is beyond the
scope of the present work.  There were very few point sources that had
counterparts in these long-wavelength catalogs.

\subsection{Merging Catalogs}
\label{sec:datamerging}

In order to merge catalogs, we started first with the largest near-IR
catalogs because we knew that they would establish a high-reliability
coordinate system to which we could link the rest of the sources, and 
that we were likely going to be primarily interested in those sources 
detected in the near-IR.  Especially since we were unlikely to be 
interested in very many sources that were detected only in the optical,
we retained (in contrast) relatively few sources that were detected 
only in the optical. Typically, we used 1$\arcsec$ as the matching
radius. All three of the first catalogs (2MASS, GLIMPSE, WISE) should
have high-quality astrometry all on the same coordinate system, and, 
when merged, provide a good anchor for merging the rest of the sources.

We merged catalogs in the following order: 
(1) 2MASS, retaining all detections;
(2) GLIMPSE360 Archive (more complete but less reliable catalog), retaining all detections;
(3) AllWISE, retaining all detections;
(4) CatWISE, retaining all detections;
(5) unWISE, retaining all detections;
(6) WISE data from PhotVis and X.~Koenig, retaining all detections; 
(7) PanSTARRS, retaining all detections whether or not there was an IR counterpart;
(8) UKIDSS, dropping sources that have no counterpart in the catalog to this point;
(9) IPHAS, dropping sources that have no counterpart in the catalog to this point;
(10) Gaia DR2 \& DR3 and associated distances from Bailer-Jones \etal\ (2018, 2021), 
dropping sources that have no counterpart in the catalog to this point;
(11) Herschel/PACS, dropping sources that have no counterpart in the catalog to this point;
(12) AKARI IRC, dropping sources that have no counterpart in the catalog to this point;
(13) MSX, dropping sources that have no counterpart in the catalog to this point;
(14) J08 data, pre-matched to 2MASS as described above, including ancillary information;
(15) J17, including ancillary information, all of which have matches 
in the assembled catalog to this point;
(16) spectral types and YSO identifications from references not already 
included were then matched by target position. 

What this process means in detail is the following.  We started with
2MASS, then looked for matches between Spitzer/GLIMPSE360 and 2MASS. Targets
that had matches in GLIMPSE360 had their GLIMPSE360 fluxes matched to their
2MASS entries, and then the GLIMPSE360-only sources were added to the
master catalog as new sources (``retaining all  detections''), using
their GLIMPSE360 positions and fluxes. Because the coordinate system is
the same between 2MASS and Spitzer, this is not a significant source
of error. Then, we looked for matches between the merged
2MASS+GLIMPSE360 master catalog and AllWISE. When a match is found, the 
AllWISE fluxes are matched to the master catalog entries. After the
matching, the AllWISE-only sources are added to the master catalog as
new sources using the WISE positions and fluxes (``retaining all
detections'').  The coordinate system is the same among 2MASS/Spitzer
and WISE. The process is repeated 
in the order as specified above, for each of the items 1-6; these are
infrared catalogs, all on the same coordinate system. Item 7 is the 
first optical catalog to be merged, PanSTARRS. For this optical catalog, the same 
process was imposed -- look for matches between PanSTARRS and the 
master catalog, copy the PanSTARRS measurements over to the counterpart's
entry, and finish by 
including the PanSTARRS-only sources into the master
catalog with their PanSTARRS positions and brightnesses. 
However, for catalogs after this point, we discovered empirically 
that the errors imposed by concatenating new sources on different coordinate
systems added more ``noise than signal'' -- e.g., the chances were much 
higher of adding false sources or sources that were offset enough from 
their `true' position such that finding their counterpart by doing blind
position matching to subsequent catalogs was substantially harder. 
Because our scientific interest is largely focused on the IR-bright sources,
we did not retain sources for which there was no match in the master
catalog after item 7 (``dropping sources that have no counterpart 
in the catalog to this point''). The exception to this is any source
that was identified independently as an `interesting' source in the 
literature. Those were explicitly included in the master catalog.

After item 16, there are $\sim$46,000 objects in the merged catalog, including
wavelengths from 0.34 to 160 \mum, with up to 58 measurements at 
up to 47 distinct wavelengths, though few objects have detections across
all bands even just 0.34 to 22 \mum, much less out to 160 \mum.

In order to get a best possible
measurement in the near-IR $JHK_s$ bands, we took first measurements
from 2MASS, then those from UKIDSS, then from MN16.  
We used (and tabulated) the best possible $JHK_s$ detections for 
calculations as described below; values from 2MASS, 
UKIDSS, and MN16 are also tabulated separately in our catalog.

As discussed in Sec.~\ref{sec:vettingseds} below, when we were
checking each source's images and SED, if there were unphysical discontinuities,
we returned to the merging process, unlinked incorrect matches, and
made correct matches where possible. In some cases, it was not obvious
which data set was wrong, and in those cases, we left them all tied to
the source; those will be obvious to the reader when inspecting the SEDs, but
they include those with rank `1r' or `4*' (see Sec.~\ref{sec:finalrankings}).

\section{Identifying YSO candidates}
\label{sec:ysocand}

\subsection{Overview}
\label{sec:ysooverview}

We assembled our list of YSOs and YSO candidates primarily in three
ways: (1) literature lists of YSOs/candidates; (2) selection by 
position in the NS; (3) selection by IR excess using 2MASS+WISE.
This section describes each of these approaches.

\begin{figure}[htb!]
\epsscale{1.0}
\plottwo{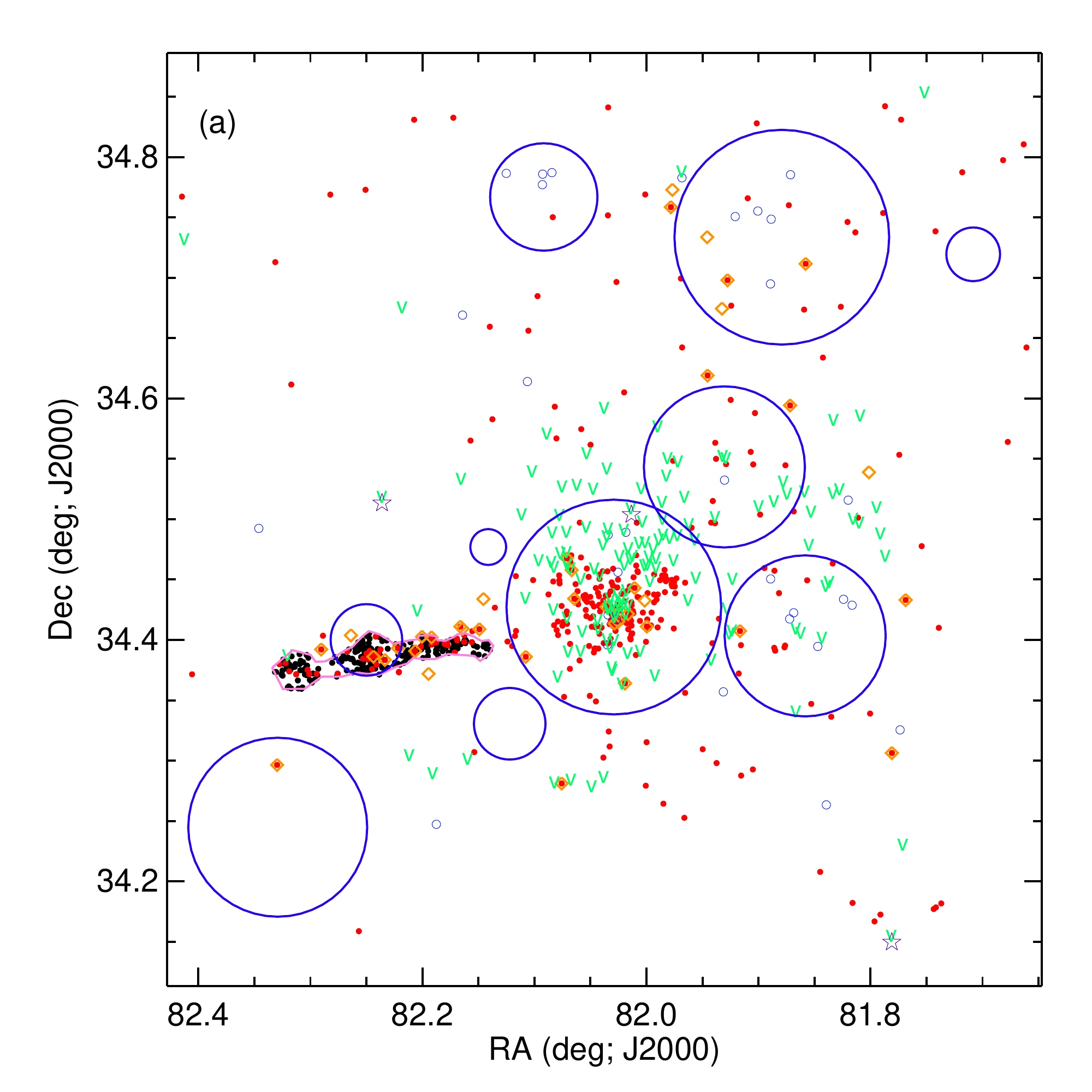}{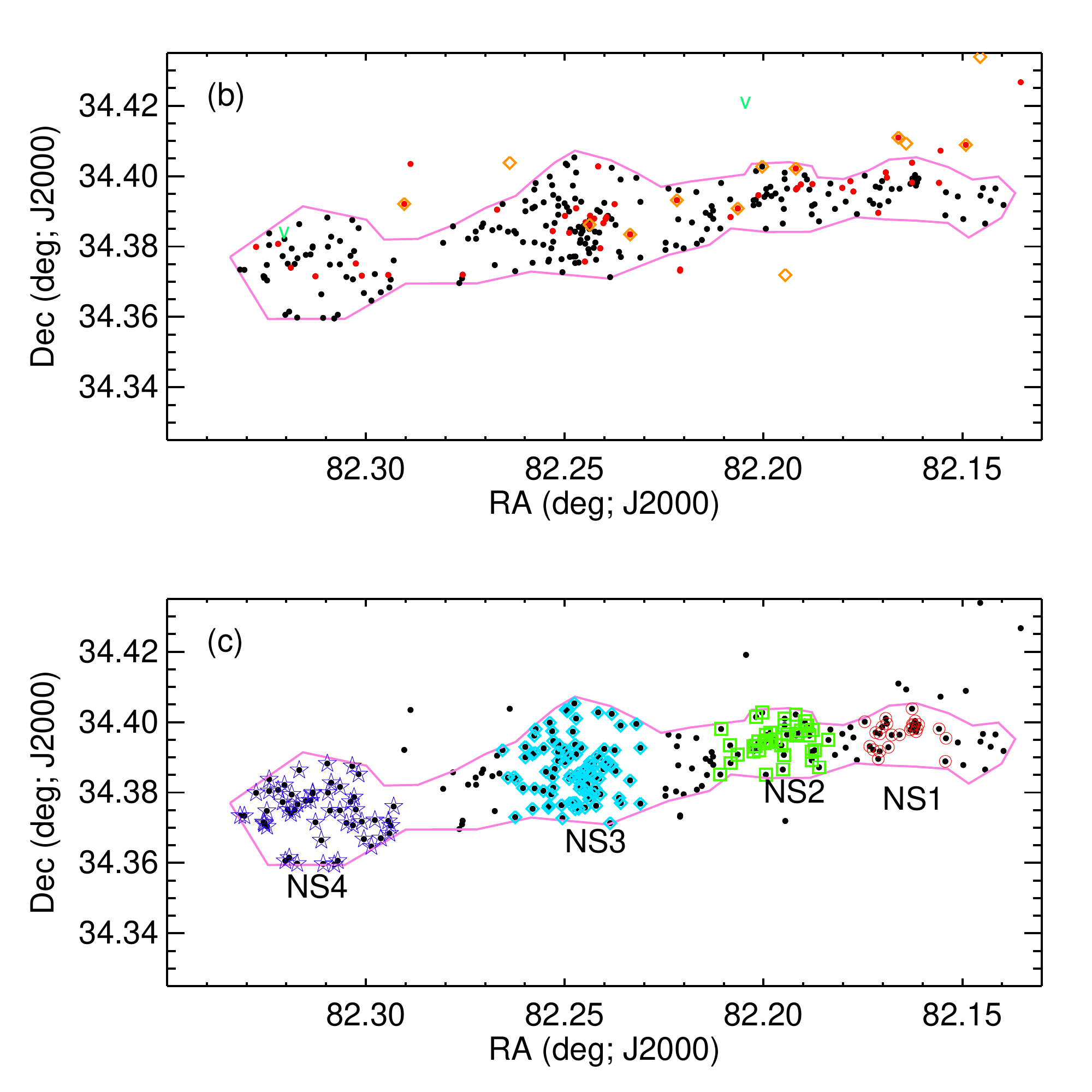}
\caption{
(a): Location of the YSO candidates on
the sky, with the clusters from Table~\ref{tab:clusters} and
Fig.~\ref{fig:where2}. 
Red filled circle: stars identified in the literature or
here via WISE as having an IR excess; blue open circle: stars identified
in the literature as an O or B star; green v: stars identified in the 
literature as variable; orange open diamond: stars identified in the
literature as H$\alpha$-bright; purple open stars: carbon stars from
the literature; filled black dots: stars identified here as part of the 
NS. 
(b): zoom-in on the NS, with same symbols as the first plot,
with the boundaries of the NS as the magenta polygon.  (c):
zoom-in on the NS, with magenta polygon, but this time different
subclusters defined as indicated, numbered in order of increasing RA.
Black dots are just stars in the catalog;  additional red
circle=subcluster 1, green square=subcluster 2,  cyan
diamonds=subcluster 3; purple star=subcluster 4.  Subcluster 3 has
substantial overlap with the BPI14 literature cluster.  Our YSO
candidate list is biased towards  stars in Stock~8 and NS because of
the way we constructed the list -- many of the literature YSOs are in
Stock~8 because that was the focus of those studies. We defined the
boundaries of the NS, so it is readily apparent in the first plot. The
right plot defines possible subclusters within the NS for use later in
the paper; note that some stars in the NS are not assigned to a
subcluster.}
\label{fig:where3sky}
\end{figure}

Table~\ref{tab:ysototals} collects all the total numbers (and
fractions) of stars in the final sample (omitting the sources ultimately 
rejected in Sec.~\ref{sec:vettingysos}). Table~\ref{tab:bigdata} has 
all of the YSOs that survived the analysis described in this paper, and has
also been delivered to IRSA (along with individual SEDs for each object). 
Figure~\ref{fig:where3sky} shows where the stars are on the sky, with
the clusters from Fig.~\ref{fig:where2} for reference. 

\begin{deluxetable}{p{4cm}ccp{4cm}}
\tabletypesize{\scriptsize}
\tablecaption{Total Number YSO Candidates In or Near IC~417\label{tab:ysototals}}
\tablewidth{0pt}
\tablehead{\colhead{Type} & \colhead{Number} & \colhead{YSO Sample Fraction} & 
\colhead{Notes} \\[-0.3cm]
& & \colhead{(out of 710)}}
\startdata
Literature H$\alpha$ excess & 40 & 6\% \\
Literature OB stars & 32 & 5\% \\
Literature carbon stars & 3 & 0.4\% \\ 
J17 (IR excess) & 159 & 23\% \\
Pandey \etal\ (2020) (IR excess) & 53 & 7\% \\
Winston \etal\ (2020)  (IR excess) & 206 & 29\% & several more rejected as not point sources\\
Any IR-selected literature & 323 & 45\% \\
Lata \etal\ (2019) (variables) & 130 & 19\% & includes all variables, not just YSO candidates\\
ASAS-SN variables & 11 & 2\% & includes all variables, not just YSO candidates \\
Any variability-selected literature & 139 & 20\% \\
Any literature & 491 & 69\% \\
Selected via position in the NS & 258 & 36\% \\
New YSO candidates selected via position in the NS & 213 & 30\% \\
New YSO candidates selected via WISE IR excess & 5 & 0.7\% & several more rejected as not point sources\\
{\bf Total YSOs or candidates in final set} & 710 & 100\% \\
\hline
Final rank 5 & 186 & 26\% \\
Final rank 4 & 95 & 13\%  \\
Final rank 4* & 9 & 1\% & everything seems ok (often even distance) but SED is odd\\
Final rank 3 & 213 & 30\% \\
Final rank 2 & 89 & 12\% \\
Final rank 1 & 32 & 5\% \\
Final rank 1d & 38 & 5\% & distance is inconsistent with IC~417\\
Final rank 1f & 39 & 5\% & too few points in SED\\
Final rank 1r & 8 & 1\% & have to be rejected (carbon stars; source confusion)\\
\hline
Final rank 5/4/4* &  290 & 41\% \\
Final rank 5/4/4*/3 & 503 & 71\% \\
\hline
SED Class I & 71 & 10\% \\
SED Class flat & 56 & 8\% \\
SED Class II & 278 & 39\% \\
SED Class III & 320 & 45\%\\
\hline
Final rank 5/4/4*/3 and SED Class I & 19 & 3\% & 4\% of final rank 3/4*/4/5 \\
Final rank 5/4/4*/3 SED Class flat & 32 & 5\% & 6\% of final rank 3/4*/4/5 \\
Final rank 5/4/4*/3 SED Class II & 199 & 29\% & 39\% of final rank 3/4*/4/5\\
Final rank 5/4/4*/3 SED Class III & 253 & 36\% & 50\% of final rank 3/4*/4/5 \\
\enddata
\end{deluxetable}

\subsection{Literature-Identified Young Stars}
\label{sec:litysos}

We started our search for YSO candidates by compiling a list of YSOs 
and YSO candidates from the literature. We kept track of these
literature-identified sources during the merging process
(Sec.~\ref{sec:datamerging} above). 

There are O and B stars identified here in the last century; Mayer \&
Macak (1971) and Savage \etal\ (1985) identify a total of 14 between
them.  MN16 identified 33 O and B stars (19 of which are new)  in this
region, all but one of which are taken to be members. Many of the MN16
OB stars were also in Chargeishsvili (1988).  Since O and B stars must
be young, we have included these in our list of literature-identified
young stars.  They are well-distributed over the field
(Fig.~\ref{fig:where3sky}).

J08 identified 25 H$\alpha$-bright stars in the heart of this region.
However, they do not report quantitative measures of H$\alpha$. 
Witham \etal\ (2008) identified stars bright in H$\alpha$ across the
sky, and we do have quantitative measures of H$\alpha$ for those.
Stars bright in H$\alpha$ could be old, chromospherically active 
stars, but they could also be young, accreting stars. We have a total
of 40 stars bright in  H$\alpha$ in our list of literature-identified
young stars or candidates. They are also well-distributed over the
field (Fig.~\ref{fig:where3sky}).

J08, J17, Winston \etal\ (2020), and Pandey \etal\ (2020) all identify
YSO candidates in their work, largely from IR selection; we included
these 323 YSOs selected by any of these authors in our set of literature YSOs. 
IR selection yields the most YSO candidates in our set; see
Table~\ref{tab:ysototals} and Fig.~\ref{fig:where3sky}. These YSO
candidates are packed most tightly in Stock~8 and the NS, but are
found  across the entire field.

Because young stars are often variable (see, e.g., Joy 1945; Herbig
1952), variability is another method for identifying YSOs.  Lata \etal\
(2019) monitored stars in the optical largely in and near Stock~8,
and relied upon supporting data from J08 and J17; we simply retained
their derived periods for all of their targets 
(all of their Table~3), tagging them all as
possible YSOs. Jayasinghe \etal\ (2018) reported variables selected
from ASAS-SN optical monitoring observations, which we also retained
as YSO candidates; there are 11 in our region. A few of these ASAS-SN
variables appear in the literature as carbon stars (see
Sec.~\ref{sec:obc}).   There are a total of 139 stars identified as
variable in our list, biased heavily towards Stock~8; see
Fig.~\ref{fig:where3sky}.

There are nearly 500 unique stars that we identified from the
literature as possible or confirmed YSOs; see
Table~\ref{tab:ysototals}. (Since we have amassed data at up
to 47 distinct wavelengths, we should be able to make a better
assessment of the YSO status of many if not most of these targets than
the literature to this point.)

Because the literature is biased towards Stock~8, the set of
YSOs/candidates pulled from the literature is also biased towards
Stock~8, and that is the main reason why Stock~8 is immediately
obvious in Fig.~\ref{fig:where3sky}. The other clusters are much  less
obvious in Fig.~\ref{fig:where3sky}, but that may be a result both of
what literature has studied until now, and how we assembled  our list
of literature YSOs/candidates. We did \textbf{not} attempt to identify
new cluster members  based on position in the sky for the
literature-identified clusters in Table~\ref{tab:clusters}; the
articles identifying clusters often use statistical arguments, as
opposed to a list of cluster members, to define the cluster. The
evidence for youth is much less clear in these clusters on their own
(see discussion in Sec.~\ref{sec:2massclusters}),  so that is why we
did not identify them by position {\em a priori} as YSO candidates. We
did, however, keep track of these possible cluster members based on
position in the sky, given the positions and radii in
Table~\ref{tab:clusters}; see Table~\ref{tab:bigdata} below.

\subsection{YSO Candidates in the NS}
\label{sec:nsysos}

The NS is identified in J08; the largest cluster within it had been 
previously identified (BPI 14; Table~\ref{tab:clusters} \& 
Fig.~\ref{fig:where2}).  The NS extended emission and some of the
point sources have been discussed in additional papers (e.g., 
Dewangan \etal\ 2018; MN16; J08). However, most of the  point source
constituents have yet to be explored in detail in the  literature at
any band.  There are four `ripples' in the NS, each  of which appears
to contain clusters of red objects in Fig.~\ref{fig:where1}; these 
clusters are most obvious in the Spitzer data, both because of the
high spatial resolution, and the transparency of the dust at these
wavelengths.  One of our primary goals of this paper is to explore
these red sources, and so we identify YSO candidates based on
projected position in the sky within the NS. Recall that our catalog 
(Sec.~\ref{sec:datamerging}) is based primarily on IR sources, so  it
is already biased towards sources detected  at 2MASS, IRAC, and/or
WISE bands.

In the regions of highest source surface density like the NS (or Stock
8), WISE simply cannot distinguish the sources, so the 2MASS+WISE
color selection (Sec.~\ref{sec:irxysos}) alone cannot identify all the
YSO candidates in the NS (or in Stock~8); this is one clear reason for
identifying YSOs in the NS using an entirely different approach.

We drew a complex polygon (see Fig.~\ref{fig:wherens}) enclosing the
nebulosity and visibly red stars in Fig.~\ref{fig:where1} in the NS.
The 258 point sources enclosed by this polygon were taken as
candidates based on position in the NS. Note that this also
encompasses both literature YSOs and YSO candidates identified from
the IR in the next section (see Fig.~\ref{fig:where2}).  Because we
defined NS membership  by position on the sky,  the NS is very obvious
by eye in Fig.~\ref{fig:where3sky}. 

We only have IRAC-1 and -2 from Spitzer, so using the available
Spitzer bands to look for IR excesses will not identify IR excesses
that start at wavelengths longer than 5 \mum. Thus, new candidate YSOs
in the NS are often identified based on IRAC-1 and -2 colors, but
assessed including optical properties. 

Based on the distribution of the point sources and the nebulosity, we
further broke the NS into four sub-clusters by eye, numbered in
direction of increasing RA; see Figure~\ref{fig:where3sky}. Note
that some NS sources are not assigned to a sub-cluster. The 
sub-cluster assignments are included in our catalog (Table~\ref{tab:bigdata}).

\subsection{YSO Candidates with an IR excess in 2MASS+WISE}
\label{sec:irxysos}

We also identified new candidate YSOs in IC~417 by looking for IR
excess sources using WISE and 2MASS data.  These IR excess sources
were identified by using a series of color cuts in various 2MASS/WISE
color-magnitude and color-color diagrams following Koenig \& Leisawitz
(2014).

As discussed in Koenig \& Leisawitz (2014), their approach makes use
of the combined 2MASS+WISE catalog; they describe the approach both in
terms of the AllWISE catalog but also Koenig's own processing approach
using his routine PhotVis (Koenig \etal\ 2012).  PhotVis run on
the WISE data in this region resulted in a list of $>$100 YSO
candidates identified from his color-selection approach, using either
AllWISE or the PhotVis data reduction. The PhotVis approach can be
tuned to be `more complete, less reliable'; as for the analogous
GLIMPSE360 data above, it was more important to get measures of every
source than it was to avoid false sources because of our vetting
process (Sec.~\ref{sec:vettingysos}).  Most of the YSO candidates so
identified are likely to be true YSO candidates, but a fraction is
likely to be image artifacts, or affected enough by image artifacts
that they are not trustworthy YSO candidates.   

Koenig \etal\ (2015) showed via spectroscopic follow-up that
$\sim$80\% of YSO candidates selected via this method near $\lambda$
and $\sigma$ Ori  are likely true YSOs. While follow-up spectroscopy
in IC~417 is beyond the scope of the present work, we here further vet
the WISE-selected IR excess sources using additional photometric data
(Sec.~\ref{sec:vettingysos}). 

Note that, while IRAC's shortest two bands are similar to WISE's
shortest two bands, the bandpasses are not identical. Even though the
IRAC bands are higher spatial resolution, Koenig's approach has been
tuned to work specifically with the WISE bands, so we do not expect to
swap in the two shortest IRAC bands for the two shortest WISE bands
and have the selection process still select YSOs as well as Koenig has
shown. We do, however, make use of the IRAC data (as well as all the
other optical data) in the analysis of the objects.

When we initiated this work, this was the first method we used for
finding YSOs. At the time, we had more than 100 new YSO candidates 
that we identified via this 2MASS+WISE approach. In the meantime, more
studies have come out using IR excesses to find YSOs (J17,
Pandey \etal\ 2020, Winston  \etal\ 2020), so most of our
IR-excess-identified then-new sources have become IR-excess-identified
literature sources. We independently identified many of them as YSO
candidates; they are noted as such in  Table~\ref{tab:bigdata}.

We did not not search for YSOs based on long wavelength detections 
(Sec.~\ref{sec:datamfir}) because the spatial resolution was just too
low and the source surface density just too high to make this a fruitful 
exercise. We retained source matches where the match was obvious, but not
where source confusion rendered this impossible.

\subsection{YSO Candidate List}

To this point, we have 726 YSO candidates, 68\% of which were from the
literature. Of the 230 new candidates, 93\% are from position  in the
NS, and just 7\% are from the WISE IR selection. Many sources are
identified via more than one approach. Now, we are ready to vet these
candidates for reliability.

\startlongtable
\begin{deluxetable}{cp{13cm}}
\tabletypesize{\scriptsize}
\tablecaption{Contents of Table: Final YSO Candidates in IC~417\tablenotemark{a}\label{tab:bigdata}}
\tablewidth{0pt}
\tablehead{\colhead{Column} & \colhead{Contents}}
\startdata
\cutinhead{Identifications (where and why)}
cat num & position-based catalog number\footnote{Assembled here to be compliant with IAU nomenclature rules, based on the best J2000 RA and Dec we have for the object. All names should start with `J' when used in the text (as per Chen \etal\ 2022), but are not listed as such here just for space considerations. } \\
why here & why this star is in our list, e.g., why this target was considered as a possible YSO. 
Possible values include: 
Carbon star=identified in the literature as a carbon star (means it will show up as having an IR excess, but is not young);
OB star = identified in the literature as an OB star (means it is young);
Witham+08 Ha bright=identified in Witham \etal\ (2008) as H$\alpha$-bright;
Jose+08 Ha excess=identified in Jose \etal\ (2008) as H$\alpha$ excess star;
Jose+17 YSO=identified in Jose \etal\ (2017) as a YSO; 
ASAS-SN variable=identified in Jayasinghe \etal\ (2018) as variable;
Lata+19 variable=identified in Lata \etal\ (2019) as variable; 
Pandey+20 YSO=identified in Pandey \etal\ (2020) as a YSO;
Winston+20 YSO=identified in Winston \etal\ (2020) as a YSO
WISE IR excess=identified here (independently) as a YSO based on WISE IR excess;
Inside NS polygon=identified here as being inside the polygon drawn on the sky encompassing the NS.\\
other name & any other common name as retrieved from Simbad \\
J08 name & name from J08\\
J17 name & name from J17\\
J08 Ha star & true (=1) if J08 identified it as an H$\alpha$ excess star\\
J08 OB star & true (=1) if J08 identified it as an OB star\\
J17 class & value copied from J17 for YSO class\\
MN16 name & name from MN16\\
Winston+20 YSO flag & true (=1) if Winston \etal\ (2020) identified it as a YSO\\
Pandey+20 YSO flag & true (=1) if Pandey \etal\ (2020) identified it as a YSO\\
Lata+19 name & name from Lata \etal\ (2019)\\
Lata+19 YSO flag & true (=1) if Lata \etal\ (2019) identified it as a YSO\\
Lata+19 period & period in days from Lata \etal\ (2019) \\
Sp Ty & spectral type from the literature \\
Sp Ty src & origin of spectral type \\
2MASS name & identifier from 2MASS catalog\\
UKIDSS name & identifier from UKIDSS catalog\\
AllWISE name & identifier from AllWISE catalog\\
CatWISE name & identifier from CatWISE catalog\\
unWISE name & identifier from unWISE catalog\\
PanSTARRS name & identifier from PanSTARRS catalog\\
IPHAS name & identifier from IPHAS catalog\\
Gaia2 name & identifier from Gaia DR2 catalog\\
Gaia3 name & identifier from Gaia DR3 catalog\\
PACS names & identifier from PACS 70 and/or 160 micron catalogs\\
AKARI name & identifier from AKARI catalogs\\
MSX name & identifier from MSX catalog\\
\cutinhead{Results of our analysis}
Nominal cluster & Based on position on the sky (see Table~\ref{tab:clusters}), is this star in the right place to be part of a cluster? \\
NS & true (=1) if it is within the NS polygon (see Figure~\ref{fig:wherens})\\
NS subcluster & equal to 1, 2, 3, or 4 if it is in the right place on the sky to be part of 
the NS subclusters 1, 2, 3, or 4 (see Fig~\ref{fig:where3sky}); note some NS stars are not part of a subcluster\\
WISE IRx & true (=1) if it has a WISE IR excess\\
Final rank & Final qualitative confidence bin (see Sec.~\ref{sec:finalrankings}), equal to 5, 4, 4*, 3, 2, 1, 1f, 1d, 1r\\
Final rank order &  Final qualitative ordering; we placed ``like with like'' such that, if the stars are sorted by this order, the stars will not only be sorted by final rank but also within each confidence bin, sorted by confidence and similar stars will be placed near each other in the list; more likely stars will be higher in the list.\\
slope 2-24 um & Slope fit to the SED to all available detections between 2 and 24 microns\\
SED Class & SED Class (I, flat, II, or III), based on SED slope\\
IRx any band & true (=1) if there is a reliable IR excess at any band\\
AV\_JHK & Reddening estimate derived from $JHK_s$ diagram (Sec.~\ref{sec:vettingcmds})\\
Chi(i1-i2) & $\chi$ calculated for [I1]$-$[I2] (Sec.~\ref{sec:vettingcmds})\\
Chi(r-Ha) & $\chi$ calculated for $r-H\alpha$ (Sec.~\ref{sec:vettingcmds})\\
large IRX flag & true (=1) if there is a large IR excess \\
JHKX flag & true (=1) if there is an IR excess likely to affect $JHK_s$ \\
HAX flag & true (=1) if there is a likely H$\alpha$ excess  \\
BlueX flag & true (=1) if there is a likely $g$-band (``blue'') excess  \\
num points SED & number points in the SED (note not necessarily same as number distinct wavelengths)\\
\cutinhead{Photometric or flux measurements}
Umag & $U$ magnitude (Vega mag; all errors taken to be 0.1 mag)\\
Bmag & $B$ magnitude (Vega mag; all errors taken to be 0.1 mag)\\
Vmag & $V$ magnitude (Vega mag; all errors taken to be 0.1 mag)\\
Icmag & $I_c$ magnitude (Vega mag; all errors taken to be 0.1 mag)\\
pangmag & PanSTARRS $g$ magnitude (AB mag)\\
pangmerr & PanSTARRS $g$ magnitude error (AB mag)\\
panrmag & PanSTARRS $r$ magnitude (AB mag)\\
panrmerr & PanSTARRS $r$ magnitude error (AB mag)\\
panimag & PanSTARRS $i$ magnitude (AB mag)\\
panimerr & PanSTARRS $i$ magnitude error (AB mag)\\
panzmag & PanSTARRS $z$ magnitude (AB mag)\\
panzmerr & PanSTARRS $z$ magnitude error (AB mag)\\
panymag & PanSTARRS $y$ magnitude (AB mag)\\
panymerr & PanSTARRS $y$ magnitude error (AB mag)\\
iphasrmag & IPHAS $r$ magnitude (Vega mag)\\
iphasrmerr & IPHAS $r$ magnitue error (Vega mag)\\
iphasimag & IPHAS $i$ magnitude (Vega mag)\\
iphasimerr & IPHAS $i$ magnitue error (Vega mag)\\
iphashamag & IPHAS H$\alpha$ magnitude (Vega mag)\\
iphashamerr & IPHAS H$\alpha$ magnitude error (Vega mag)\\
gaia2gmag & Gaia DR2 $G$ magnitude (Vega mag) \\
gaia2gmerr & Gaia DR2 $G$ magnitude error (Vega mag)\\
gaia2bpmag & Gaia DR2 $G_{RP}$ magnitude (Vega mag) \\
gaia2bpmerr & Gaia DR2 $G_{RP}$ magnitude error (Vega mag)\\
gaia2rpmag & Gaia DR2 $G_{BP}$ magnitude (Vega mag) \\
gaia2rpmerr & Gaia DR2 $G_{BP}$ magnitude error (Vega mag)\\
gaia2plx & Gaia DR2 parallax (mas) \\
gaia2bjdist & Gaia DR2 distance from Bailer-Jones \etal\ (2018), in pc\\
gaia2bjdistup & Gaia DR2 distance from Bailer-Jones \etal\ (2018), upper limit, in pc\\
gaia2bjdistdwn & Gaia DR2 distance from Bailer-Jones \etal\ (2018), lower limit, in pc\\
gaia3gmag & Gaia DR3 $G$ magnitude (Vega mag) \\
gaia3gmerr & Gaia DR3 $G$ magnitude error (Vega mag)\\
gaia3bpmag & Gaia DR3 $G_{RP}$ magnitude (Vega mag) \\
gaia3bpmerr & Gaia DR3 $G_{RP}$ magnitude error (Vega mag)\\
gaia3rpmag & Gaia DR3 $G_{BP}$ magnitude (Vega mag) \\
gaia3rpmerr & Gaia DR3 $G_{BP}$ magnitude error (Vega mag)\\
gaia3plx & Gaia DR3 parallax (mas) \\
gaia3dist & Gaia DR3 distance (pc) \\
gaia3bjdist & Gaia EDR3 distance from Bailer-Jones \etal\ (2021), in pc\\
gaia3bjdistup & Gaia EDR3 distance from Bailer-Jones \etal\ (2021), upper limit, in pc\\
gaia3bjdistdwn & Gaia EDR3 distance from Bailer-Jones \etal\ (2021), lower limit, in pc\\
gaia3ruwe & Gaia DR3 RUWE \\
jose08umag & J08 $U$ magnitude (Vega mag; all errors taken to be 0.1 mag) \\
jose08bmag & J08 $B$ magnitude (Vega mag; all errors taken to be 0.1 mag) \\
jose08vmag & J08 $V$ magnitude (Vega mag; all errors taken to be 0.1 mag) \\
jose08icmag & J08 $I_c$ magnitude (Vega mag; all errors taken to be 0.1 mag) \\
jose17vmag & J17 $V$ magnitude (Vega mag) \\
jose17vmerr & J17 $V$ magnitude error (Vega mag) \\
jose17imag & J17 $I$ magnitude (Vega mag) \\
jose17imerr & J17 $I$ magnitude error (Vega mag) \\
marcosumag  & MN16 $u$ magnitude (Stromgren $u$ mag) \\
marcosumerr & MN16 $u$ magnitude error (Stromgren $u$ mag) \\
marcosvmag  & MN16 $v$ magnitude (Stromgren $v$ mag) \\
marcosvmerr & MN16 $v$ magnitude error (Stromgren $v$ mag) \\
marcosbmag  & MN16 $b$ magnitude (Stromgren $b$ mag) \\
marcosbmerr & MN16 $b$ magnitude error (Stromgren $b$ mag) \\
marcosymag  & MN16 $y$ magnitude (Stromgren $y$ mag) \\
marcosymerr & MN16 $y$ magnitude error (Stromgren $y$ mag) \\
marcosbeta  & MN16 $\beta$ (Stromgren $\beta$) \\
lataumag & Lata \etal\ (2019) $U$ magnitude (Vega mag; all errors taken to be 0.1 mag) \\
latabmag & Lata \etal\ (2019) $B$ magnitude (Vega mag; all errors taken to be 0.1 mag) \\
latavmag & Lata \etal\ (2019) $V$ magnitude (Vega mag; all errors taken to be 0.1 mag) \\
lataimag & Lata \etal\ (2019) $I$ magnitude (Vega mag; all errors taken to be 0.1 mag) \\
bestjmag  & best $J$ magnitude available (Vega mag)\\
bestjmerr & best $J$ magnitude error available (Vega mag)\\
besthmag  & best $H$ magnitude available (Vega mag)\\
besthmerr & best $H$ magnitude error available (Vega mag)\\
bestkmag  & best $K_s$ magnitude available (Vega mag)\\
bestkmerr & best $K_s$ magnitude error available (Vega mag)\\
tmjmag  & 2MASS $J$ magnitude (Vega mag)\\
tmjmerr & 2MASS $J$ magnitude error (Vega mag)\\
tmhmag  & 2MASS $H$ magnitude (Vega mag)\\
tmhmerr & 2MASS $H$ magnitude error (Vega mag)\\
tmkmag  & 2MASS $K_s$ magnitude (Vega mag)\\
tmkmerr & 2MASS $K_s$ magnitude error (Vega mag)\\
ukidssjmag  & UKIDSS $J$ magnitude (Vega mag)\\
ukidssjmerr & UKIDSS $J$ magnitude error (Vega mag)\\
ukidsshmag  & UKIDSS $H$ magnitude (Vega mag)\\
ukidsshmerr & UKIDSS $H$ magnitude error (Vega mag)\\
ukidsskmag  & UKIDSS $K_s$ magnitude (Vega mag)\\
ukidsskmerr & UKIDSS $K_s$ magnitude error (Vega mag)\\
jose17jmag  & J17 $J$ magnitude (Vega mag)\\
jose17jmerr & J17 $J$ magnitude error (Vega mag)\\
jose17hmag  & J17 $H$ magnitude (Vega mag)\\
jose17hmerr & J17 $H$ magnitude error (Vega mag)\\
jose17kmag  & J17 $K$ magnitude (Vega mag)\\
jose17kmerr & J17 $K$ magnitude error (Vega mag)\\
marcojmag   & MN16 $J$ magnitude (Vega mag)\\
marcojmerr  & MN16 $J$ magnitude error (Vega mag)\\
marcohmag   & MN16 $H$ magnitude (Vega mag)\\
marcohmerr  & MN16 $H$ magnitude error (Vega mag)\\
marcokmag   & MN16 $K$ magnitude (Vega mag)\\
marcokmerr  & MN16 $K$ magnitude error (Vega mag)\\
latajmag & Lata \etal\ (2019) $J$ magnitude (Vega mag; all errors taken to be 0.1 mag) \\
latahmag & Lata \etal\ (2019) $H$ magnitude (Vega mag; all errors taken to be 0.1 mag) \\
latakmag & Lata \etal\ (2019) $K$ magnitude (Vega mag; all errors taken to be 0.1 mag) \\
irac1mag  & best IRAC-1 magnitude (Vega mag)\\
irac1merr & best IRAC-1 magnitude error (Vega mag)\\
irac2mag  & best IRAC-2 magnitude (Vega mag)\\
irac2merr & best IRAC-2 magnitude error (Vega mag)\\
glirac1mag  & GLIMPSE360 IRAC-1 magnitude (Vega mag) \\
glirac1merr & GLIMPSE360 IRAC-1 magnitude error (Vega mag) \\
glirac2mag  & GLIMPSE360 IRAC-2 magnitude (Vega mag) \\
glirac2merr & GLIMPSE360 IRAC-2 magnitude error (Vega mag) \\
jose17irac1mag  & J17 IRAC-1 magnitude (Vega mag) \\
jose17irac1merr & J17 IRAC-1 magnitude error (Vega mag) \\
jose17irac2mag  & J17 IRAC-2 magnitude (Vega mag) \\
jose17irac2merr & J17 IRAC-2 magnitude error (Vega mag) \\
wise1flim & limit flag for WISE-1, in the sense of flux; that is $<$ means that the measure given is an upper limit in flux, but a lower limit in magnitudes -- the true brightness of the source is fainter than the number given in the next column\\
wise1mag  & WISE-1 magnitude (Vega mag)\\
wise1merr & WISE-1 magnitude error (Vega mag)\\
wise2flim & limit flag for WISE-2 (same sense as that for WISE-1)\\
wise2mag  & WISE-2 magnitude (Vega mag)\\
wise2merr & WISE-2 magnitude error (Vega mag)\\
wise3flim & limit flag for WISE-3 (same sense as as that for WISE-1)\\
wise3mag  & WISE-3 magnitude (Vega mag)\\
wise3merr & WISE-3 magnitude error (Vega mag)\\
wise4flim & limit flag for WISE-4 (same sense as as that for WISE-1)\\
wise4mag  & WISE-4 magnitude (Vega mag)\\
wise4merr & WISE-4 magnitude error (Vega mag)\\
catwise1flim & CatWISE limit flag for WISE-1 (same sense as as that for WISE-1)\\
catwise1mag  & CatWISE WISE-1 magnitude (Vega mag)\\
catwise1merr & CatWISE WISE-1 magnitude error (Vega mag)\\
catwise2mag  & CatWISE WISE-2 magnitude (Vega mag)\\
catwise2merr & CatWISE WISE-2 magnitude error (Vega mag)\\
unwise1mag  & unWISE WISE-1 magnitude (Vega mag)\\
unwise1merr & unWISE WISE-1 magnitude error (Vega mag)\\
unwise2mag  & unWISE WISE-2 magnitude (Vega mag)\\
unwise2merr & unWISE WISE-2 magnitude error (Vega mag)\\
pacs70flux & PACS-70 flux in Jy \\
pacs70ferr & PACS-70 flux error in Jy\\
pacs160flux & PACS-160 flux in Jy \\
pacs160ferr & PACS-160 flux error in Jy \\
akari9flux & AKARI 9 \mum\ flux in Jy\\
akari9ferr & AKARI 9 \mum\ flux error in Jy\\
akari18flux & AKARI 18 \mum\ flux in Jy\\
akari18ferr & AKARI 18 \mum\ flux error in Jy\\
msxaflux & MSX A flux in Jy\\
msxaferr & MSX A flux error in Jy\\
msxb1flux & MSX B1 flux in Jy\\
msxb1ferr & MSX B1 flux error in Jy\\
msxb2flux & MSX B2 flux in Jy\\
msxb2ferr & MSX B2 flux error in Jy\\
msxcflux & MSX C flux in Jy\\
msxcferr & MSX C flux error in Jy\\
msxdflux & MSX D flux in Jy\\
msxdferr & MSX D flux error in Jy\\
msxeflux & MSX E flux in Jy\\
msxeferr & MSX E flux error in Jy\\
\enddata
\tablenotetext{a}{Also delivered to IRSA.}
\end{deluxetable}

\section{Vetting of YSO Candidates}
\label{sec:vettingysos}

Since the data span a wide range of spatial resolutions ($\lesssim
1\arcsec$ to 12$\arcsec$; see Table~\ref{tab:datalist}),  survey
depths, and survey reliability, individual inspection of each
candidate YSO is important, especially since we made decisions about
which catalogs to include knowing that we would be checking each
source. Toward that end, we vetted each of these sources manually in
at least three different ways: image inspection, SED inspection, and
location in color-color and color-magnitude diagrams. To first order,
we wanted to explore whether there was a legitimate point source at
each star's  location, with multi-wavelength photometry; secondarily,
we used  the collected information about each  star to place the YSO
candidate into qualitative confidence bins, ranked 1-5. We now discuss
each of these steps in turn.   Table~\ref{tab:ysototals} summarizes
the numbers of sources surviving this vetting, and
Table~\ref{tab:bigdata} has all of the YSOs that survived the vetting
process and their final quality ranking estimate. Sample SEDs are provided
in the figures here, but a full set of SEDs has been delivered to IRSA.

\subsection{Image Inspection}
\label{sec:vettingimages}

Optical imaging has been shown (see, e.g., Rebull \etal\ 2010) to be
important in assessing whether or not a YSO candidate is an isolated
point source (and therefore likely a star), or actually $>$1 source,
or a background galaxy.  Because WISE has relatively low spatial
resolution, and because star formation has the same colors whether it
is in our Galaxy or a nearby galaxy, a point source in WISE can be
revealed to be a nearby star-forming galaxy when viewed in
high-spatial-resolution optical images.  Moreover, especially in
regions of high, and highly-structured, background emission like
star-forming regions, the WISE pipeline sometimes struggles with channels 3 and 4
(12 and 22 \mum).  Checking the images is the best way to determine if
the source is really there, and really a point source.

\begin{figure}[htb!]
\epsscale{1.0}
\plotone{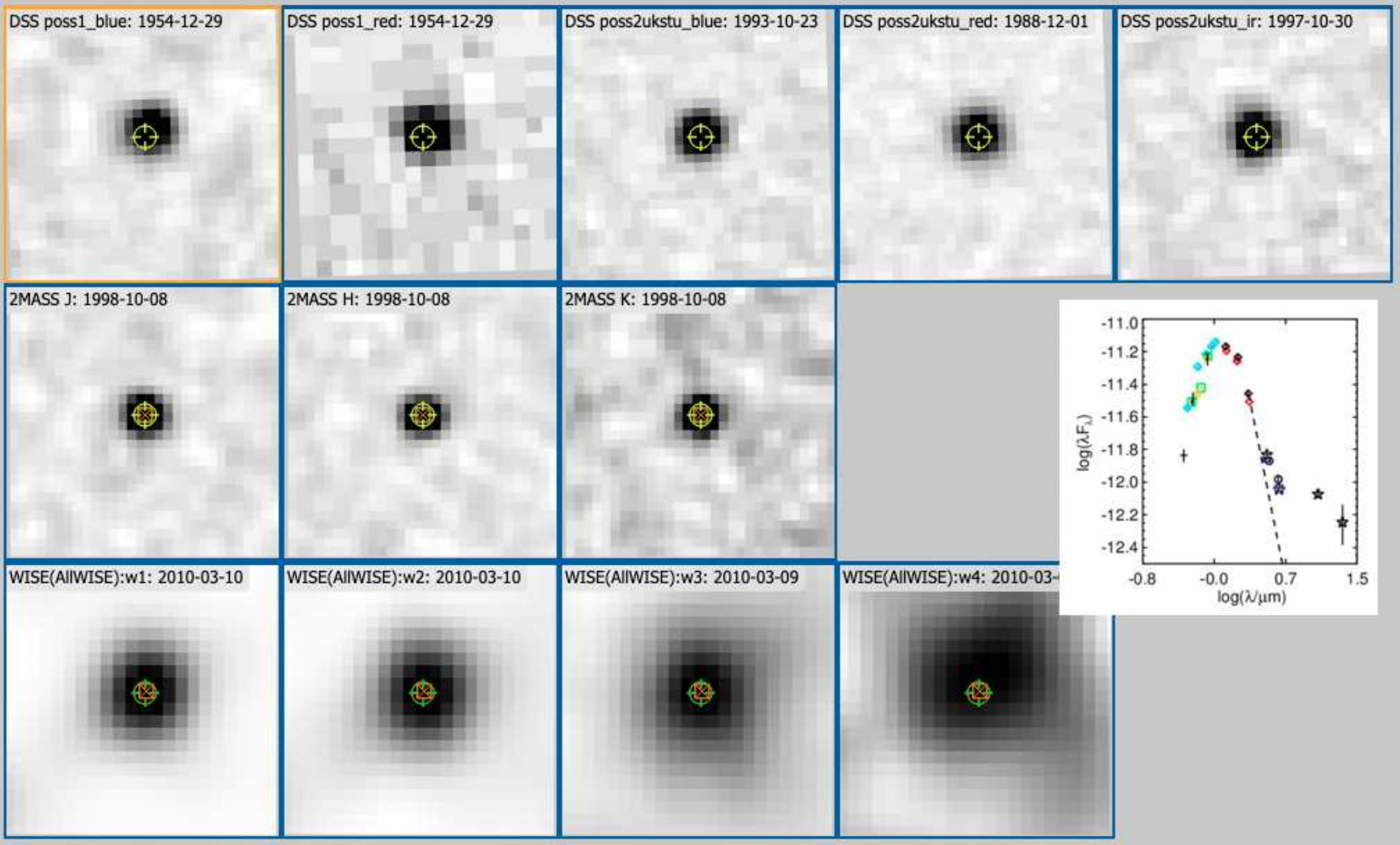}
\caption{Reverse greyscale images of J052704.46+342559.0, obtained via
FinderChart, in DSS (top row), 2MASS (middle row) and WISE (bottom
row); images are 30$\arcsec$ across.  The target position is given by
the crosshairs; the additional colored symbols on top of that position
are positions of corresponding catalog entries (the 2MASS and WISE
catalog counterparts are exactly on top of  this source).  The inset
is an SED using all the available data with symbols as defined in
Table~\ref{tab:datalist}.  Error bars are vertical black lines, most
obvious here in WISE-4. Dashed line is the expected flux density from
photosphere assuming \ks\ is on the photosphere.  This source was
identified by IPHAS as bright in H$\alpha$ and by Winston \etal\
(2020) as having an IR excess.  This is a well-behaved point source,
clearly and cleanly detected in the images, with a nice, YSO-like SED
showing a clear IR excess; we accept this source as a high-quality 
YSO candidate. }
\label{fig:sedgoodexample}
\end{figure}

The IRSA tool Finder Chart\footnote{http://irsa.ipac.caltech.edu/applications/finderchart/; \dataset[https://doi.org/10.26131/IRSA540]{https://doi.org/10.26131/IRSA540} }
provides easy access to the same patch of sky in several different
optical and IR  surveys, including the Digitized Sky Survey (DSS),
which is a digitization of the photographic sky survey plates from
Palomar (the Palomar Observatory Sky Survey, POSS) and UK Schmidt
telescopes, 2MASS, Spitzer (just cryo-era data currently, from the 
Spitzer Enhanced Imaging Products, Capak 2013) and WISE (just AllWISE
currently). This tool provides an easy way to check, for any given
point source, the point source quality and verify that the matching of
the source across wavelengths has been done correctly. Since the
Spitzer images of IC~417 were taken after Spitzer's cryogen had run
out, our Spitzer data are not available in Finder Chart. However, 
these data, along with unWISE images, are available via IRSA
Viewer\footnote{https://irsa.ipac.caltech.edu/irsaviewer/}. These
tools also overlay catalogs on the images so that it is easier to
assess if the sources are blended. We used both Finder Chart and IRSA
Viewer to inspect the available images for each target. We also
considered the SED (see next section,  Sec.~\ref{sec:vettingseds})
during this inspection process. Figure~\ref{fig:sedgoodexample} is an
example of a YSO candidate which looks good in the images and has a
nice, YSO-like SED.

\begin{figure}[htb!]
\epsscale{1.0}
\plotone{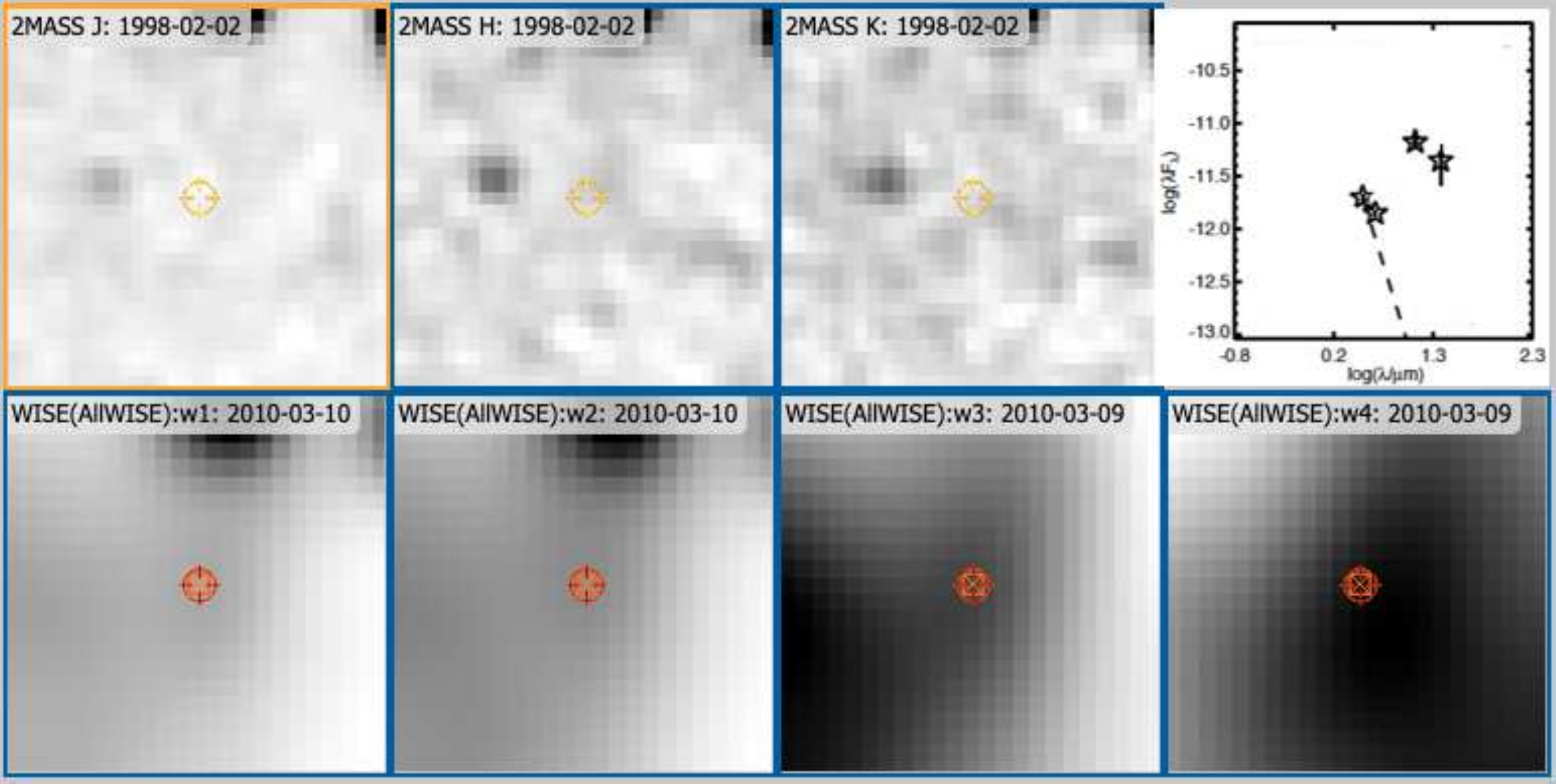}
\caption{Reverse greyscale images of J052828.16+342411.7, obtained via
FinderChart, in 2MASS (top row) and  WISE (bottom row); images are
30$\arcsec$ across.  The target position is given by the crosshairs. 
The top right is an SED using the existing photometric measurements
for this source, four WISE bands (black stars); the dashed line is the
expected photosphere if WISE-1 is on the photosphere. This source
appears in the AllWISE point source catalog (red squares in the WISE
images) as a high-quality detection; it also appears as a YSO
candidate in Winston \etal\ (2020).  However, as can be seen, there is
no counterpart at 2MASS (or any other) bands, and the emission at 22
\mum\ is significantly offset from the target position. The SED is 
inconsistent with that of a YSO. We dropped this source as likely 
to be a nebular knot, not a YSO.}
\label{fig:rejectexample}
\end{figure}

In this fashion, we rejected 16 sources from the candidate YSO list,
largely based on false WISE detections, which fall into two
categories. PhotVis in particular is known to find false sources
within diffraction spikes around some of the  brightest stars. There
were also sources that nominally have reliable detections in all four
WISE bands but no other counterpart at any other wavelength. That in
itself is suspicious; given the depth and diversity of catalogs
included here (and the distance of IC~417), a counterpart from at least 
one other survey is
expected. When the WISE images are inspected in these cases, especially in
conjunction with the SED shape (below,  Sec.~\ref{sec:vettingseds}), 
it becomes clear that the source is really a nebular knot, not a point
source. Several sources from Winston \etal\ (2020)  were rejected on
that basis. Figure~\ref{fig:rejectexample} is an example of such a
rejected source; note the SED shape as well.

\begin{figure}[htb!]
\epsscale{1.0}
\plotone{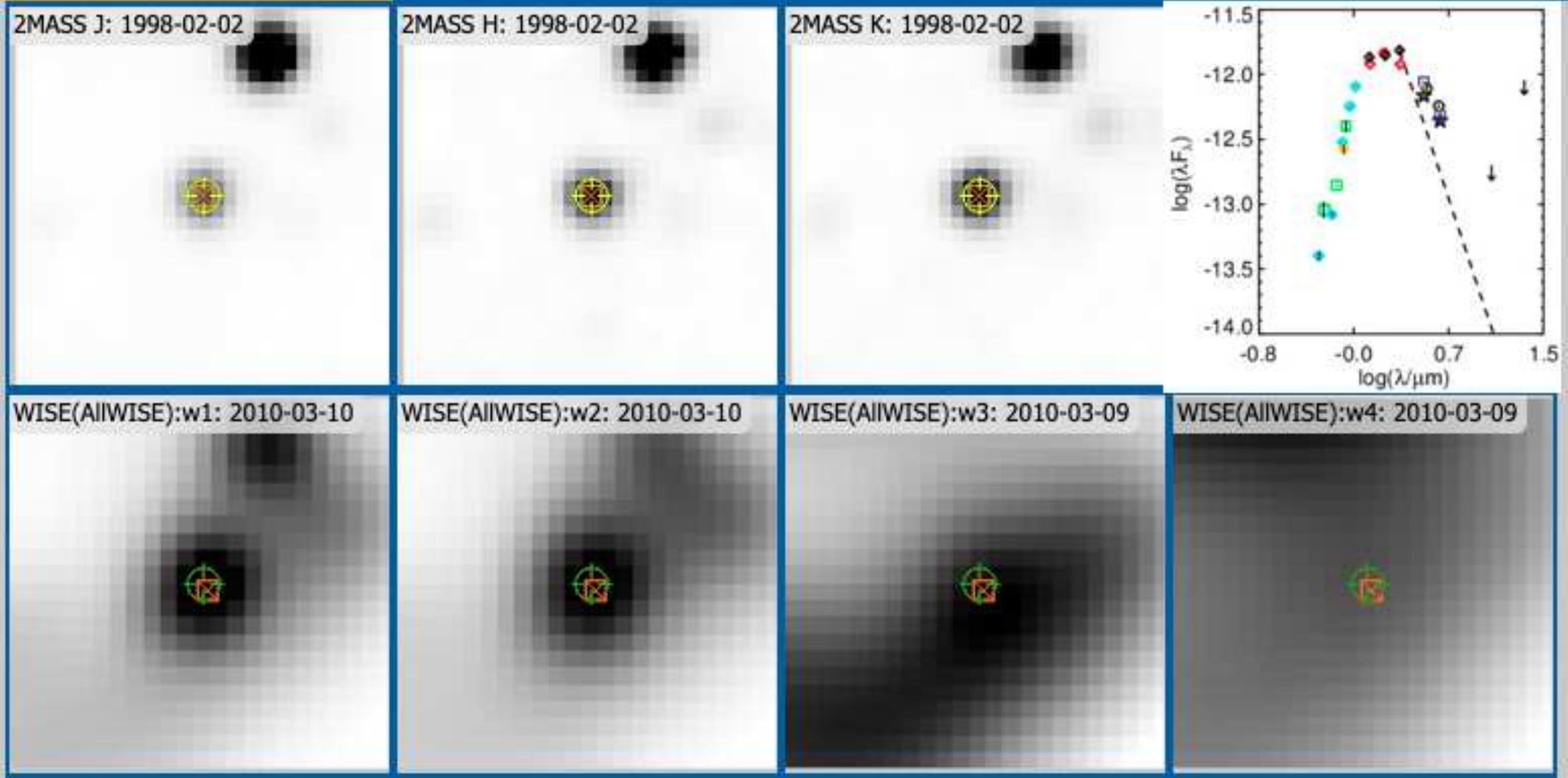}
\caption{Reverse greyscale images of J052724.69+342049.1, obtained via
FinderChart, in 2MASS (top row) and WISE (bottom row); images are
30$\arcsec$ across. The target position is given by the crosshairs,
with an additional symbol for the corresponding catalog source.  The
top right is an SED with symbols as defined in
Table~\ref{tab:datalist}.  WISE-3 and 4 are both plotted here as
limits. Dashed line is the expected flux density from photosphere assuming
\ks\ is on the photosphere.  This source appears in the AllWISE point
source catalog with high-quality detections in all four WISE bands,
but inspection of the image  calls WISE-3 and -4 into question. The
SED suggests that WISE-3 could be a detection and still be consistent
with the rest of the SED; WISE-4 isn't physically reasonable. We
turned both the WISE-3 and WISE-4 detections into limits based on this
image assessment. We retained this source as a YSO candidate; it was in the
list of literature YSOs because it appears in Winston \etal\ (2020).}
\label{fig:limitexample}
\end{figure}

There are several cases where the WISE pipeline identified high
signal-to-noise ratio (SNR) sources in the 12 and 22 \mum\ images, but
individual inspection of the images suggests that the SNR was
overestimated, and the detections should instead have been limits.  In
those cases, we were guided by the morphology in the images
themselves, in addition to the SED.  Figure~\ref{fig:limitexample} is
an example of this sort of source.

Where source confusion was clearly very important, we also checked 
the optical images of our targets by pulling corresponding images 
from the IPHAS or PanSTARRS archives.

\subsection{SED Inspection}
\label{sec:vettingseds}

After merging the available data (Sec.~\ref{sec:data}), we created
SEDs for all the sources, combining all available data. The units of
these plots (used in figures here, such as in
Figs~\ref{fig:sedgoodexample}-\ref{fig:limitexample}) are cgs units for the $\lambda
F_{\lambda}$  axis, erg sec$^{-1}$ cm$^{-2}$, and the wavelength axis
is in microns.  Symbols used for the various data sets are listed in
Table~\ref{tab:datalist}. Sample SEDs are provided here, but a full set of 
SEDs has been delivered to IRSA along with the data from this paper.

\begin{figure}[htb!]
\epsscale{1}
\plottwo{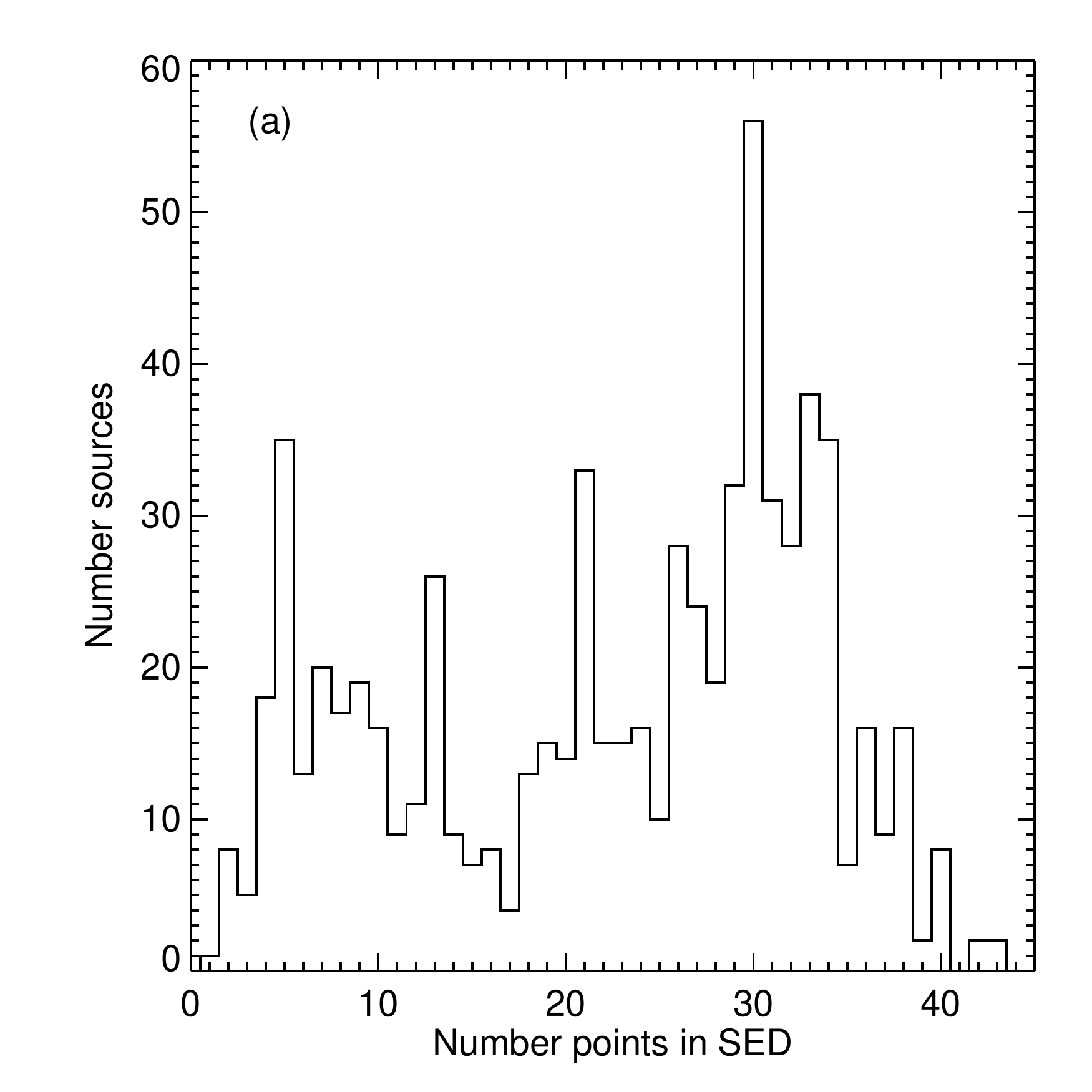}{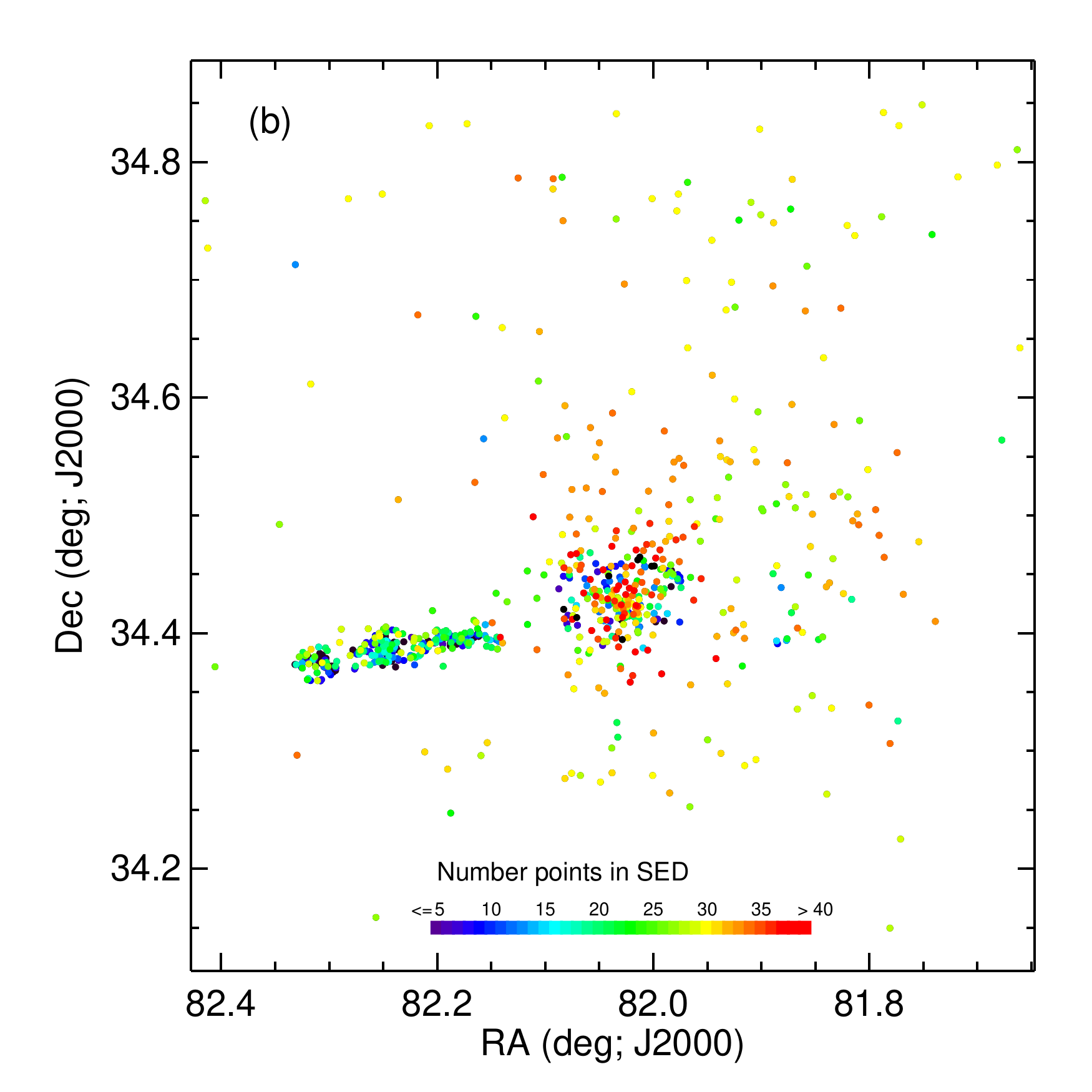}
\caption{(a): Histogram of number of points in the SED created
from the catalog merging. Nearly two-thirds of the sources have more than 20
points defining the SED, and just $\sim$10\% have 5 or fewer points
defining the SED. For most sources, we have enough photometry to
characterize the object fairly well. (b): Distribution of
points on the sky, where the color corresponds to number of points in
the SED, where black=5 or fewer points per SED and red=35 or more points per
SED (color scale shown in plot). Most of the best-populated SEDs are 
in the best-studied portion
of this region, Stock~8. Most of the least-well-populated SEDs are in
the NS, where SEDs can be only two IRAC points.   (Compare to
Figs.~\ref{fig:where2} \& \ref{fig:where3sky}; see Sec.~\ref{sec:s8ns}
for zoom-in on just the NS.) Given how many points we have in most of the
SEDs here, we should be able to make some well-founded assessments
of the status of most of these candidate YSOs.}
\label{fig:sednpts}
\end{figure}

We have photometry ranging from 0.34 to 160 \mum, but no stars have
complete coverage over that whole range.  The Koenig color selection
requires at least the first three bands of WISE and 2MASS $H$ and \ks,
so all of the sources so selected must, by definition, have at least 5
points delineating their SEDs between 1 and 12 \mum.  For the
worst-characterized sources (all in the regions of highest source
surface density where source confusion is rampant, e.g., the heart of
Stock~8 or in the NS), we may have only one or two points from
Spitzer, or multiple detections but all at one or two bands (WISE-1
and -2 from AllWISE, CatWISE, \& unWISE). Figure~\ref{fig:sednpts} has
the distribution of points per SED. Nearly two-thirds of the sources have
more than 20 points defining the SED, so for most sources, we have
enough photometry to characterize the object fairly well. Just
$\sim$10\% have 5 or fewer points defining the SED. Later in the 
process (Sec.~\ref{sec:finalrankings}), objects that have fewer points
in their SED are ranked as less confident YSO candidates.  Most of the
least-well-populated SEDs are in the NS, which is unsurprising given
that we defined the NS by all objects encompassed by the polygon in
Fig.~\ref{fig:where3sky}, a region of high reddening and high 
source surface density (see more discussion below). Most of the
best-populated SEDs are in Stock~8, which is  the best-studied portion
of this region (e.g., J17, Lata \etal\ 2019).  

We reviewed all the SEDs in conjunction with the image inspection
(Sec.~\ref{sec:vettingimages}). We found sources likely to be nebular
knots (Fig.~\ref{fig:rejectexample}). We identified cases where the
position-based source matching across bands had clearly failed 
(Sec.~\ref{sec:datamerging}), betrayed by an unphysically
discontinuous SED; in those cases, we returned to image inspection and
catalog merging, and checked to make sure the bandmerging across
catalogs had been done correctly, finding and resolving any errors
where possible.  Some of the O and B stars are very bright, and
unphysical SED shapes were a result not necessarily of source mismatch
so much as saturation; in those cases, the counterpart at that
saturated band was removed from the catalog.  Some sources had [12] or
[22] values that were unphysically discontinuous with the rest of the
SED; we returned to the images and checked the AllSky (rather than
AllWISE) catalogs to decide what measured flux value was most
appropriate to use. In some cases where a point source was not
apparent in the [12] or [22] images, we converted the values reported
as detections in the AllWISE  catalog for WISE-3 and/or WISE-4 into
upper limits (see  Fig.~\ref{fig:limitexample}). 

\begin{figure}[htb!]
\epsscale{1}
\plotone{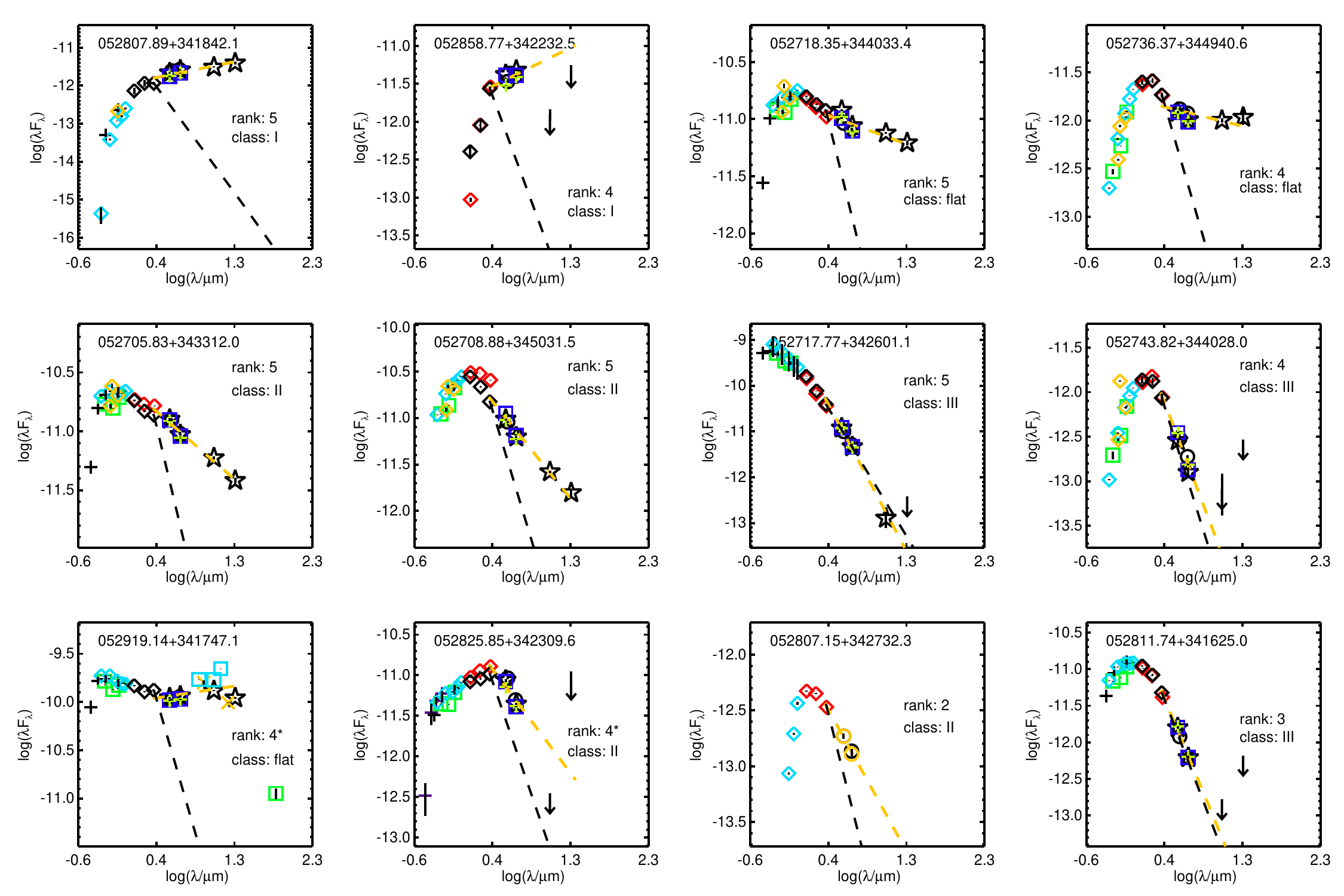}
\caption{Twelve example SEDs, with plot symbols as defined in 
Table~\ref{tab:datalist}. The $y$-axis is the log of the spectral 
energy density in cgs units, e.g., ergs s$^{-1}$ cm$^{-2}$; the $x$-axis
is the log of the wavelength in microns. The black dashed line is a
Rayleigh-Jeans line extended from $K_s$, assuming that $K_s$ is on
the photosphere of the star. The yellow dashed line is a fit to all
available detections between 2 and 25 \mum. The sources are, from 
top to bottom, left to right, are:
J052807.89+341842.1, final rank 5, SED Class I;
J052858.77+342232.5, final rank 4, SED Class I;
J052718.35+344033.4, final rank 5, SED Class flat;
J052736.37+344940.6, final rank 4, SED Class flat;
J052705.83+343312.0, final rank 5, SED Class II;
J052708.88+345031.5, final rank 5, SED Class II;
J052717.77+342601.1, final rank 5, SED Class III; 
J052743.82+344028.0, final rank 4, SED Class III;
J052919.14+341747.1, final rank 4*, SED Class flat;
J052825.85+342309.6, final rank 4*, SED Class II;
J052807.15+342732.3, final rank 2, SED Class II; and
J052811.74+341625.0, final rank 3, SED Class III. 
See the text (and Appendix~\ref{app:briefexamples}) 
for more discussion.}
\label{fig:sampleseds}
\end{figure}

Figure~\ref{fig:sampleseds} provides twelve example SEDs, representing a
range of YSOs. The reasons for their final
rankings are discussed in more detail in Appendix~\ref{app:briefexamples}.

Despite some sources appearing point-like in the images, their SEDs do
not look like textbook YSOs in that they could be more like quasars or nearby
star-forming galaxies or even giants, but could also be consistent with 
highly variable YSOs. Follow-up spectroscopy will be required to
determine the nature of these sources. We retained them as somewhat
lower-confidence YSO candidates, largely on the basis of a Gaia
distance that is in the right regime to be part of IC~417. 

Following Wilking \etal\ (2001) and, e.g., Rebull (2015), we define
the near- to mid-IR (2 to 25 \mum) slope of the SED, $\alpha = d \log
\lambda F_{\lambda}/d \log  \lambda$, where  $\alpha > 0.3$ for a
Class I, 0.3 to $-$0.3 for a flat-spectrum  source, $-$0.3 to $-$1.6
for a Class II, and $<-$1.6 for a Class III.  For each object,  we
performed a simple least squares linear fit to all available
photometry (just detections, not including limits) as observed between
2 and 25 $\mu$m, inclusive. These classes are included in
Table~\ref{tab:bigdata}.

\subsection{Color-Magnitude and Color-Color Diagrams}
\label{sec:vettingcmds}

Koenig's color selection cuts only use 2MASS and WISE, and selection
of sources by position in the NS will only weakly constrain the YSO
nature of the candidates. But we have considerable ancillary data
(Sec.~\ref{sec:data}).  We can therefore further cull sources by
making color-color and color magnitude diagrams and investigating
whether each source appears in positions consistent with a YSO
status.  This approach follows, \eg, Guieu \etal\ (2010) or Rebull
\etal\ (2011).

\begin{figure}[htb!]
\epsscale{1.1}
\plotone{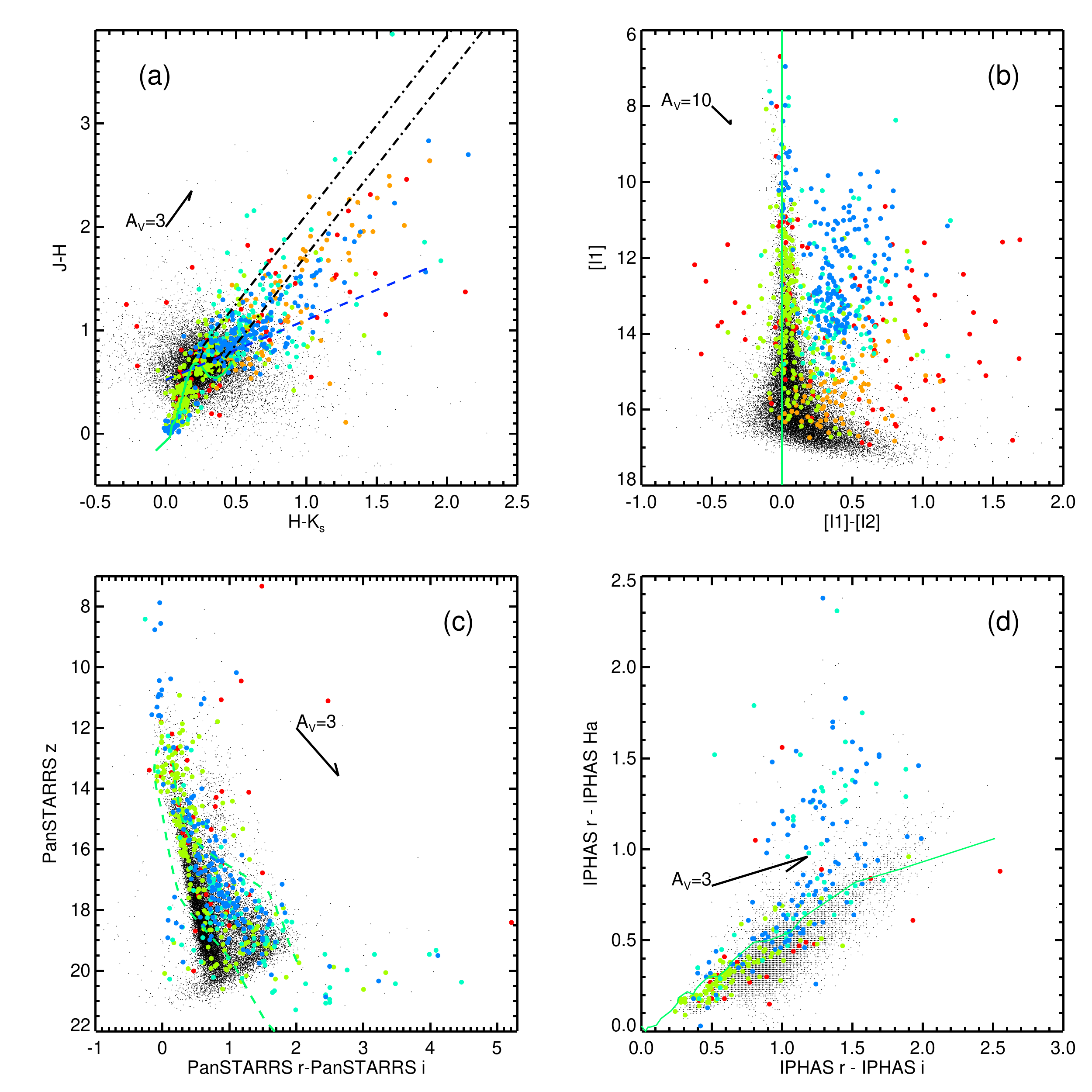}
\caption{Sample color-color and color-magnitude diagrams used to
assess further the quality of the YSO candidates. In each plot, small
black dots are the ensemble catalog, and larger dots are the YSO
candidates. Diagrams like this were used to assess the confidence we
had that the YSO candidates were actually YSOs. Anticipating
discussion in Sec.~\ref{sec:finalrankings}, the colors of the large
dots follow the final YSO rankings. Targets ranked 5  (highest) are
blue, 4 are cyan, 3 are green, 2 are orange, and  1 (lowest) are red;
the bluer the symbol, the more reliable a YSO candidate it is.  Many of
the outliers are lower-ranked YSO  candidates. The highest-ranked YSOs
include both early-type stars and stars without too much $JHK_s$
reddening or a possible $JHK_s$ excess  (a; upper left), stars with an
IRAC excess (b; upper right), stars clustered near the 6 Myr model
isochrone (c; lower left), and stars with an H$\alpha$ excess (d; lower
right). Reddening vectors (following the reddening law from Indebetouw
\etal\ 2008 and Mathis 1990) are as shown. Green solid lines are the
expected (empirical) ZAMS relationship.  In the $JHK_s$ plot, the ZAMS
is taken from Pecaut \& Mamajek (2013), the dashed blue line is the
Meyer \etal\ (1997) T~Tauri  locus, and the dash-dot lines are
reddening vectors extending roughly from the  green ZAMS relation to
give an indication of which of these stars could be reddened MS stars.
The green dashed lines in the PanSTARRS plot are 6 Myr and 9 Myr
isochrones from PARSEC models (Bressan \etal\ 2012). The IPHAS ZAMS is
from Drew \etal\ (2005); the IPHAS data appear quantized due to the
precision with which the magnitudes are reported in H$\alpha$. }
\label{fig:cmds2}
\end{figure}

Our process included identifying each star separately in several
different color-color and color-magnitude diagrams; because we have
so much available photometry, we have a lot of diagrams to choose from. 
The diagrams we used primarily included $J-H$ vs.~$H-K_s$, 
[I1] vs.~[I1]$-$[I2], [W3]$-$[W4] vs.~[W1]$-$[W2], Pan-STARRS 
$z$ vs.~$r-i$, Pan-STARRS $g-r$ vs.~$i-z$, IPHAS $r-H\alpha$ vs.~$r-i$, 
and Gaia DR3 $G$ vs.~$G_{BP}-G_{RP}$ observed and absolute.
Figure~\ref{fig:cmds2} shows four sample color-color and
color-magnitude diagrams out of the several we used. In each case, 
points from the ensemble catalog are shown in addition to the YSO
candidates.  Reddening vectors as shown are calculated following the
reddening law  from Indebetouw \etal\ (2008) and Mathis (1990).  The
expected zero-age main sequence (ZAMS) in the near-IR is taken from
Pecaut \& Mamajek (2013), and the T~Tauri locus as shown is from Meyer
\etal\ (1997).  The model isochrones in the PanSTARRS plot are 6 Myr
and 9 Myr isochrones from PARSEC models (Bressan \etal\ 2012), shifted
to 2 kpc. The IPHAS ZAMS is from Drew \etal\ (2005). 

If a given object was an outlier in any diagram, we returned to its
SED and even its images to determine if the data causing the outlying
location was erroneous. For objects that appear as outliers, such as
the apparently too blue sources in the IRAC color-magnitude diagram,
checking the SEDs show that indeed there is something ``off'' for one
or both IRAC channels for that object given the rest of its SED, but
it is not severe enough, given the image and SED as a whole, to merit
unhooking the star from the IRAC counterpart.  Therefore, too blue in
IRAC doesn't exclude a source if the rest of the information we have
about it suggests it is still a YSO candidate,  but it may lessen the
confidence we have that it is young and a member.  We were able to
notice patterns, such as the  well-known feature that a faint measure
in the Gaia blue band is often `off' given the rest of the SED, so a
Gaia $G$ vs.~$G_{BP}-G_{RP}$ color-magnitude diagram presents many
apparent faint outliers, whereas $G$ vs.~$G-G_{RP}$ for those outliers
may be fine.  The faintest stars in PanSTARRS often have considerable
scatter (unsurprisingly), which is readily apparent in both the SEDs
and the color-magnitude diagrams. Sources that are outliers in more
than one plot received more scrutiny and were demoted depending on the
source's properties.

\begin{figure}[htb!]
\epsscale{1}
\plottwo{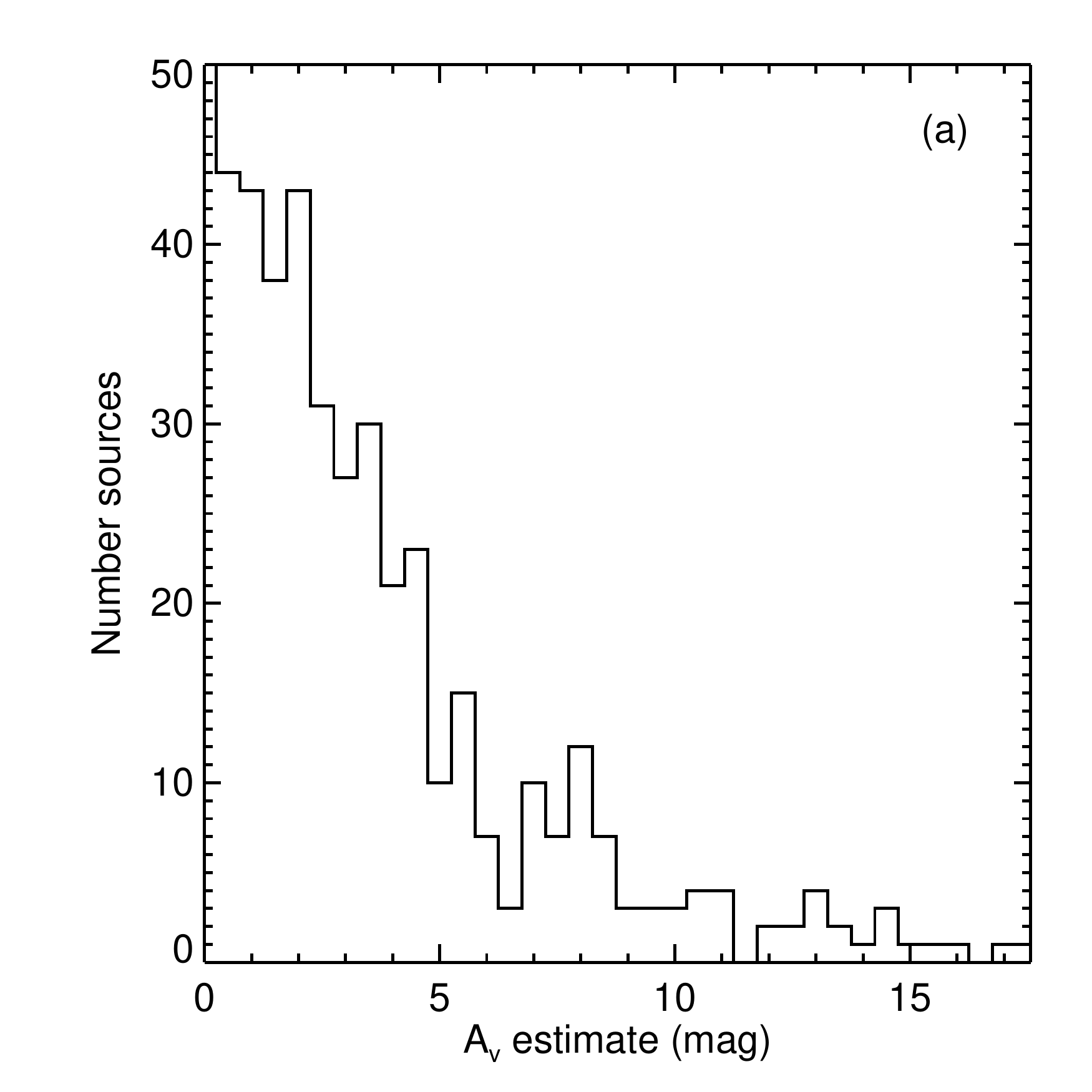}{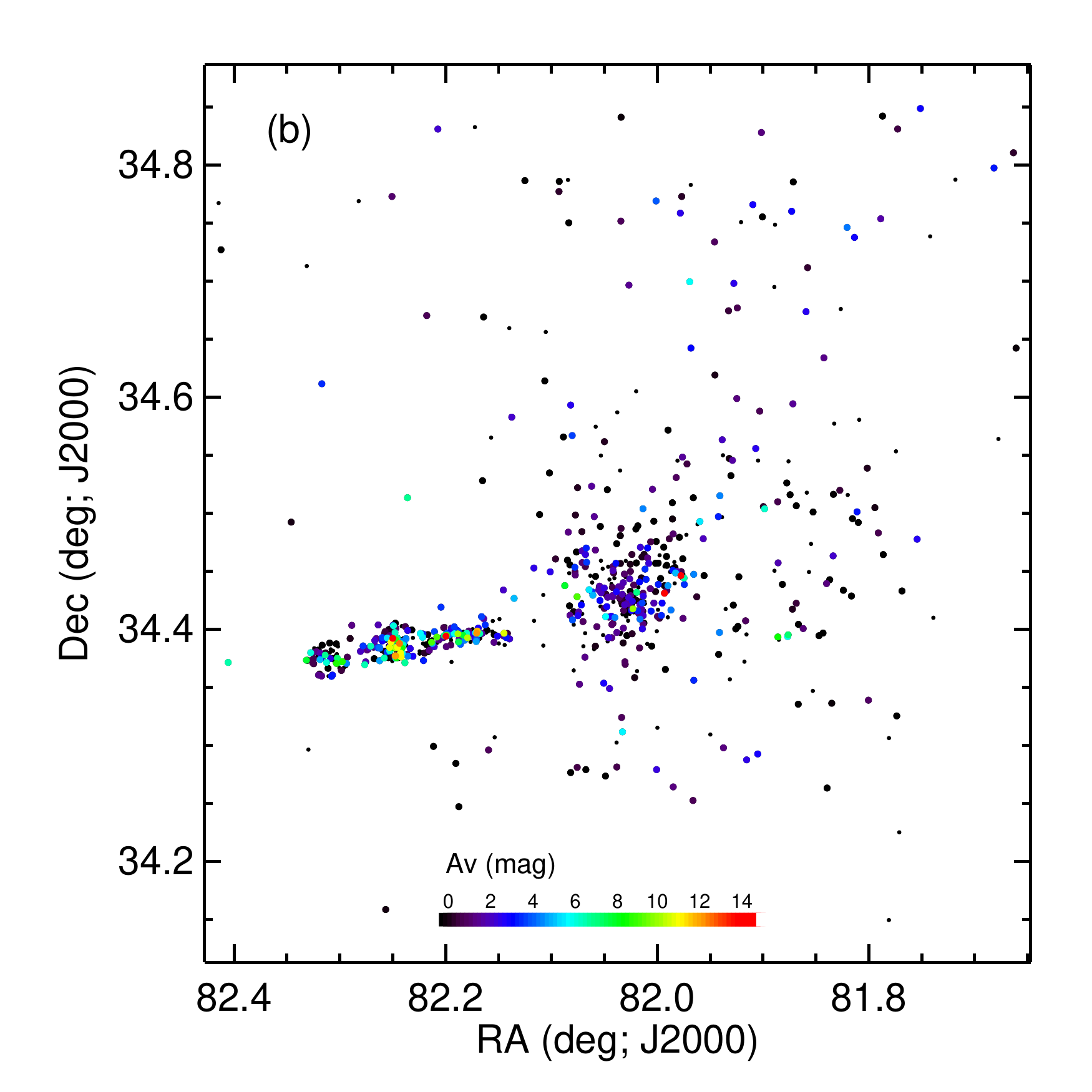}
\caption{(a) Histogram of $A_v$ estimates, derived from $JHK_s$  as
discussed in the text. The histogram peaks at $\sim$100 stars with
$A_v\sim$0 (offscale here). The distribution falls off with $A_v$,
with a max($A_v$) of $\sim$17 mag. (b) distribution of sources on
the sky, with color corresponding to $A_v$ (redder point is more
$A_v$). Small dots: sources for which we could  not derive an $A_v$
estimate; larger circles are sources for which we could derive an
estimate, and the redder the symbol, the higher the $A_v$ estimate.
Most of the stars have little reddening. Most of the high-$A_v$
sources are in the NS, but the two with the highest estimate are in
Stock~8 (compare to Figs.~\ref{fig:where2} \& \ref{fig:where3sky}; see
Sec.~\ref{sec:s8ns} for zoom-in on just the NS). }
\label{fig:avhisto}
\end{figure}

Because we are leveraging the position of each source in various
color-color and color magnitude diagrams, reddening could be a 
significant factor in the placement of the star in said diagrams,
particularly those using shorter wavelengths.  To account for,
estimate, and limit the influence of reddening, we wished to make
plots where each star in question had been dereddened.  In cases where
we had $JHK_s$ for the target (93\% of the YSO candidates), we were
able to use the same dereddening approach as in Rebull \etal\ (2022; 
2020) where we create a $J-H/H-K_s$ diagram and slide the observed
$JHK_s$ back along a reddening vector (Indebetouw \etal\ 2005) to
expected colors from Pecaut \& Mamajek (2013) or the T~Tauri locus
(Meyer \etal\ 1997). This process results in merely an estimate of the
reddening, but this is an efficient way to get an estimate. It fails
outright for about a third of the targets, because, even given the
best available $JHK_s$, there is no way to deredden and still end up
in a reasonable location (see outliers in Fig.~\ref{fig:cmds2}).  As
seen in Fig.~\ref{fig:avhisto}, about 100 ($\sim$15\%) of the stars
end up with an estimate  of $A_v\sim$0, and the distribution falls off
steeply with $A_v$.  There are more high-$A_v$ sources in the NS than
any other place, but the largest estimates ($\sim$17 mag!) are for two
very reddened (and embedded) stars in Stock~8.

For each target with an $A_v$ estimate, we plotted the color-color and
color-magnitude diagrams with the observed and dereddened position
indicated. In this fashion, we tried to distinguish (in optical diagrams) 
between reddened background giants and red YSOs, and included this 
consideration in the final ranking of the YSO candidates.

Additionally, for two diagrams, we explicitly calculated the 
significance of the excess, following, e.g., Mizusawa \etal\ (2012).  
For the IRAC data (e.g., Fig.~\ref{fig:cmds2}), we calculated $\chi_{{\rm IRAC}}$, 
where 
\begin{equation}
\chi_{{\rm IRAC}} = \frac{([I1]-[I2])_{\rm observed} - [I1]-[I2])_{\rm expected}}{\sqrt{\sigma_{[I1]}^{2} + \sigma_{[I2]}^{2}}}
\end{equation}
For the mass ranges we are likely to detect in IC~417 
(earlier than mid-M), $([I1]-[I2])_{\rm
expected}$ is 0. We took there to be a significant excess in the IRAC
bands when $\chi_{{\rm IRAC}}>$3; an IRAC excess suggests a dusty disk, making it
more likely that a star is young.  Similarly, for the IPHAS
color-color diagram, 
\begin{equation}
\chi_{{\rm IPHAS}} = \frac{(r-H\alpha)_{\rm observed} - (r-H\alpha)_{\rm expected}}{\sqrt{\sigma_{r}^{2} + \sigma_{H\alpha}^{2}}}
\end{equation}
The expected $(r-H\alpha)$ is taken from the IPHAS ZAMS 
and is calculated assuming that $r-i$ is not subject to
reddening, a poor assumption in general. However, the reddening
vector is largely parallel to the ZAMS (see Fig.~\ref{fig:cmds2}), so
even large errors in reddening are unlikely to create a false
H$\alpha$ excess. Here, again, we took there to be a significant
H$\alpha$ excess when $\chi_{{\rm IPHAS}}>$3. An H$\alpha$ excess can arise from
accretion in young stars, or from stellar activity. Stellar activity
is generally higher in young stars (which are, on average, rotating
faster than main sequence stars), so an H$\alpha$ excess can be
indicative of youth. Because an H$\alpha$ excess need not uniquely
identify youth, the influence of any H$\alpha$ excess on the final
ranking of the star was less than the influence of any IRAC excess. 
Rarely, some stars appeared to have a $g$-band excess. The $\chi_{{\rm IRAC}}$ and 
$\chi_{{\rm IPHAS}}$ values,
as well as an indication of whether or not stars had an IR excess, an H$\alpha$
excess, or a blue ($g$-band) excess, are all included in Table~\ref{tab:bigdata}.

For those stars with measured Gaia DR3 parallaxes and distances
from Bailer-Jones \etal\ (2021), we also looked to see if the star was
between 1 and 3 kpc away (or had a distance that was within 1$\sigma$ of
1-3 kpc away), which is the expected range of distances we took
to be associated with IC~417 (also see Appendix~\ref{app:distances}
on distances). 
We investigated whether or not proper 
motions would be helpful in selecting members of IC~417 (or for
any of the clusters described in Sec.~\ref{sec:intro}); there
is nothing obviously helpful to be found among the proper motions, 
likely as a result of the significant distance.

\subsection{Final Rankings}
\label{sec:finalrankings}

For each one of the sources that survived the vetting, we assessed 
whether or not the star, given all the information we had amassed, was
consistent with being a YSO candidate. Each reviewer ranked the
targets and then results were combined for a final net grade. We
placed each in one of basically five bins, where 1 is less likely to
be a YSO and 5 is more likely to be a YSO. We had to create additional
major subdivisions in a few specific cases, discussed below. Within
each of these ranks, we grouped apparently similar kinds of objects
together, and then ordered them within each rank roughly by
confidence, such that lower ordering within the rank was less
confident. For example, in the lowest ranks,  all of the YSOs that
have only 2 IRAC points in their SED are grouped together, followed by
those that have 2 IRAC and 1 2MASS point, then 2 IRAC, 1 2MASS, and 1
other point, and so on; in the higher ranks, stars with H$\alpha$
excesses as well as IR excesses can be found together, the OB stars
can be found together, etc. The SEDs in the IRSA delivery are 
provided in this rank
order so that paging through the SEDs is easy, and apparently
similar objects can be found near each other in the list. 

The rank 1 stars, based on the information we currently have, are
least likely to be true YSOs at the distance of IC~417. There are
several major subcategories within the rank 1 bin, which we now describe.

The ``1r'' (r for reject) stars are least likely to be young, 
consisting of stars that have to be rejected due to irreconcilable
source confusion or because they are confirmed or likely carbon
stars.  There are three known carbon stars here, and we believe
we have identified a fourth -- see Sec.~\ref{sec:obc}.

The next major category within the rank 1 bin is ``1d'' (d for
distance), which means that the SED is well-defined, with many points,
and it seems fine but has no obvious indications of youth, and
moreover the Gaia DR3 distance is unambiguously too far or too close.
With no good reason to retain it as young (besides the criterion/a that
placed it in our YSO candidate list initially), we therefore put 
the star in the rank 1d bin.

The ``1f'' (f for few) means that the SED really has too few
unique wavelengths to reliably assess it (typically $\lesssim$8-10
points). The 1f stars will need much more information before we can
determine whether or not they are young and at the distance of
IC~417.  These stars could be young members of IC~417; we just don't
know yet.

The rest of the rank 1 stars are, in general, relatively
unremarkable, relatively sparse SEDs (particularly by comparison to
the rank 2-5 stars), with few indicators of youth or available
distances.  Rank 2 stars typically have more points in their SEDs than
rank 1 stars, but still not a lot of indications of youth or
available distances. 

Ranks 3, 4, and 5 are where the most likely YSO members of IC~417 can
be found. Typically, all of these stars have many points defining 
their SEDs. (Rank 3, 4, or 5 stars have on average $\sim$24 points in
their SED, to be compared with $\sim$7 for rank 1 or 2.)  In general,
all the stars of rank 3 are sort of mid-grade in that there is no
compelling reason to drop them (they are in places consistent with
youth in all the relevant diagrams), but also no strong reason to keep
them either, such as an excess of any sort, or even necessarily 
significant reddening. (Rank 3 stars have mean \av$\sim$1 mag, and
rank 4 or 5 stars have mean \av$\sim$3-4 mag; see
Fig.~\ref{fig:cmds2}.)   All the rank 4s and 5s have an excess of some
sort -- most commonly a significant IR excess, but H$\alpha$,
and/or blue excesses can also be found (see Fig.~\ref{fig:cmds2}). If
there is a DR3 distance estimate for the rank 5s, it is in the right
range, $\sim$1-3 kpc, to be a member of IC~417, at least within
1$\sigma$. If there are obvious and/or multiple signs of youth but
the distance is not quite right, then the star is set as a rank 4. The
subcategory ``4*'' is reserved for those that seem to have the right
distance, and within a given survey, the colors seem ok (e.g., within
PanSTARRS, the CMD placement is ok), but the complete, net SED shape
is unusual or confusing, not necessarily suggestive of a YSO, and 
so follow-up spectroscopy is particularly needed to check on the 
status of these targets.  

\begin{figure}[htb!]
\plottwo{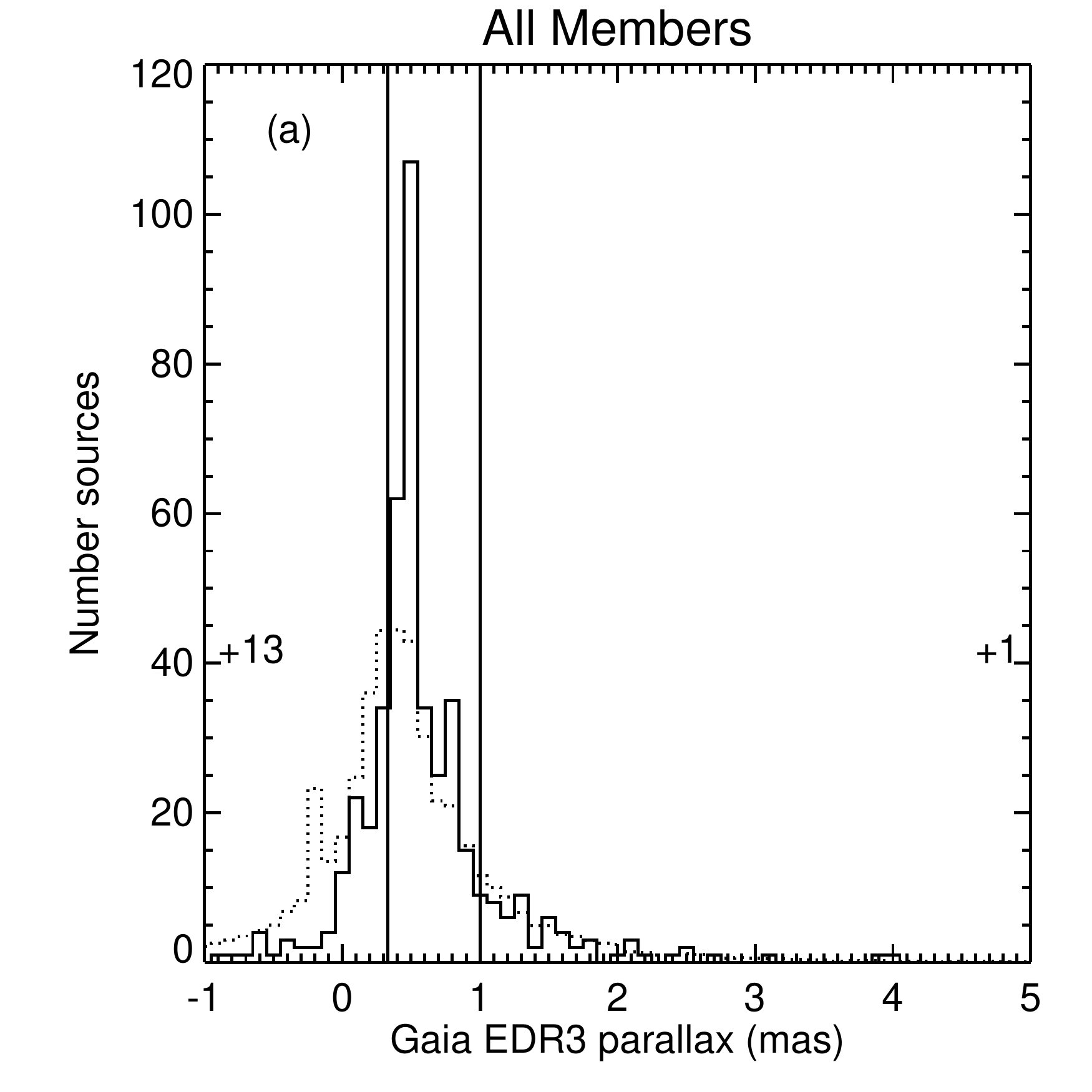}{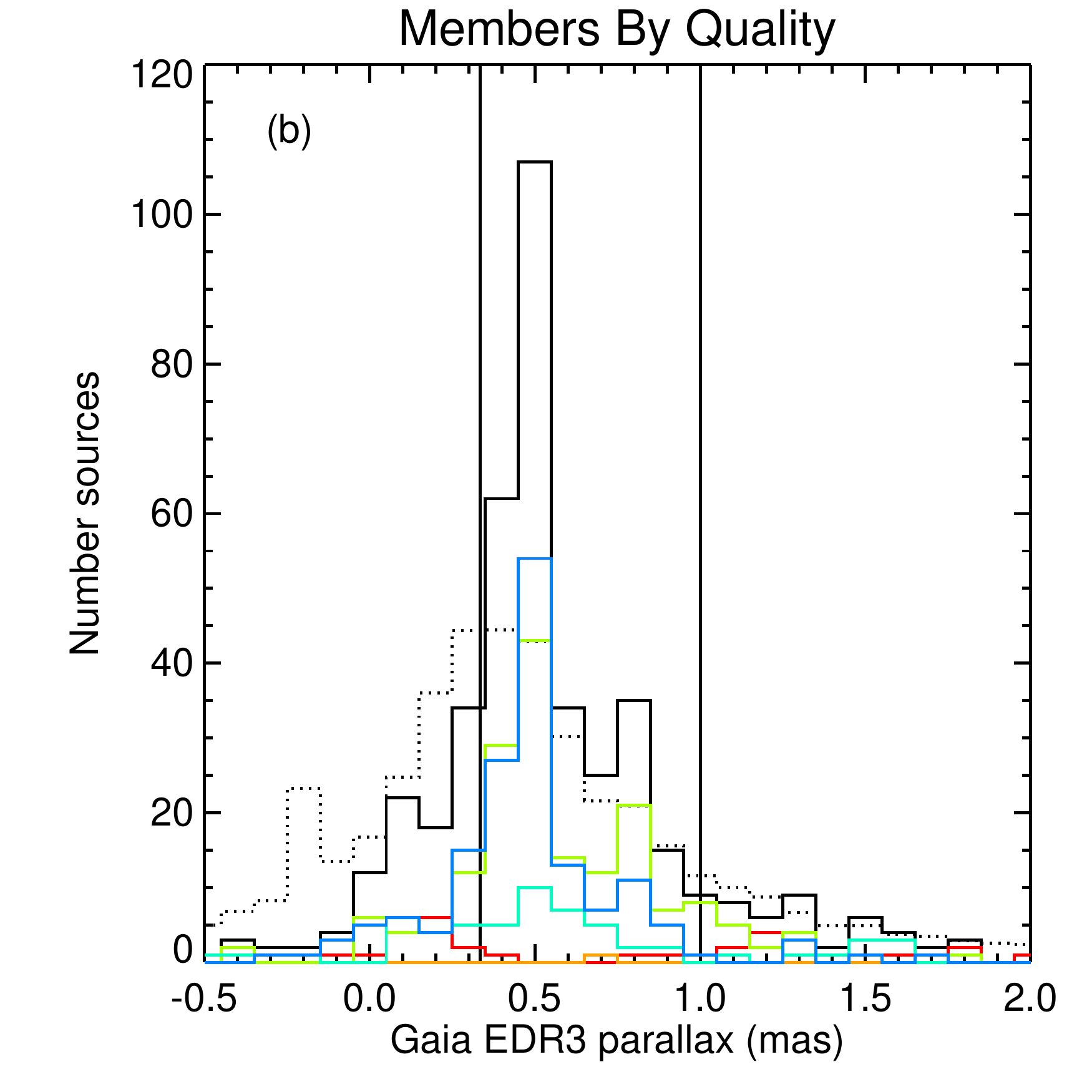}
\caption{ Histograms of Gaia DR3 parallaxes.  (a): Solid line
is all YSO candidates; dotted line is the rest of the catalog, scaled
to have the same sample size as the YSO candidate sample.   The solid
vertical lines are at 1 and 3 kpc, e.g., the range of distances we took 
to be consistent with IC~417.  The
numbers on the right and left of the plot indicate how many objects in
the member sample fall beyond the ranges of the plot (e.g., there are
13 stars with  DR3 parallaxes $<-1$).  The member sample is more
tightly biased towards objects between 1 and 3 kpc; the rest of the
catalog is slightly farther away, on average, consistent with it
being largely the rest of the Perseus arm population.  (b):
zoom in on a smaller range of parallax, with the YSO rank plotted
separately, color-coded as in earlier figures (5=blue or highest
quality, 4=cyan, 3=green, 2=orange, 1=red or lowest rank).  The
bulk of the YSO candidates that are at the correct distance are
largely rank 3, 4, or 5; the YSO candidates that are too close or too
far away are largely rank 1 or 2. }
\label{fig:disthisto}
\end{figure}

Figure~\ref{fig:disthisto} shows a histogram of the parallaxes for
the entire YSO sample, in context with a scaled histogram of
parallaxes to everything in the catalog  ($\sim$19,000). It also shows
the breakdown of the parallaxes for each ranked sample of YSO
candidates.

There is a strong peak among the YSO candidates at $\sim$2 kpc, which
is approximately the expected distance of IC~417. We took 1-3 kpc as
the range of distances we accepted as consistent with IC~417, and that
range is indicated on the plot.  The entire catalog is on average 
slightly farther away than IC~417, consistent with IC~417 being on
the  near side of the Perseus arm, as suggested by MN16. However, the
peak of the background distribution is within the 1-3 kpc range we
accepted as more likely to be members. This suggests that more work
will be needed to identify objects securely as being part of IC~417.
(Also see Appendix~\ref{app:distances} on distances.)

The sources that are outside the 1-3 kpc range are more often
lower-ranked YSO candidate quality, as a result of our selection
approach.  We accepted some YSO candidates as higher quality if the
error bars on distance (as provided in Bailer-Jones \etal\
2021) brought the star within range, and the SED/colors/excesses were
consistent with youth, so some high-quality YSO candidates populate
the bins outside the 1 and 3 kpc limits.

The rankings (the coarse rank and the ordering within each rank) are
included in Table~\ref{tab:bigdata}. The colors of the points in
Fig.~\ref{fig:cmds2} reflect these 1-5 YSO rankings. For the remainder
of the paper, the ``entire YSO candidate sample''  means the 710 that
have made it this far. The highest-quality YSO sample is made up of
the 503 stars that are ranked 5, 4, 4*, or 3.

\section{The YSO Candidates}
\label{sec:ensembleysos}


Figure~\ref{fig:cmds2} above shows several color-color and
color-magnitude diagrams (also see Appendix~\ref{app:textbook})
with the points colored corresponding to
final rankings.  The $JHK_s$ and IRAC diagrams are the best-populated
diagrams of all those that we constructed, because nearly all the
sources (92\%) have  $JHK_s$ as well as both IRAC-1 and -2.  The
higher-ranked YSO candidates include the early-type stars, which can
be seen distinctly in at least three of the four diagrams. Stars
bright in IRAC with large IRAC excesses, as well as those with large
H$\alpha$ excesses, tend to be highly ranked YSO candidates. Stars
hugging the 6 Myr model isochrone also tend to be highly ranked YSO
candidates. Stars that are likely FGK stars with little reddening,
clustered around the ZAMS relation in the $JHK_s$ diagram, are
typically rank 3, and they are also those stars that have no IRAC or
H$\alpha$ excess, and appear roughly where expected, between the
isochrones in the optical CMD.  As described above, many of these rank
3 stars have nothing very obvious to reject or recommend them as YSO
candidates, other than the property that placed them in the YSO
candidate list initially.  Many of the outliers in all the plots are
lower-ranked YSO candidates. Many of the stars that appear to be
likely giants in the optical CMD are rejected, as are many of the
very red and faint stars in the IRAC CMD.

\begin{figure}[htb!]
\epsscale{1.1}
\plotone{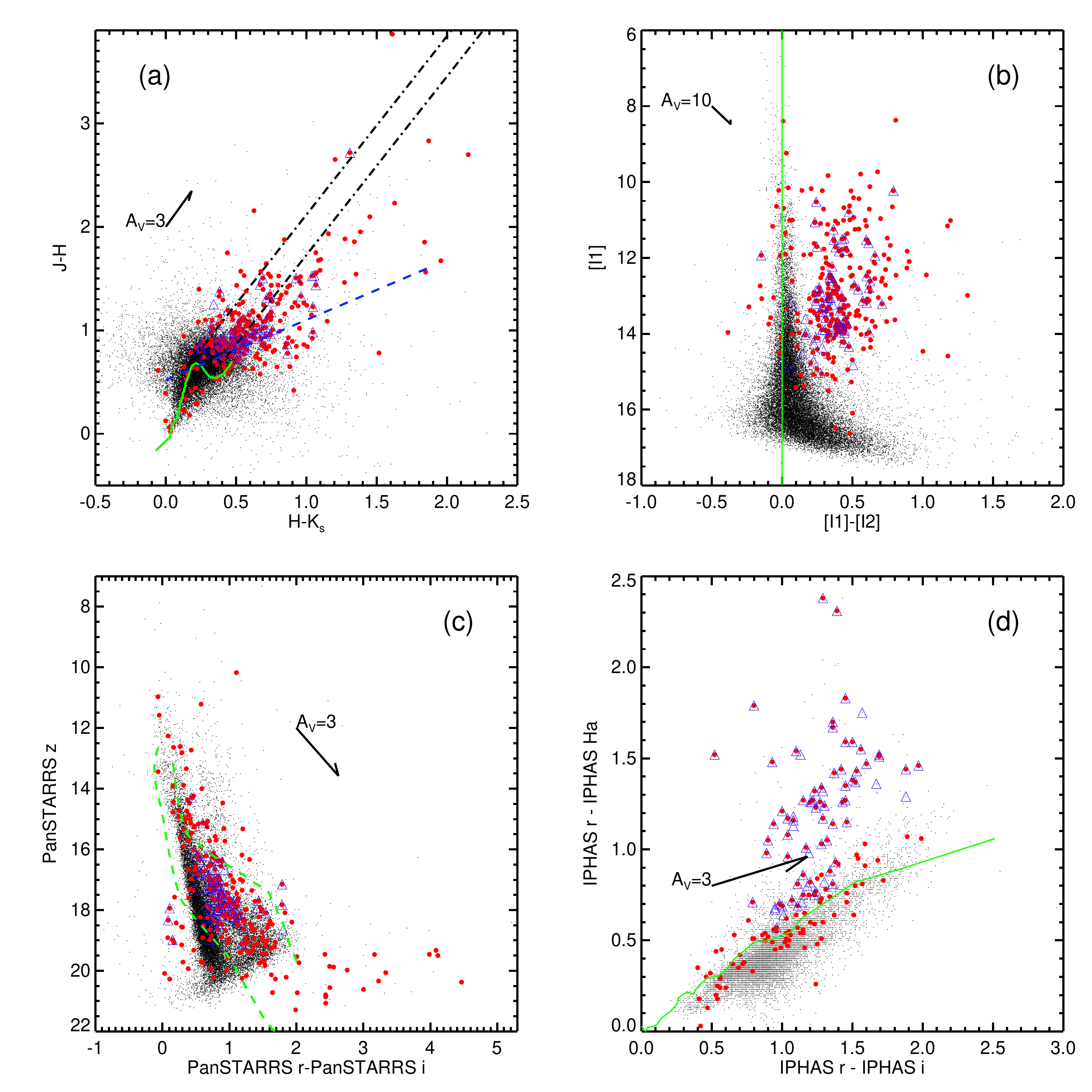}
\caption{Color-color and color-magnitude diagrams as in prior figures;
black dots are the entire catalog and  colored symbols are limited to
YSO candidates with rank 3-5. The stars highlighted with red dots are
those with IR excess at any band (SED Class $<$ III plus Class IIIs
that seem to have an excess at long wavelengths). Additional blue
triangles are those with H$\alpha$ excess. Most YSO candidates with
H$\alpha$ excess also have an IR excess, but relatively few of the IR
excess stars have an H$\alpha$ excess. The stars with IR and/or
H$\alpha$ excesses largely fall where expected in these diagrams --
where YSOs are found (also see App.~\ref{app:textbook}).}
\label{fig:cmds3}
\end{figure}

Figure~\ref{fig:cmds3} includes the same four color-color and
color-magnitude diagrams as considered above, but now, among the rank
3-5 YSO candidates, highlighting those that have an H$\alpha$ excess,
or a clear IR excess at any band.  Most of the stars with an H$\alpha$
excess also have an IR excess, but as far as we can tell (given that 
our H$\alpha$ data does not cover our entire region), relatively few
of the stars with an IR excess also have an H$\alpha$ excess, though
we very well may be limited by data availability.  On the whole,
though, these points largely fall where we expect them to fall.  Many
IR excess sources are reddened in the $JHK_s$ diagram, with many
having large enough IR excesses to influence the NIR. Most of these
sources have significant IRAC excesses.  Most cluster near the 6 Myr
isochrone.  Recall that our H$\alpha$ calculation 
(Sec.~\ref{sec:vettingcmds}) made assumptions, but this diagram shows
that our selection of H$\alpha$ excess stars is reasonably robust and
hopefully indicative of youth.

We note that limiting this plot to the highest quality YSOs still
includes several objects faint in PanSTARRS, resulting in large
scatter. Looking at the SEDs for these stars, it is clear that
PanSTARRS $r$ or $i$ is different than expected given the rest of the
SED. Different optical color-magnitude diagrams show these stars in a
more physically reasonable location. Additionally, some of the
brightest stars highlighted in this diagram are suggestive of giants,
and may in fact be older than YSOs; they persist here in rank 3 or 4 
because they appear to have an IR excess, and we think they
are more likely to be YSOs than giants based on the ensemble of data
accumulated.

Stars with very high accretion rates will not only have ultraviolet (UV)
excesses, but may also produce a blue excess. Because we have $g$ from
PanSTARRS, we explored whether selection of stars with an apparent blue
excess would work for identifying YSO candidates. However, it does
not work well in this sample for several reasons -- there is a lot of
scatter, especially at the short wavelengths where the stars are faint;
there is a lot of uncertainty introduced by reddening; and these
stars may not, on the whole, be young enough to have accretion rates
sufficiently high as to make many large $g$-band excesses. We noted where 
$g$-band excesses happened to occur and used that information to bolster
evidence for youth, but did not select new objects based on $g$-band excesses.

\begin{figure}[htb!]
\plottwo{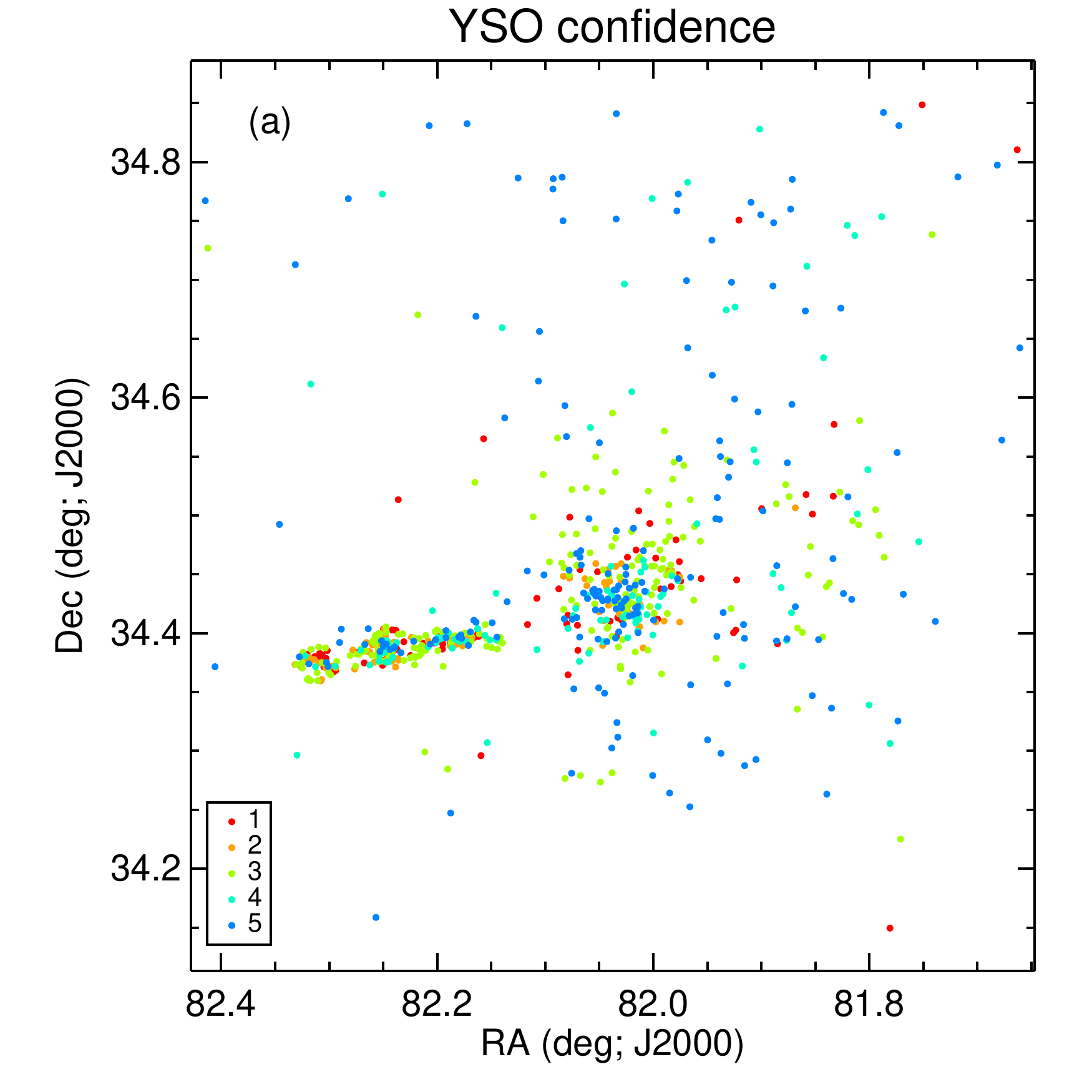}{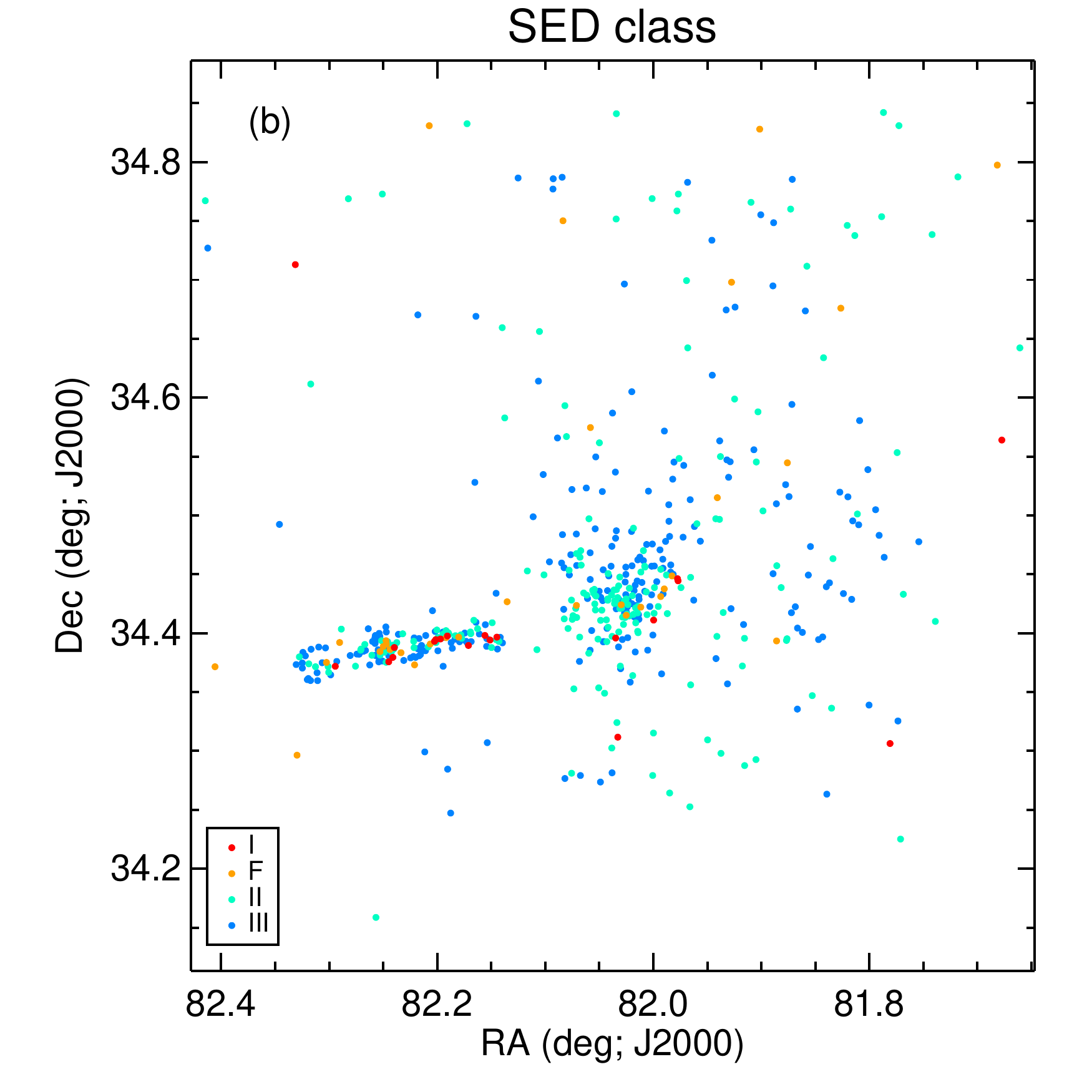}
\caption{Distribution on the sky of YSO candidates. (In both cases,
see Sec.~\ref{sec:s8ns} for zoom-in on just the NS.)  (a): The
YSO quality is color-coded as in earlier figures  (5=blue or highest
quality, 4=cyan, 3=green, 2=orange, 1=red or lowest quality). The
clustering in Stock~8 and the NS is obvious by eye; this is  a result
of how we constructed our candidate list. Many of the lowest-ranked
YSOs are in the highest surface density areas (NS and Stock~8) 
because they have the fewest points in their SEDs.  (b): Just
the rank 3/4/5s, and the SED slope is color coded (Class I=red,
flat=orange, II=cyan, III=blue). Most of the embedded things are in
the NS or Stock~8, but there are some scattered throughout the field.
Most of the objects, however, are less embedded.}
\label{fig:sky2}
\end{figure}

Figure~\ref{fig:sky2} shows the distribution on the sky of the
YSOs/candidates, color-coded by final rank. As noted earlier,  (see
Fig.~\ref{fig:where2} and \ref{fig:where3sky}), Stock~8 and the NS are
both immediately obvious, because of the way our target list was
constructed. High and low ranking YSO candidates are found throughout
the region. The many low-ranking candidates in the NS and Stock~8
arise because these regions have such a high surface density of
sources, and the NS has such high reddening, that many of the
resultant SEDs are poorly sampled, and as a result are low ranking.
This is a larger problem in the NS than Stock 8.  Fig.~\ref{fig:sky2}
also plots the highest-quality candidates on the sky with colors
corresponding to SED class. Most of the sources with very large IR
excesses are also in the NS or Stock~8, though some are found
throughout the field. (Of the 19 Class I sources that are rank 3-5,
11 are in the NS, and 4 are in Stock 8; of the 32 Flat Class sources
that are rank 3-5, 10 are in the NS, and 8 are in Stock 8.) The
majority of the sources in general, however, are Class III or II (see
Table~\ref{tab:ysototals}). 

We note, however, that our sample is likely incomplete.  Our Class III
sample is least likely to be complete, given the generally IR-centric
methods we (and those in the  literature) have used to assemble our
YSO candidate list. Our sample of YSO candidates not in Stock~8 or the
NS is also probably less likely to be complete, because the wider
field area has not been subject to as careful detailed scrutiny as
Stock~8 or the NS. 

Pandey \etal\ (2020) find fewer YSOs within one of the large bubbles
that they identify, where IC~417 is on the southern edge of a bubble.
Given that, we might expect to find fewer YSOs/candidates north of 
IC~417 than south of it. We do not see anything obvious consistent with
that, but we may not be sampling a large enough area to 
adequately test this.

\section{The Obvious Clusters}
\label{sec:obvclusters}

\subsection{Stock~8 and the NS}
\label{sec:s8ns}

In Figs.~\ref{fig:where1} and \ref{fig:where2}, Stock~8 and the 
NS are immediately obvious, 
Stock~8 because the cluster stands out against the background and the
NS because of its nebulosity and red (or reddened) stars. In the 
distribution of YSO candidates (e.g., Fig.~\ref{fig:where3sky}),
Stock~8 and the NS are immediately apparent just because of how we
picked the YSO candidates.  In this section, we explore what, if
anything, we can understand about the NS as a whole, comparing to
Stock~8 (and to some extent the rest of the field).

J17 place Stock~8 at 3 Myr old, and the NS as younger still. Given the
inventory of YSO candidates we have amassed, can we constrain these
ages? One way we can do this is to assess disk fractions as a proxy
for relative age. We can consider those that have a clear IR excess at
any band -- this includes objects with SED Class I, flat, or II, plus
those of Class III whose SED has an unambiguous excess (either much
smaller excesses than IIs, or emerging at longer wavelengths than can
significantly influence the SED slope between 2 and 25 \mum).   The
resultant overall disk fraction in the NS is $\sim$51\%, in Stock~8 is
$\sim$56\%, and outside of either of those clusters is $\sim$56\%.
Assuming Poisson statistics, those numbers are identical, and it seems
possible that all the stars here may indeed be on average about the
same age.  However, as noted above, it is likely, given our IR-centric
methods of finding YSOs, that our Class III sample is incomplete, and 
perhaps we should work only with stars that have IR excesses.
Considering the ratio of (Class I+Flat)/(Class I+Flat+Class II), the
NS is 34$\pm$8\%, Stock~8 is 12$\pm$4\%, and the rest of the field is
21$\pm$6\%. By this metric, then, the NS is the youngest, and Stock~8
is the oldest.

\begin{figure}[htb!]
\epsscale{1.0}
\plotone{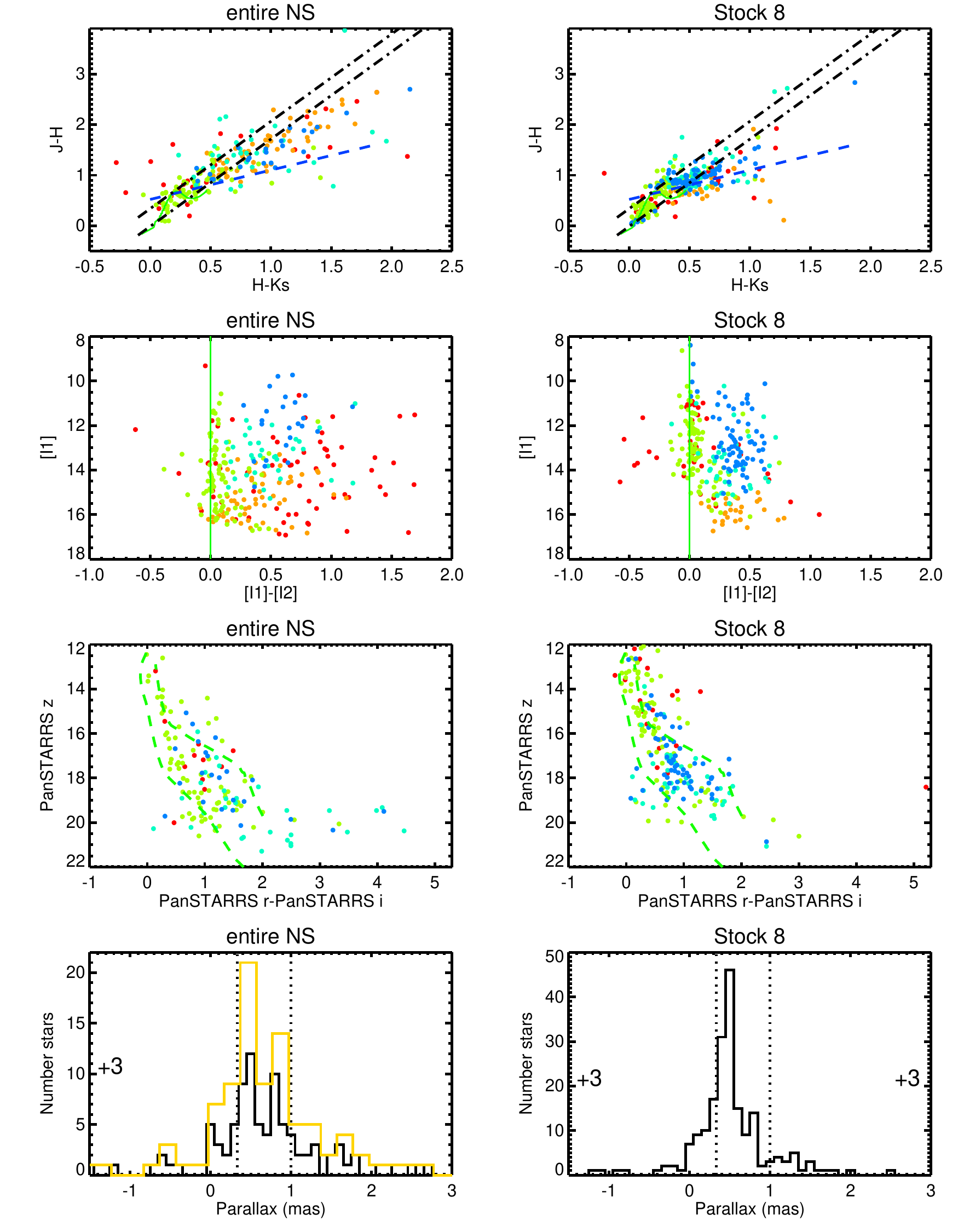}
\caption{$JHK_s$ color-color diagrams (top row), the IRAC
color-magnitude diagram (second row), and the PanSTARRS
color-magnitude diagram (third row), and a histogram of parallaxes
(fourth row) for the entire NS (left column), and Stock~8 (right 
column). Colors in the top three rows correspond to YSO ranks (as in
Fig.~\ref{fig:cmds2} or  Fig.~\ref{fig:nsexp}), so bluer is better. 
Note that the limits of the $JHK_s$ plot are the same as earlier in
the paper, but the limits are adjusted for the other two diagrams. 
The yellow line in the lower left histogram is with a broader bin
size, because there are double the number of stars in Stock~8 (200) 
compared to the NS (92). The vertical dotted lines correspond to
1 and 3 kpc, the range of distances we took to be consistent with 
IC~417. The numbers at the left and right sides of the plot indicate
how many outliers there are (e.g., there are 3 stars with parallax $<-1.5$
mas and 3 stars with parallax $>$3 mas in Stock~8). 
The NS is likely younger than Stock~8, and they are
likely at the same distance.}
\label{fig:nsstock8}
\end{figure}

Figure~\ref{fig:nsstock8} compares the NS to Stock~8 in color-color
and color-magnitude diagrams. There are twice as many low-ranking YSO
candidates in the NS as in Stock~8, but the distributions even without
the low-ranking YSO candidates are distinctly different. There are 71
YSO candidates of rank 5 in Stock~8 (and 39 of 4/4*, and 93 of rank 3
for a total of 203) and just 24 YSO candidates of rank 5 (and 33 of
4/4*, and 79 of rank 3, for a total of 136) in the NS. There is more
reddening and IR-excess-influenced NIR seen in the $JHK_s$ diagrams 
in the NS compared to Stock~8. Also based on the $JHK_s$ diagram,
Stock~8 has more higher mass stars, but the uncertainties in
reddening, particularly in the NS, may be limiting the number of
higher-mass stars we can identify there based solely on photometry.
There are more stars with larger IRAC excesses in the NS. The Stock~8
IRAC colors for the highest rank YSO candidates hover near
[I1]$-$[I2]$\sim$0.5, whereas the highest rank YSO candidates in the
NS are dispersed out to $\sim$1.5, though admittedly there are fewer
of them.  The optical color-magnitude diagram is the most uncertain in
the NS because of the \av\ uncertainties, but there is less scatter in
the Stock~8 distribution, which is constrained more tightly between
the 6 and 9 Myr isochrones.  The distance distribution is also
markedly different perhaps due in part to reddening (there are about
twice as many stars available in Stock~8 with Gaia distances than in
the NS) -- the histogram of distances to the few stars with distances
in the NS is rather flat, and the distribution for Stock~8 is sharply
peaked at $\sim$2 kpc. If the bin size is broadened for the
less-well-populated NS, then there is a broad peak in the NS centered
on 2 kpc.

It seems reasonably likely from our analysis here that the NS is
younger than Stock~8, and both are at about the same distance.  This
agrees with some literature -- J08 suggested that the NS is at about
the same  distance as Stock~8 (MN16 disagrees), and J17 found that the
NS is likely younger than  Stock~8.

\subsection{Within the NS}
\label{sec:insidens}

The NS is one of the most striking things in Fig.~\ref{fig:where1}, 
and the clumping of the red (or reddened) stars within it is
immediately  obvious. In Figure~\ref{fig:where3sky}, we broke the NS
distribution into those four sub-clusters just based on a by-eye
assessment. NS 3 (defined there) is coincident with BPI~14, though the
nominal definition of BPI~14 is larger than our NS~3. NS~3 is the
best-populated sub-cluster in the NS, consistent with BPI~14 being
obvious enough that it has been already identified as a cluster. 

If there are differences across the NS, it would shed light on the
star formation process happening here.  The spacing of the clusters on
the sky is one of the things that draws the eye. Measuring the cluster
spacing by eye in a variety of ways and comparing the mean RA and dec
for the stars defined for each cluster yields spacing of NS1 to NS2
of $\sim90-100\arcsec$, NS2 to NS3 of $\sim130-150\arcsec$, and NS3
to NS4 of $\sim200\arcsec$. At 2 kpc, and assuming an isothermal sound
speed of 0.3 km s$^{-1}$ (sound speed for dense hydrogen gas at 20 K),
the sound crossing time for {\em each} clump is on the order of
$\sim$200-400 Myr. If this is triggered star formation, and if the
trigger is coming from Stock~8, we should be able to see a gradient
across the chain, correlated with distance from Stock~8; if the
trigger is coming instead from the north, as suggested by J08, then
no age gradient will likely be apparent. Either way, however, there
is no immediately obvious explanation for the semi-regular clumping
of the clusters within the NS.

\begin{figure}[htb!]
\epsscale{0.8}
\plotone{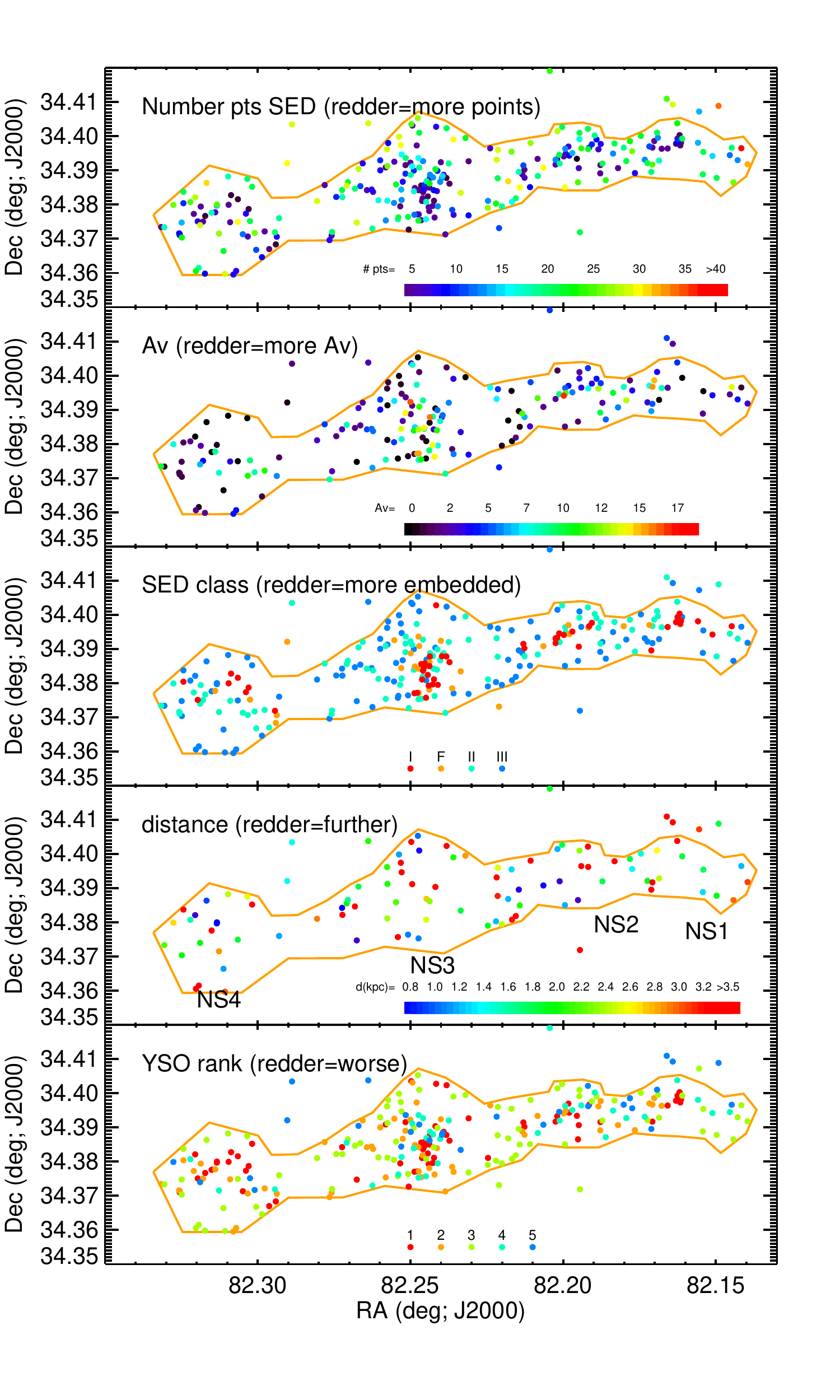}
\caption{Zoom-in on the NS for several properties discussed earlier
for the sample as a whole -- number of points in the SED, amount of
reddening (\av) estimated from the $JHK_s$ diagram,  SED Class
(I/flat/II/III), distance from Gaia DR3 as derived by Bailer-Jones
\etal\ (2021) (with additional annotations  of the subclusters), and
our final YSO confidence rank (1-5).  Color scales are as indicated in
each panel.  Points are only plotted if there is a corresponding value
-- there are fewest points in the fourth panel because there are fewer
distances known than anything else here. Because the YSO candidates in
the NS were identified by position on the sky, many objects in the NS
have very few points in their SED, which results in many low-ranked
SEDs. There can be a lot of reddening (obvious in
Fig.~\ref{fig:where1}), which results in many sparsely populated
SEDs, but also steep SEDs and few distances. NS3 has the most stars,
the most \av, and the most embedded sources.}
\label{fig:exploremore2}
\end{figure}

Because the sources in the NS are so close together, in all the plots
above of the stars on the sky, it is hard to see what is happening in
the NS. Figure~\ref{fig:exploremore2} shows a zoom-in on the NS for
several parameters discussed above for the entire sample.  As a result
of how we selected stars in the NS, many have very few points in their
SED; some of the worst-sampled SEDs of the entire sample are in the
NS. For those that have $JHK_s$, we can derive an \av, from which we
know that some of the most highly reddened sources in the sample are
in the NS, specifically NS~2 and NS~3. There are objects with SED
slopes suggesting very embedded sources throughout the NS, with the
most embedded ones in NS~3.  Distances are relatively rare in the NS,
a result of the high \av\ and frequently embedded sources.  Because
there are so many poorly sampled SEDs, a direct result is that there
are many low-ranking YSO candidates in the NS. 

Low-ranking YSOs can be found both too close and too far away (because
the distance is incorporated into the source assessment) -- for those
$\sim$90 objects in the NS that do have DR3 distances,  $\sim$40\% have a
distance from Bailer-Jones \etal\ (2021) that is $<$1000 or $>$3000
pc, e.g.,  nominally too close or too far away. However, the distance
to IC~417 is not well known, and there are uncertainties on the
stellar distances. Taking the errors provided by Bailer-Jones \etal, 
just 15\% of the NS objects with distances are unambiguously too
close or too far away -- for example, their distance is $>$3 kpc and
their lower limit on distance is also $>$3 kpc.

\begin{figure}[htb!]
\epsscale{1.0}
\plotone{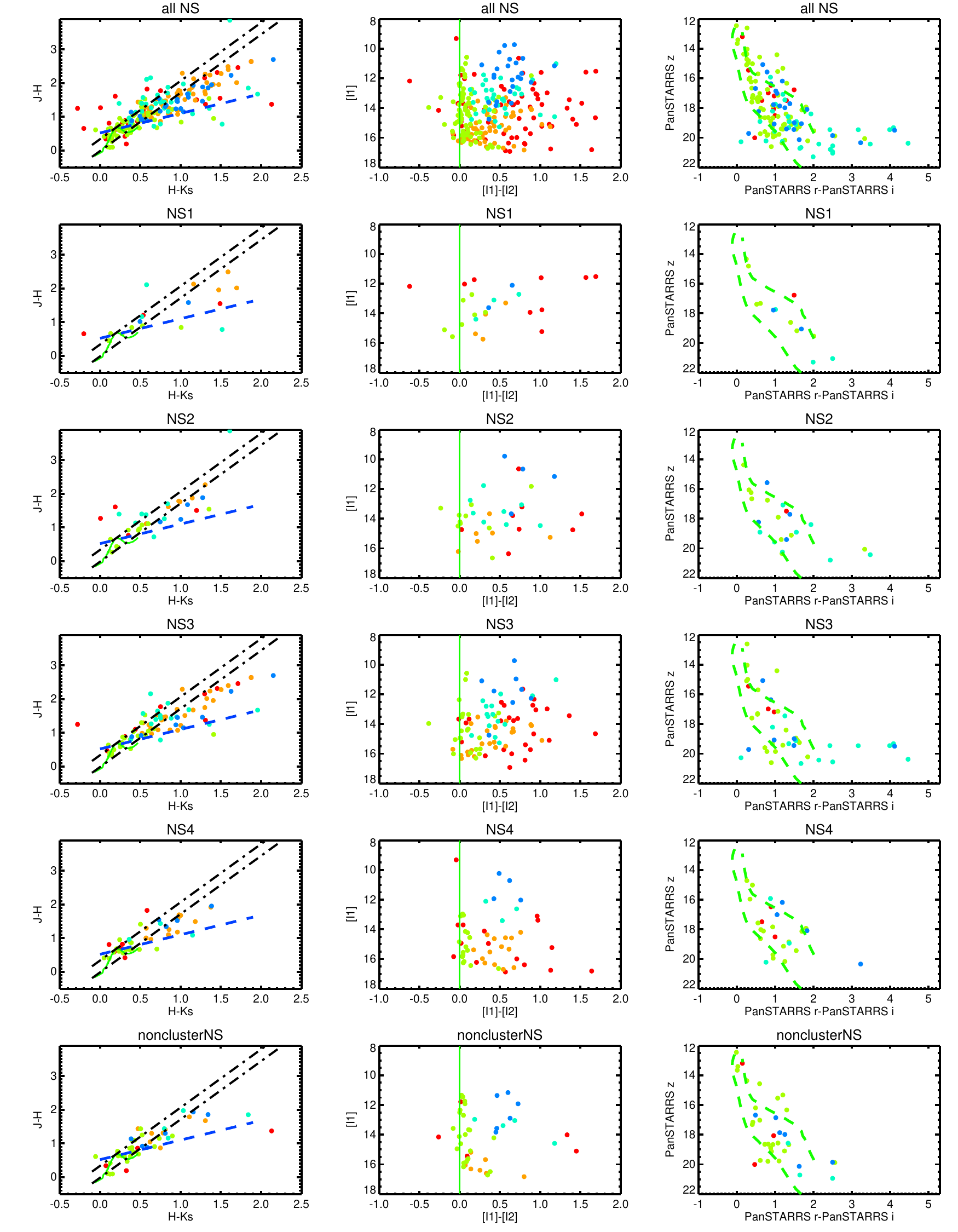}
\caption{$JHK_s$ color-color diagrams (left column),  the IRAC
color-magnitude diagram (center column),  and the PanSTARRS
color-magnitude diagram (right column) for various parts of the NS:
the entire NS (first row), just NS1 (second row), just NS2 (third
row), just NS3 (fourth row), just NS4 (fifth row), and those NS
stars not assigned to a subcluster (sixth row). Colors correspond to
YSO ranks (as in Fig.~\ref{fig:cmds2}), so bluer is better.  Note that
the limits of the $JHK_s$ plot are the same as earlier in the paper,
but the limits are adjusted for the other two diagrams. There are no
obvious differences among the clusters, other than NS~3 being the
best-populated.}
\label{fig:nsexp}
\end{figure}

Figure~\ref{fig:nsexp} shows the $JHK_s$ color-color diagram,  the
IRAC color-magnitude diagram, and the PanSTARRS color-magnitude 
diagram for various parts of the NS. Note that these are three of the
same diagrams shown earlier and are using the same notation as
Fig.~\ref{fig:cmds2}, but the axes in the IRAC and PanSTARRS are
optimized for this sample.  There are no obvious differences  across
these clusters in these diagrams, having tried various tests 
including a 2D Kolmogorov-Smirnov (KS) test, including for subsets of
the data  (e.g., just ranks 3/4/4*/5 for each box in
Fig.~\ref{fig:nsexp}).  It is likely that either they are all about
the same age (and mass, and reddening, and IR excess) distribution, or
the uncertainties in age are comparable to the age spread, or that
small number statistics are masking any differences.

The overall disk fraction is a function of how one defines the disk
fraction. In terms of all YSOs, IR excesses at any band, the overall
disk fraction in the NS is 64$\pm$6\% (assuming  Poisson statistics).
However, we can take only the high quality YSOs (3/4/4*/5) and any IR
excess and get 51$\pm$8\%. Further, taking only the high quality YSOs
(3/4/4*/5)  and only large IR excesses (SED Classes I and flat), the
overall fraction is 16$\pm$4\%. Looking across the NS, among each  of
the 4 clusters and the non-cluster NS stars, and assuming Poisson
statistics, in each case, comparable disk fractions are obtained. So
there is no discernible difference across the NS when  considering
disk fraction. Again, all evidence seems to suggest that they are all
about the same age.

We conclude that it is indeed most likely that any star formation 
trigger is coming from the north, not from Stock 8, as suggested by J08.


As noted above, Dewangan \etal\ (2018) concluded that most of  the
star formation was happening in what they termed ns1, and none in ns2.
We did not select any stars from the region consistent with their ns2,
because there were not any clumps of red stars in
Fig.~\ref{fig:where1}, and we agree that their ns2 is far more
``boring'' than their ns1 (our NS).  Dewangan \etal\ estimate that
there are $\sim$80 YSOs forming here. We started with $\sim$260 YSO
candidates in the NS; in the end, we have 24 that are  rank 5, 33 that
are rank 4 or 4*, 79 that are rank 3, 56 that are  rank 2, and 65 that
are any kind of rank 1.  Taking all the rank 3/4/4*/5 YSO candidates, that is
136, which is about 70\% more than Dewangan \etal\ estimated.
Additional work will be required to estimate a total mass in YSOs.

J08 find that the NS is likely excited by an O8 star from the north,
and that the ionization front from Stock~8 has not yet reached the NS.
They also find that the stars in what we have called NS3 are likely
younger than Stock~8. Our findings are consistent with this. MN16
further identify BD+34$\arcdeg$1058 as the ionizing source for the NS;
this is one of the stars featured in the next section.

\section{Specific Stars of Interest}
\label{sec:obc}

There are 32 known OB stars in this region and three known
carbon stars. As part of our analysis here, we believe 
we have identified one new carbon star. The OB stars are 
young and the carbon stars are old, but all are bright.  

\begin{figure}[htb!]
\epsscale{1.2}
\plotone{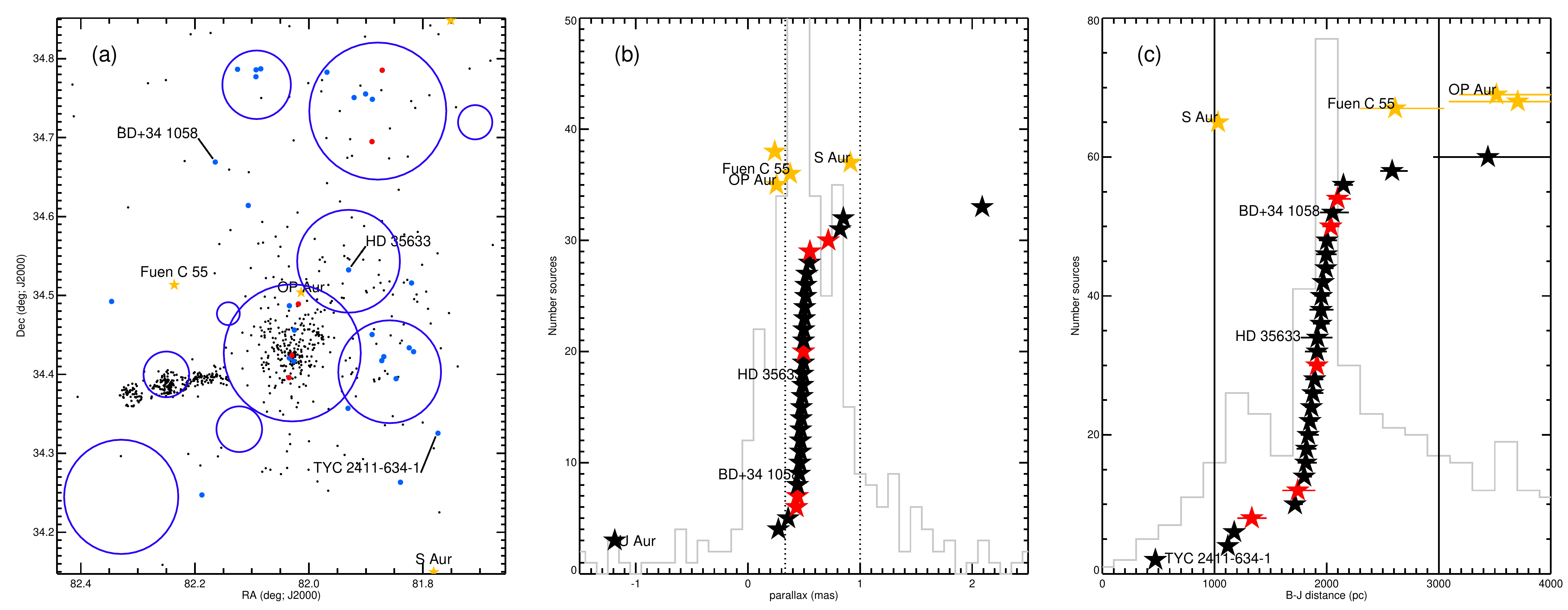}
\caption{(a): Plot on the sky of the YSO candidates (small black dots),
with the cluster circles from Fig.~\ref{fig:where3sky} (blue circles;
unlabeled here just for clarity).  The OB stars are highlighted
(larger blue dots); those OB stars with apparent IR excesses are
larger red dots.  In all plots, three OB stars are highlighted. HD
35633 is thought to be the powering source for the nebulosity around
Stock~8 (MN16). BD+34$\arcdeg$1058 is thought to be the driving source
for the NS (MN16). TYC 2411-634-1 is too close to be a member of
IC~417. The four carbon stars are also indicated by the yellow stars
in both plots; known carbon stars have their names indicated, and the
unlabeled gold point is the newly identified carbon star.  
(b) and (c): Histograms of parallaxes and distances to the final set
of YSOs in grey, with the locations of the OB and carbon stars
highlighted at arbitrary $y$-axis positions as 5-pointed stars (with
the uncertainties from Bailer-Jones \etal\ 2021 indicated by
horizontal lines).  Red symbols are stars with disks. The carbon stars
are indicated in yellow.  Certain stars are called out. One 
has a parallax so negative as to not appear in panel b
(J052729.32+342502.6=[JPO2008] IC 417 3); and that one, plus one more
(J052752.39+344658.1=IU Aur), have distances too far to appear in panel c.  
The OB stars are distributed across the sky and in distance,
but may preferentially appear in the clusters. They are certainly
clustered at the distance of IC~417 (1-3 kpc). The carbon stars are
not young and are unlikely to be members of IC~417.}
\label{fig:obstars}
\end{figure}

Figure~\ref{fig:obstars} shows the location of the OB and C stars on
the sky. Figure~\ref{fig:obstars} also  includes the parallaxes and
distances  to the stars, which are available for all the OB and C
stars. A few of the published parallaxes are negative, which is why we
have also plotted the distances from Bailer-Jones \etal\ (2021). 
Several things are of note in this figure. 

Most of the OB stars are (in projection) within the literature
clusters from  Fig.~\ref{fig:where3sky}; just 8 of them are not within
the clusters. There are some clusters without known OB stars, however;
there are no OB stars known in the NS.   HD 35633 is called out in
the Figure because it is thought to be the powering source for the
nebulosity around Stock~8 (MN16); similarly, BD+34$\arcdeg$1058 is
thought to be the driving source for the NS (MN16). 

Most of the O and B stars are at the correct distance to be members
of IC~417. The two stars that MN16 proposes as powering the NS
and Stock~8 are both within 2-3 kpc.

One nominal OB star, J052740.98+344502.5=TYC 2411-634-1, at $\sim$470
pc, is too close to be a member of IC~417. Ness \etal\ (2016)
identifies it as a red giant, which is quite inconsistent with the B8V
type listed for it in MN16. The object's optical colors could be
consistent with a giant, but given the distance, it is not very far
above a 6 Myr isochrone, and does not seem to be subject to very much
reddening. It is unclear what is going on with this source.

Two other OB stars (J052729.32+342502.6=[JPO2008] IC 417 3, and
J052752.39+344658.1=IU Aur), appear to have distances in excess of 6
kpc, too far to be members of IC~417.

Neither one of the stars powering the NS or Stock 8's nebulosity
appears  to have an IR excess, but there are 5 other  OB stars that
appear to have IR excesses, three of which are in Stock~8, and two of
which are in FSR 777, and all of which are at the right distance to be
part of IC~417.  Because the long wavelength surveys we have are
relatively poor resolution, if there is a nearby (in projection)
cooler star to the OB star in question, or surrounding nebulosity, it
can manifest as a long-wavelength IR excess in the SED. A lower-mass
binary companion could also result in a small IR excess. 

We collect a few notes on specific OB IR excess sources here. The
sources J052729.13+344707.2 (LS V +34 17) and J052733.43+344141.6 (LS V
+34 20) are both in FSR 777. The available spectral types in the
literature are B2V (MN16) or B2III (Chargeishsvili 1988), and B1.5V
(MN16) or B0III (Chargeishsvili 1988), respectively. They both appear
to only have an excess at 22 \mum. There are nearby red sources, but
in both cases, the possible polluting source is far enough away that
WISE can resolve the target.  The remaining sources are all within
Stock~8,  where the source surface density is high, but the images
seem clean enough that we retained these sources. Source
J052804.41+342921.0  (LS V +34 27) is relatively isolated and has a
clear IR excess at 12 and 22 \mum; it is a B2IV (MN16). Source
J052807.13+342526.8 (BD+34 1054) is an O9.7IV (MN16) or B0V
(Chargeishsvili 1988), and it has an IR excess measured by both WISE
and MSX. The SED suggests, however, that the detections at 7.7 \mum\
and longer might be contaminated by nebular emission. The images
indicate that the photometry is relatively uncontaminated.  Finally,
source J052808.37+342345.2 (BD+34 1056) is a B0.5V (MN16) or a B0I
(Chargeishsvili 1988). It has an IR excess measured by both WISE and
AKARI. Like the previous source, the SED might be contaminated by
nebulosity, but the images look good.  Further assessment of the
origin and reliability of the IR excess in  these sources will require
additional data, beyond the scope of this work.



The carbon stars are not young, but have an unambiguous IR excess in
their SEDs -- their SEDs are well-populated and have an IR excess at
multiple bands from multiple instruments.   There are three known
carbon stars here -- J052707.44+340858.6=S Aur, Nassau \& Blanco
(1954), a C-N5+ according to Barnbaum \etal\ (1996); 
J052803.23+343013.8=OP Aur, a C0-C1e according to Kohoutek \& Wehmeyer
(1997); and J052856.65+343048.1=Fuen C 55, a C D according to Fuenmayor
(1981).   We identified one new likely carbon star here,
J052700.29+345055.0. It was brought to our attention initially because
it was identified as variable in ASAS-SN.  
S Aur is too bright to have data from the Zwicky
Transient Facility (ZTF; Bellm \etal\ 2019), but the other three have
$i$ and $r$ light curves suggestive of carbon stars. (Detailed
discussion of the ZTF light curves of the entire  IC~417 sample is
beyond the scope of the present paper.) Looking at the SED, placement
in CMDs and color-color diagrams, and even its variability properties
(in ASAS-SN and ZTF), we strongly suspect that it is also a carbon
star; follow-up spectroscopy is warranted. (This new candidate carbon
star is in the 1r bin with the other known carbon stars.) OP Aur is
apparently in projection within Stock~8; Fuen C 55 is close to it in
declination but not within a cluster.  S Aur is barely in the field
under consideration in this work, on the south side.  The star we
believe to be a newly identified carbon star (J052700.29+345055.0)  is
also barely in the field, but this time on the north side. Two of the
C stars have Bailer-Jones \etal\ distances between 1-3 kpc; the other 
two are a bit further away. 

Because the carbon stars are nearing the ends of their lives, they are
not expected, necessarily, to be clustered with any ostensibly young
star clusters that may (or may not) be making up the constituents of
IC~417. However, the OB stars are young, and cannot have moved far
from their formation site, and thus are expected to be clustered. 
Several clusters (beyond Stock~8 and the NS) are identified in the
literature here (Fig.~\ref{fig:where2}), and most of the OB stars
appear, in projection at least, to be coincident with these clusters
(Fig.~\ref{fig:obstars}).  In the next section, we consider whether or
not we can distinguish these clusters from the rest of our sample.

\section{The 2MASS Clusters}
\label{sec:2massclusters}

\begin{deluxetable}{lcccc}
\tabletypesize{\scriptsize}
\tablecaption{Some Information on the Literature Clusters from our Catalog \label{tab:2mclusters}}
\tablewidth{0pt}
\tablehead{\colhead{Name} & \colhead{\# YSO candidates (all)} & \colhead{\# YSO candidates (3/4/4*/5)}& \colhead{\# srcs in catalog}&
\colhead{srcs/sq.asec} }
\startdata
CBB 7   &   0 &  0 &   189 & 0.0074  \\
FSR 777 &  22 & 21 &  3891 & 0.0096  \\
FSR 780 &  35 & 31 &  2061 & 0.0090  \\
CBB 3   &  34 & 31 &  1602 & 0.0070  \\
Kronberger 1 & 6&6 &  1210& 0.0119  \\
CBB 4   &   0 & 0 & 258 & 0.0057  \\
CBB 5   &   0 & 0 &  57 & 0.0050  \\
CBB 6   &   1 & 1 & 2464 & 0.0087 \\
\hline
Stock~8 & 279 & 208 & 4220 & 0.0104  \\
BPI 14  & 125 & 66 &   275 & 0.0061  \\
\enddata
\end{deluxetable}

Much of the early analysis done in the IC~417 region used 2MASS star
counts to identify several clusters (see Sec.~\ref{sec:intro}  and
Fig.~\ref{fig:where2}). If they are really $\lesssim$10 Myr, as
claimed, then there should be plenty of young stars in them waiting
to be found.  However, in Fig.~\ref{fig:sky2}, which does not have the
cluster circles overlaid, one would be hard-pressed to identify the
locations of the clusters besides Stock~8 and the NS (which includes
BPI~14), hereafter referred to as `the 2MASS clusters'; they do not
obviously stand out against the background when just the YSO 
candidates are plotted on the sky.  However, it is true
(Fig.~\ref{fig:obstars}) that most of the OB  stars are coincident
with the 2MASS clusters. 

The literature did not publish detailed lists of candidate members of
these 2MASS clusters; the analyses were based on statistics and star
counts, and the clusters were defined via a circle's position and
radius on the sky, and so this is what we have done as well. This
approach does not result in enough YSOs or candidate YSOs per cluster
from our final list to do much analysis; Table~\ref{tab:2mclusters}
collects information about the clusters as they appear in our catalog,
including the numbers of stars from our YSO candidate catalog. Aside
from Stock~8 and BPI~14 (in the NS), there just are not very many
stars in most of the 2MASS clusters. That could be because the
clusters are of different sizes on the sky, or a result of our
selection process. Our selection process is IR-biased, in
that a reasonable fraction of our YSO candidates  were identified via
IR excesses, and our catalog is biased towards IR catalogs. However,
the process for identifying  these clusters in the first place is
based on 2MASS star counts, so is also  IR-biased. If we cannot find
obvious evidence for the clusters in our  YSO catalog, it is worth
looking in our larger catalog. Table~\ref{tab:2mclusters} notes the
total numbers of sources (likely overwhelmingly stars) in our catalog
in each cluster, even though it is incomplete and/or biased; since
the clusters are different sizes on the sky, the number of sources
(in projection) per square arcsecond is also noted.  Among the 2MASS
clusters, FSR~777 has the most sources, but Kronberger~1 has the
highest surface density of sources, comparable to Stock~8.  

We tested several different parameters and parameter combinations, 
including \ks\ histograms, distance histograms, color-color and 
color-magnitude diagrams such as those earlier in this paper. 
(Gaia proper motions here may shed additional light eventually, 
but are not any material help at this time.)
For the most part, these diagrams reveal what we already know --
Stock~8 and BPI~14 are different than the rest of the clusters, 
with BPI~14 being most obviously apparently younger. Even
without clean membership lists, it is clear that these
clusters are not the same age as Stock~8 and BPI~14.

Another approach we can take is to compare circles of the same size as
these clusters placed randomly in the field, and compare  color-color
and color-magnitude diagrams to see if the diagrams from the purported
clusters are different than those randomly placed. For each of the
2MASS clusters, we randomly dropped between 200 and 300 pseudo-cluster
circles of the same size elsewhere in the field, making sure not to 
overlap with the NS or Stock 8.  We compared $J$
vs.~$H-K_{\rm s}$,  $z$ vs.~$r-i$ (from PanSTARRS), and [i1]
vs.~[i1]$-$[i2] for stars in each pseudo-cluster with the true
cluster using a 2-dimensional 2-sided Kolmogorov-Smirnov (KS2d2s)
test, and looked at the distribution of probabilities that resulted.
We also did this for Stock~8 and BPI~14 as a `control' of sorts; both
of those clusters were obviously different than the field for every
test we implemented.  For CBB~4, CBB~5, CBB~6, and CBB~7, the results
of our tests suggested these clusters were indistinguishable from the
field. For Kronberger~1, it was less obvious that the results were
different than the field, so we experimented with smaller cluster
radii, since an overdensity of point sources can be seen in this
area in the IRAC and WISE images. These experiments did not result in more
clarity.  CBB~3 seemed to be different than the field in PanSTARRS $z$
vs.~$r-i$ but perhaps not in the two IR CMDs. FSR~777 and FSR~780 were
both significantly different than the randomly placed field circles. 
More work is needed to assess whether or not these clusters are
real and, if so, extract their members from the background.

We conclude that the additional clusters identified here from 2MASS
star counts are evidently not $\lesssim$10 Myr; they are not the same
age as Stock~8 and the NS or BPI~14. The 2MASS clusters, as far as we
can tell, are older than Stock~8, if they are real. They may or may
not be legitimate clusters; FSR~777 and FSR~780 seem to stand the best chance
of being legitimate clusters, in that they seem 
significantly different than randomly selected fields in the vicinity.

\section{Summary}
\label{sec:summary}

IC~417 is in the Galactic plane, about 2 kpc away, with at least
portions of it thought to be young, $\lesssim$10 Myr. We started by 
assembling YSO candidates from the literature, consisting of O and B 
stars, H$\alpha$-bright stars, IR excess stars, and variable stars. We 
defined a polygon on the sky encompassing the nebulosity and red point
sources seen in Fig.~\ref{fig:where1}, and identified all of the
sources in the catalog within that polygon as potential new YSO
candidates in the Nebulous Stream (NS). We identified new YSO 
candidates using IR excesses  identified using WISE+2MASS data. We
then vetted each of these  YSO candidates by inspecting images where
possible, constructing and inspecting SEDs, and constructing and
inspecting various color-color and color-magnitude diagrams. There are
710 YSOs or YSO candidates that made it through this process, nearly
two-thirds of which have more than 20 points defining their SEDs. We
placed those 710 YSOs or candidates into ranked bins, from which 503
were in the higher (more confident) rank 3-5 bins. Of those 503, half
are SED Class III, and $\sim$40\% are SED Class II.  The lowest
ranking bins (1 and 2) include stars less likely to be YSOs; some 
of the stars in bin 1 include stars that are irreconcilably subject
to source confusion or likely carbon stars.

We rediscover three literature carbon stars (S Aur, OP Aur, Fuen C
55), and  identify one new candidate carbon star
(J052700.29+345055.0). These objects are retained in the YSO 
candidate list in the lowest ranked bin, subgroup 1r (``r'' for reject).

Follow-up spectroscopy will be necessary to confirm or refute the
young-star status of the YSO candidates (and the new carbon star
candidate) presented here.  The variability properties of these stars
may also help constrain their young-star status.

Because of the way in which we constructed our YSO candidate list, two
clusters are very obvious in the YSO candidate distribution. Stock~8,
which is well-studied in the literature, is well-represented in the
YSO catalog. We identified all stars within the NS as candidate YSOs,
so that region is also immediately apparent in the YSO catalog.
(BPI~14, a literature-identified cluster, is the most densely
populated  part of the NS.) Our results agree with some of the
literature (J08, J17) in that we find that the NS and Stock~8 (and the
rest of the YSO candidates for  which we have distances) are at about
the same distance, $\sim$2 kpc,  and the NS is the youngest region,
with Stock~8 a little older. We do not find any evidence for an age
spread within the NS, consistent with the idea that the trigger for
the star formation event came from the north (J08), but our 
information may be limited by reddening and small-number statistics.
We do not find that the other literature-identified 2MASS clusters
here are as young as either the NS or Stock~8; at best they are older
than Stock~8.  They may not all even be legitimate clusters. 
There are some literature-identified OB stars that are
coincident with some of those literature-identified clusters,
however.


\facility{Spitzer, 2MASS, IRSA, WISE, Simbad, Vizier}

\begin{acknowledgments}

This work was conducted as part of the NASA/IPAC Teacher Archive 
Research Program (NITARP; https://nitarp.ipac.caltech.edu),  which
receives funding from the NASA ADAP program. Teachers (and their
students) from two teams (2015 and 2020/2021)  worked on this project.
We acknowledge the additional students not on the author list  who
contributed their time and energy to this work in both large and
small ways.

This research has made use of the NASA/IPAC Infrared Science Archive
(IRSA), which is operated by the Jet Propulsion Laboratory, California
Institute of Technology, under contract with the National Aeronautics
and Space Administration.  This research has made use of NASA's
Astrophysics Data System (ADS) Abstract Service, and of the SIMBAD
database, operated at CDS, Strasbourg, France.  This research has made
use of data products from the Two Micron All-Sky Survey (2MASS), which
is a joint project of the University of Massachusetts and the Infrared
Processing and Analysis Center, funded by the National Aeronautics and
Space Administration and the National Science Foundation. The 2MASS
data are served by the NASA/IPAC Infrared Science Archive, which is
operated by the Jet Propulsion Laboratory, California Institute of
Technology, under contract with the National Aeronautics and Space
Administration. This publication makes use of data products from the
Wide-field Infrared Survey Explorer, which is a joint project of the
University of California, Los Angeles, and the Jet Propulsion
Laboratory/California Institute of Technology, funded by the National
Aeronautics and Space Administration.

\end{acknowledgments}

\appendix

\section{On the Distance to IC~417}
\label{app:distances}

As discussed above, there have been a variety of distances determined to IC~417:
 2.97 kpc (Mayer \& Macak 1971); 
 2.3 $\pm$0.7 kpc (Fich \& Blitz 1984);
 1.897 kpc (Malysheva 1990);
 2.68 kpc (Mel'Nik \& Efremov 1995);
 2.05$\pm$0.1 kpc (J08);
 2.7 kpc (Camargo \etal\ 2012); 
 2.80$^{+0.27}_{-0.24}$ kpc (MN16); and 
 2.8 kpc (Dewangan \etal\ 2018).
We adopted a distance of $\sim$2 kpc, and accepted as likely members 
anything between 1 and 3 kpc. 
We did not set out to determine a distance to IC~417, and moreover, 
we did not impose a tight constraint on the distances for each 
candidate YSO; if the 1$\sigma$ errors on the distance 
brought the star within 1-3 kpc,
we allowed the star to remain a candidate YSO. In this Appendix, we 
briefly summarize why we made this decision.

There are many different available distance estimates based on Gaia
data, and evaluating the details of all of them is beyond the scope of
this paper. As noted above, some parallaxes in DR2 and DR3 are
negative, leading us at least initially to turn to Bailer-Jones \etal\
(2018, 2021) for distance estimates. In Bailer-Jones \etal\ (2021),
two distances are provided, geometric and photogeometric. In DR3
itself, a distance is provided from the General Stellar Parameterizer
from Photometry (``GSP-Phot''; Andrae \etal\ 2022).  All of these
methods have limitations, particularly when the fractional uncertainty
in the parallax is large, and they may not do well much past 2 kpc in
general.  

\begin{figure}[htb!]
\plottwo{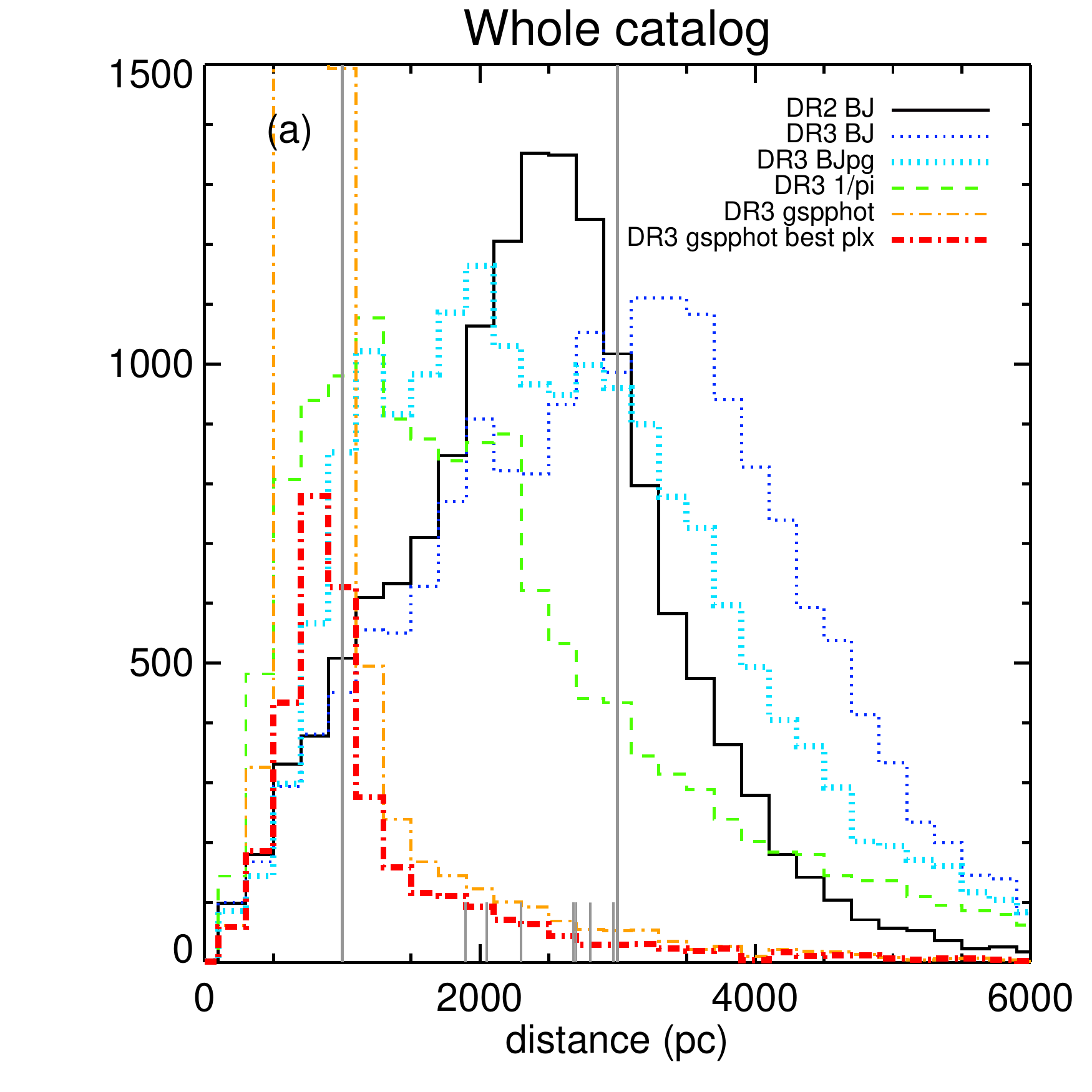}{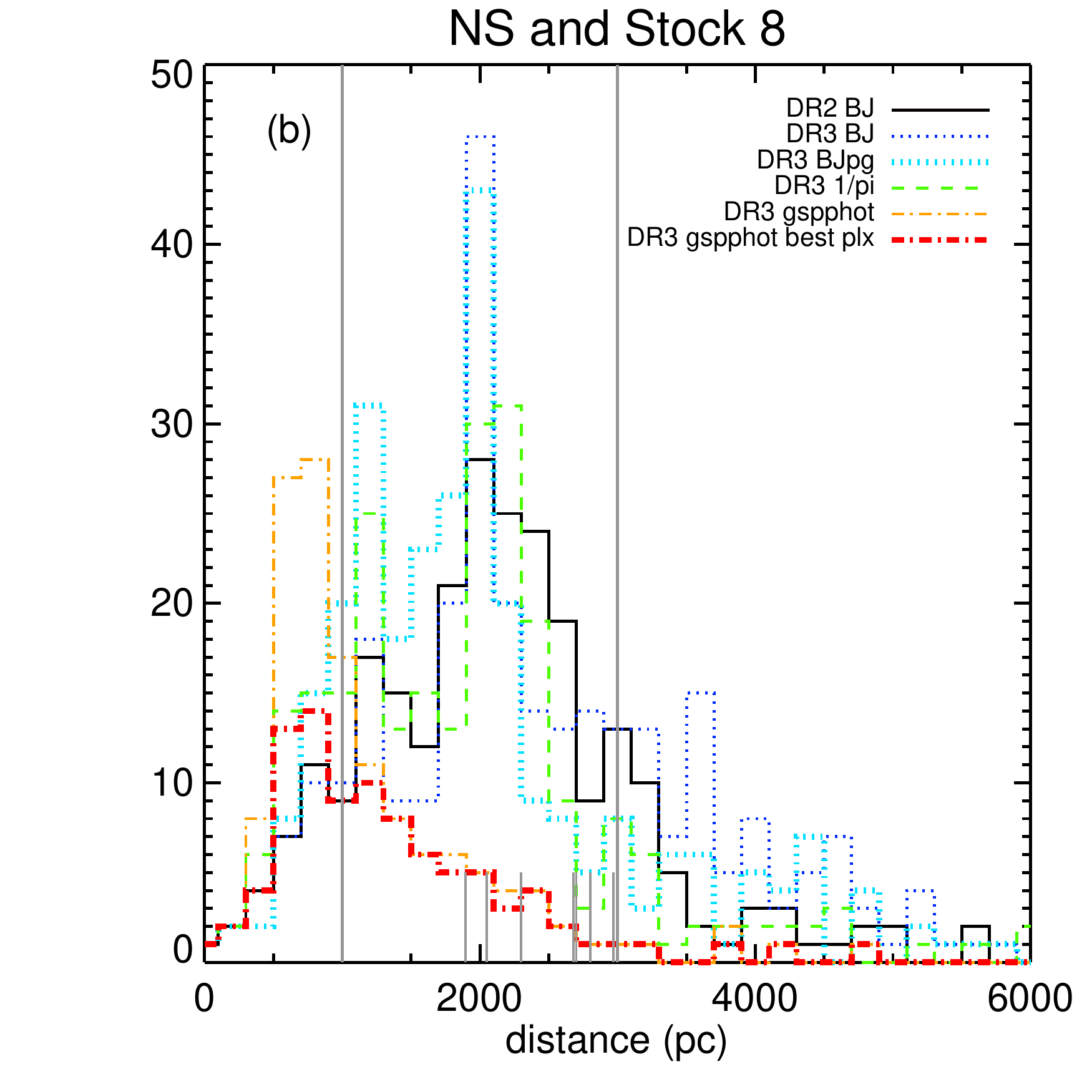}
\caption{Histograms of distances from the (a) entire catalog in this region,
and (b) just the NS and Stock 8.
Legend in the upper right, e.g.: 
solid black line (DR2 BJ) = DR2 Bailer-Jones \etal\ (2018) distances;
dotted thin dark blue line (DR3 BJ) = DR3 Bailer-Jones \etal\ (2021) geometric distances;
dotted thick turquoise line (DR3 BJpg) = DR3 Bailer-Jones \etal\ (2021) photogeometric distances;
dashed thin green line (DR3 1/pi) = distance from inverted parallax from DR3;
dash-dot thin orange line (DR3 gspphot) = distance from GSP-Phot from DR3 (note that 
the peak is up well beyond the top of the plot, at 2345);
dash-dot thick red line (DR3 gsppphot best plx) = distance from GSP-Phot from DR3
but only for those stars where the error in parallax/parallax $<$0.2 (e.g., those
where Andrae \etal\ 2022 say the distances are likely valid).
The grey vertical lines going from the top to the bottom of the plot
are at 1 and 3 kpc, e.g., the range of distances we accepted as 
members of IC~417. The short, thick grey lines at the bottom are the
literature distances given for IC~417 summarized in the text. Given
these bulk distributions on the left, it is clear that we are
operating near the outer range of where these distances are available,
let alone where they may be valid.
The distributions on the right show that the  distances of stars in
the NS and Stock 8 are peaked at 2 kpc, which is the main reason that
we took stars with a distance of 2 kpc, with  a range of 1-3 kpc, as
members of IC~417.}
\label{fig:appdistance1}
\end{figure}

Figure~\ref{fig:appdistance1} shows the distribution of all of the
available distances for all of the (presumed) stars in our catalog. Given these
raw distributions, it is clear that we are operating near the outer
range of where these distances are largely available, let alone where 
they may be valid. Most of the available
distances peak within 2-3 kpc; the DR3 Bailer-Jones geometric
distances peak just past 3 kpc, but according to Bailer-Jones \etal\
(2021), these may not be the best distances to use for our situation,
where the fractional uncertainties in parallax may be large. The DR3
GSP-Phot distances are apparently the most inappropriate for us to
use, as they are most biased towards $<$1 kpc, even when dropping all
the parallaxes with the largest uncertainties.

Stock~8 and the NS are the best-populated clusters in IC~417.
Figure~\ref{fig:appdistance1} also has a histogram of just the
distances to stars in those clusters. Ignoring the DR3 GSP-Phot
distances, now all of the remaining distance distributions peak at 2
kpc.  This is the main reason that we took stars with a distance of 2
kpc, with  a range of 1-3 kpc, as members of IC~417.

Note that the literature distance values are primarily slightly
farther than 2 kpc. A more secure (and less circular) 
distance determination will require
work beyond the scope of this paper, including spectroscopy. Later
Gaia data releases may also shed light on this subject.

\section{On Source Matching and Correcting Coordinates}
\label{app:srcmatch}

Sec.~\ref{sec:dataoptical} mentions that we had to correct the 
coordinates provided in J08 and J17 to match those in 2MASS before we 
could sensibly and accurately match them to the rest of the catalog. A
few more details on this process are provided here. In Rebull (2015),
a similar goal was to reconcile coordinates of YSOs from NGC~1333 over
20 years of literature. Here in IC~417, there is less of a time
baseline, but similar fundamental issues of errors in the warp and
weft of the world coordinate system (WCS) as applied to observations
of small regions, resulting in distortions of the derived coordinates
for any given source.  We took the experience gained from Rebull
(2015) and other similar work and applied it here.

The IRSA Firefly-based tools (called IRSA Viewer, Finder Chart, and the
Catalog Search Tool, among others) provide interactive ways to overlay
catalogs on images, and easy access to 2MASS images and catalogs, where
those coordinates are robustly tied to J2000 (Skrutskie \etal\ 2006),
and to Spitzer data and catalogs (both 2MASS and Spitzer are 
registered to the same coordinate system; IRAC Instrument and 
Instrument Support Teams 2021). The IRSA Firefly-based tools provide
ways to click on sources and have the images zoom to center the source
in question, the catalogs highlight the source in question, and the
plots highlight the source in question. They can make plots, e.g., of
position on the sky or attempt to make matches by position and plot
offsets in position. 

By overlaying catalogs from any given literature (such as, but not
limited to, J08 or J17) on 2MASS images and trying to match these
source catalogs to the 2MASS catalogs, one can easily identify pattern
offsets.  For example, it becomes obvious that
in one portion of the sky, all the sources from a given literature
catalog are offset from the sources in the 2MASS image by
0.73$\arcsec$ with a position angle of 30$\arcdeg$, but in another
portion of the same catalog, the sources are offset
from those in 2MASS by 1.58$\arcsec$ with a position angle of
102$\arcdeg$, or that in this strip of sky which must be an overlap
region between two tiles in the observations that generated the
catalog, there are two copies of every source in the catalog where one
of each pair is offset by 1.3$\arcsec$ in the same direction from the
other of each pair.

For the present work, specifically for J08 and J17, we did automatic
position matching between those catalogs and 2MASS where possible, but in both
cases, we discovered large, systematic, location-dependent offsets and
frequent duplicates within the catalogs. We took care of these matches
by hand using IRSA tools to overlay catalogs, identify sources, solve
duplicate issues, and make matches between catalogs where necessary.
If a match could not be found with 2MASS, it was checked against
Spitzer in case the counterpart was too faint for 2MASS but appeared
in Spitzer.  The fact that the resultant SEDs make sense (e.g., that
the J08 and/or J17 points are consistent with the rest of the points
in the SED) is assurance that we have made the match correctly.

\section{Background Information on YSOs}
\label{app:textbook}

This section provides an explicit discussion of each of the 
color-color and color-magnitude diagrams that we most frequently used,
including a demonstration of where young stars and background objects
are expected to be found. Since this kind of material was developed as
part of the NITARP learning process for the high school teachers and
students involved in this project, we have included a summary of this
here in the hopes that other learners can benefit.

\subsection{Definitions}

A magnitude is really a flux ratio, defined as:
\begin{equation}
M_1 - M_2 = 2.5 \log\frac{F_2}{F_1}
\end{equation}
The magnitude system is traditionally defined to be referenced to 
Vega; that is, Vega is defined to be zero magnitude, and 
magnitudes of everything implicitly reference that flux:
\begin{equation}
M = 2.5 \log\frac{F_{\rm Vega}}{F}
\end{equation}
We note for completeness that calibration in the IR is complicated
because Vega has an IR excess, and this references Vega's photospheric
flux (see, e.g., Cohen \etal\ 2003). Colors are the differences of two
magnitudes measured in two filters of the same object, which is
another representation of the ratio of the fluxes through those two
filters. Magnitudes depend on distances; colors do not.  Reddening due
to dust grains integrated along the line of sight between the detector
and the target affect shorter wavelengths more than longer
wavelengths. This is typically denoted as \av, and expressed (or
perhaps parameterized is a better word) in units of magnitudes of
extinction in the $V$ band, but can be calculated for any wavelength
according to reddening laws, e.g., Indebetouw \etal\ (2008) and Mathis
(1990) as we did here.

Following Wilking \etal\ (2001) and, e.g., Rebull (2015), we define
the near- to mid-IR (2 to 25 \mum) slope of the SED, $\alpha = d \log
\lambda F_{\lambda}/d \log  \lambda$, where $\alpha > 0.3$ for a
Class I, 0.3 to $-$0.3 for a flat-spectrum source, $-$0.3 to $-$1.6
for a Class II, and $<-$1.6 for a Class III.  These classes are empirically
defined based on the SED overall shape, and deliberately
do not capture any substructure in the SED between 2 and 25 \mum. These
classes are often assumed to be roughly mappable to age, where Class I is
the youngest and Class III is the oldest. However, projection effects
may be important in that, for example, an edge-on Class II may look 
like a Class I; also see Evans \etal\ (2009) and references therein.

As is noted in Evans \etal\ (2009), nomenclature can be confusing. 
In our work, we use the term YSO to refer to all stages of star formation, 
from Class I (or even Class 0, not that we have any here in this work) 
through a star's early life on the ZAMS after H ignition. 

\subsection{Finding YSOs}

YSOs are different from older main sequence stars in many subtle and 
not-so-subtle ways, and many different methods can be used to 
identify YSOs from out of the set of foreground/background objects.
Note that our definition of YSOs is broad enough to encompass the
very youngest, still-embedded phases through those ZAMS stars that
have begun burning hydrogen but are still young, as they are newly on
the main sequence and still manifest characteristics of youth. Note
that typically, several indicators of youth are necessary, including
follow-up spectroscopy, before it is accepted that a star is a
genuine YSO, but even having follow-up spectroscopy may not be
sufficient in some cases.

{\bf X-rays and outflows.}
YSOs can be very active and have strong, bright flares in X-rays. In
the present paper, we did not use X-ray emission because we do not
have sensitive enough X-ray data here, but one literature example of
this is, e.g., Alcala \etal\ (1996).  Very young stars can have
outflows; we did not look for outflows here because our stars are not
young enough to still have outflows, but see, e.g., Ogura \etal\
(2002) or Walawender \etal\ (2006) for examples of using
outflows to identify young stars. 

{\bf UV excess and H$\alpha$ emission.}
YSOs can be actively accreting and thus bright in emission lines,  and
have an UV excess. Just one example of using  UV excess to look for
YSOs is Rebull \etal\ (2000); a paper that uses H$\alpha$ to look for
YSOs is Ogura \etal\ (2002). In the present paper, we did not have UV
data, but for stars with a high enough accretion rate, the UV excess
can spill over into the blue bands, which is why we looked for
$g$-band excesses. In the present paper, we also looked for stars
bright in H$\alpha$ emission specifically to look for accretors. 
Stars that are young enough to still be rotating quickly can be very
active and thus be bright in H$\alpha$ from activity, not accretion.
Stars that are old but anomalously active can contaminate a sample of
YSOs selected solely based on H$\alpha$, but those levels of H$\alpha$
are generally far lower than that from accretion and can be
judiciously eliminated with careful H$\alpha$ limits (see, e.g.,
Slesnick \etal\ 2008); spectroscopy is particularly needed in these
borderline cases.

{\bf Variability.}
YSOs are variable -- indeed, variability was one of the original
defining  characteristics of young stars (Joy 1945, Herbig 1952). Two
examples of investigations using variability to identify YSOs include
Carpenter \etal\ (2000) and Rebull \etal\ (2014). In the present
paper, we have not used variability ourselves to identify YSOs, but we
have taken from the literature variables that others have identified
and investigated them as potential candidate YSOs.  Future work we
have planned include delving into the ZTF light curves for our YSO
candidates. Stars that have convective outer zones (e.g., mid-F and
later) have starspots and therefore will be variable, but older stars
are generally variable at much lower levels than young stars, even
those without disks (see, e.g., Fischer \etal\ 2022 or Rebull \etal\
2018). Stars that still have circumstellar disks often have
stochastic variability (see, e.g., Cody \etal\ 2022, Fischer \etal\
2022). Contaminants in a sample of YSOs selected based on
large-amplitude variability could include background giants.

{\bf IR excess.}
YSOs can have circumstellar disks and therefore can have IR excesses. 
Many people have used IR excesses, particularly from Spitzer (e.g.,
Gutermuth \etal\  2008,2009) or WISE (e.g., Koenig \& Leisawitz 2014),
to find YSO candidates. These techniques use cuts in multiple 
color-color and color-magnitude diagrams to select likely YSO
candidates out of the general population. Contaminants for a YSO
candidate sample selected in this way include primarily background
asymptotic giant branch (AGB) stars and active galactic nuclei (AGN).
Having optical data helps tremendously in weeding out these
contaminants, both in terms of direct imaging at higher spatial 
resolution and fleshing out the short-wavelength side of the SED (see,
e.g., Rebull \etal\ 2010). We have used IR excesses here in the
present paper to identify YSO candidates.

{\bf Location on the sky.}
YSOs are often found in regions of high extinction or associated
with nebulosity. When used in conjunction with other data, making an
initial selection of YSO candidates based on position on the sky is a
common approach (see, e.g., Kiss \etal\ 2006, Ogura \etal\ 2002, 
Padgett \etal\ 2004, Rebull \etal\ 2007, among many others). 
In the present paper, we have used position on the sky to make 
an initial guess at stars belonging to the NS, and we have used
substantial additional analysis to continue to refine that list.

When one has a wealth of data, it becomes powerful to combine 
as many of these methods as possible, such as Kuhn \etal\ (2021), who
used IR excess, location, and variability, or Getman \etal\ (2017), who 
combined IR excess, location, and X-ray detections. It is also important
to note that having many photometric bands, three-dimensional
space motions from Gaia, high-resolution spectroscopic data,
even multi-wavelength monitoring, can still result in confusing and
contradictory information about any given star, and judgement
calls still need to be made in order to move forward. Even for stars 
in the Taurus Molecular Cloud, only 140 parsecs away and studied 
for more than 100 years, there is room for debate (see 
discussion in Luhman 2023). IC~417, discussed here, is at $\sim$2 kpc
and far less well studied. All of this is a primary driver
behind our publishing our entire list of YSO candidates, so that
subsequent investigators can make their own judgement calls 
about what is a YSO candidate.

\subsection{Exploring locations in color-color and color-magnitude 
diagrams}

Many people have used color-color and color-magnitude diagrams in the
optical and infrared (and ultraviolet) to identify and characterize
YSOs. One efficient way of identifying the expected locations of YSOs in
various parameter spaces is to compare to a catalog of
confirmed YSOs, and/or a catalog of things not likely to be YSOs.  In
the context of the figures in this section, we compare the catalog
from this paper both to a catalog of known YSOs and likely
background. We used Taurus members from Luhman (2023; Table 2 from
that paper) as the known YSO sample, and matched those sources by
position to all the same catalogs as we used here. Only four stars
from this Taurus catalog were observed by IPHAS, but essentially all
of the stars have counterparts in all of the other surveys.
Specifically because so few Taurus members were observed by IPHAS, we
sought a comparable background sample, so we looked nearby on the sky
and assembled a catalog of objects from a half degree cone centered on
a nearby position  (05:33:29.60, +33:02:11.3 J2000) that was selected 
essentially randomly with the only real constraint being that it be
nearby but have no IR nebulosity. Because this position is relatively
close on the sky, it has all the same data available and comparable
stellar surface densities from the Galactic Plane in this
direction. 

\begin{figure}[htb!]
\epsscale{1.1}
\plotone{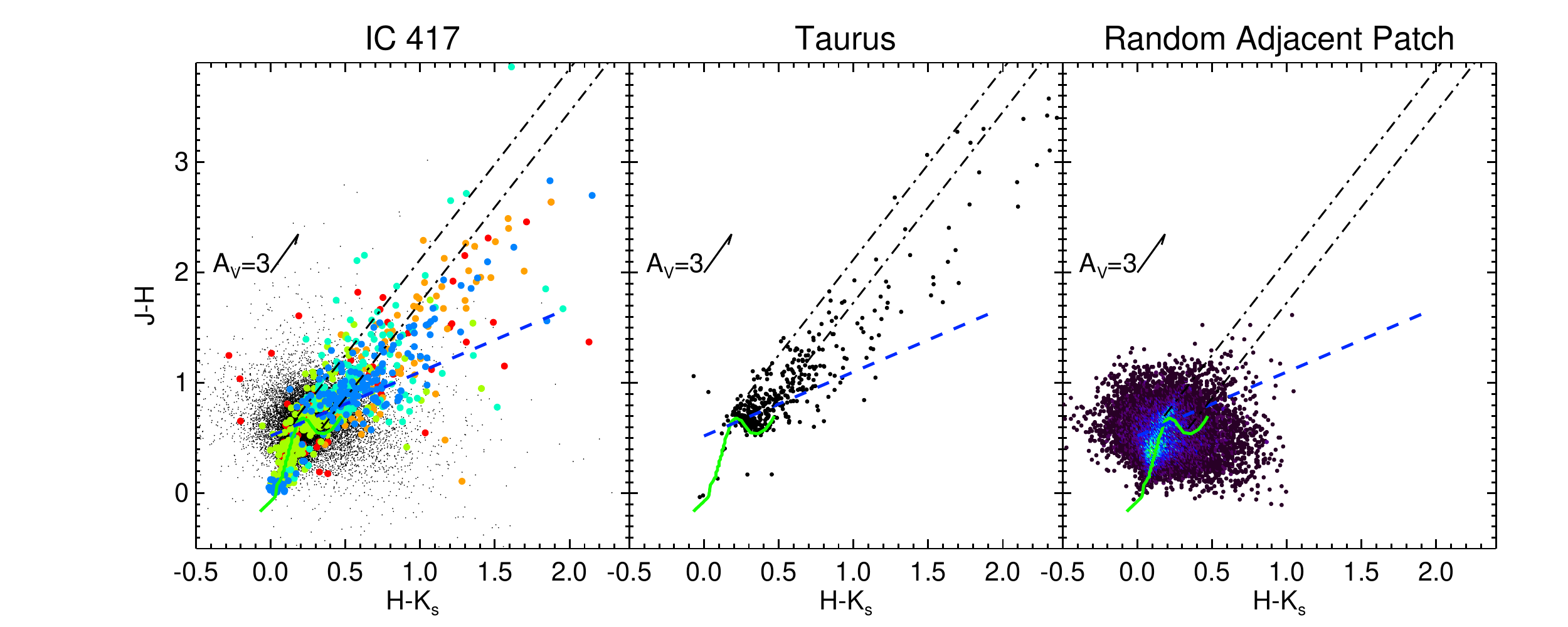}
\caption{$JHK_s$ color-color diagrams. Left: the IC~417 sample, as  in
Figure~\ref{fig:cmds2} above. Small black dots are the ensemble 
catalog, and larger dots are the YSO candidates.  Targets ranked 5 
(highest) are blue, 4 are cyan, 3 are green, 2 are orange, and  1
(lowest) are red; the bluer the symbol, the more reliable a YSO
candidate it is.  Center: Taurus sample of confirmed YSOs. Right:
randomly selected adjacent patch of sky near IC~417 -- e.g., there are
unlikely to be YSOs here. In this panel, colors correspond to point
density with black/purple being the lowest number of points and
orange/red being the highest (the highest density bins are in this
case hidden under the green ZAMS line). For all three plots, reddening
vectors (following the reddening law from Indebetouw \etal\ 2008 and
Mathis 1990) are as shown. Green solid lines are the expected
(empirical) ZAMS relationship.  The ZAMS is taken from Pecaut \&
Mamajek (2013), the dashed blue line is the Meyer \etal\ (1997)
T~Tauri locus, and the dash-dot lines are reddening vectors extending
roughly from the green ZAMS relation to give an indication of which of
these stars could be reddened MS stars. Young stars should be
clustered near the T~Tauri locus and up and to the right, and the
Taurus sample is found there. The IC~417 YSO candidate sample is
clearly more like the Taurus sample than the targets found in the
adjacent patch of sky.}
\label{fig:appendixjhk}
\end{figure}

\begin{figure}[htb!]
\epsscale{1.1}
\plotone{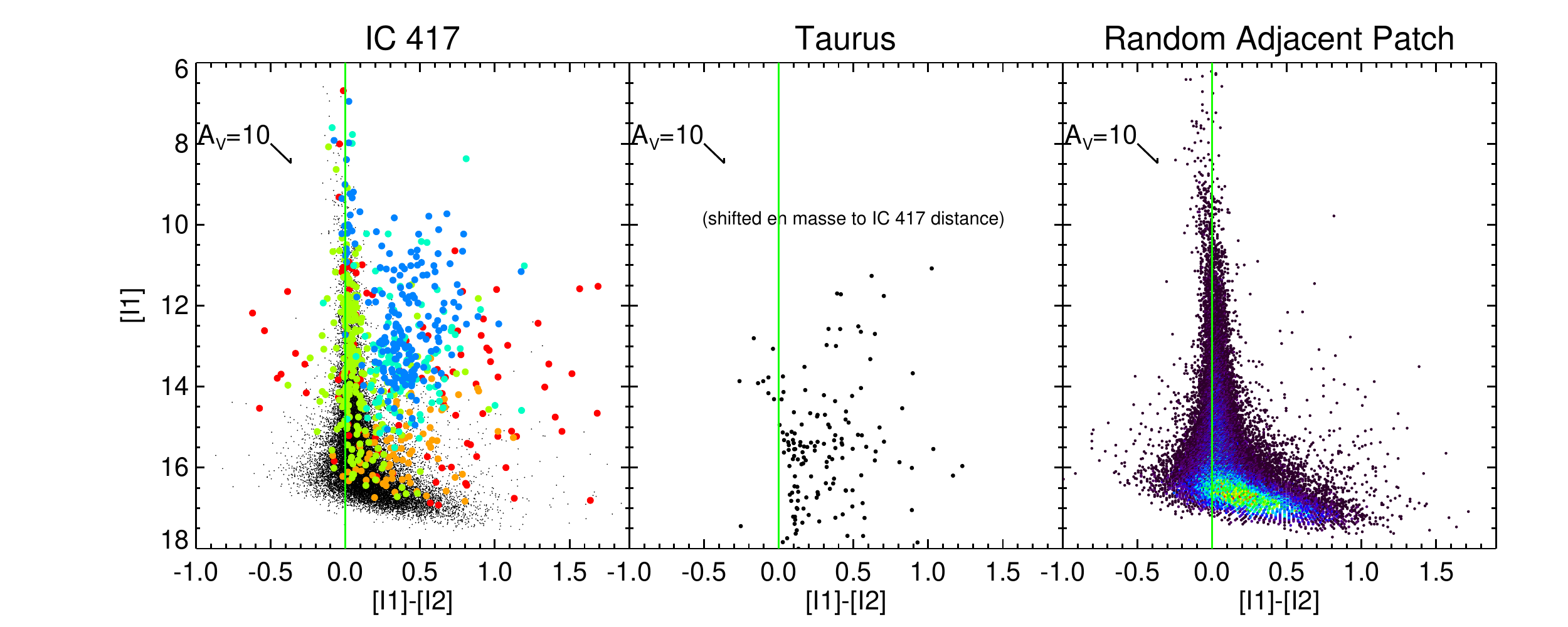}
\caption{IRAC color-magnitude diagrams, with notation as in 
Figure~\ref{fig:appendixjhk}.  Left: the IC~417 sample.  Small black
dots are the ensemble catalog, and larger dots are the YSO
candidates, where the bluer the symbol color, the more reliable a YSO
candidate it is.  Center: Taurus sample of confirmed YSOs, shifted to
be at 2 kpc, the distance of IC~417. Right: randomly selected adjacent
patch of sky near IC~417 -- e.g., there are unlikely to be YSOs here.
In this panel, colors correspond to point density with black/purple
being the lowest number of points and orange/red being the highest.
For all three plots, reddening vectors (following the reddening law
from Indebetouw \etal\ 2008 and Mathis 1990) are as shown. 
The green solid line is the expected color for main sequence stars.  Young stars
should generally be found significantly to the right of the green
line, and the Taurus sample is found there. The IC~417 YSO candidate
sample is clearly more like the Taurus sample  than the background
stars found in the adjacent patch of sky. }
\label{fig:appendixi1i1i2}
\end{figure}

\begin{figure}[htb!]
\epsscale{1.1}
\plotone{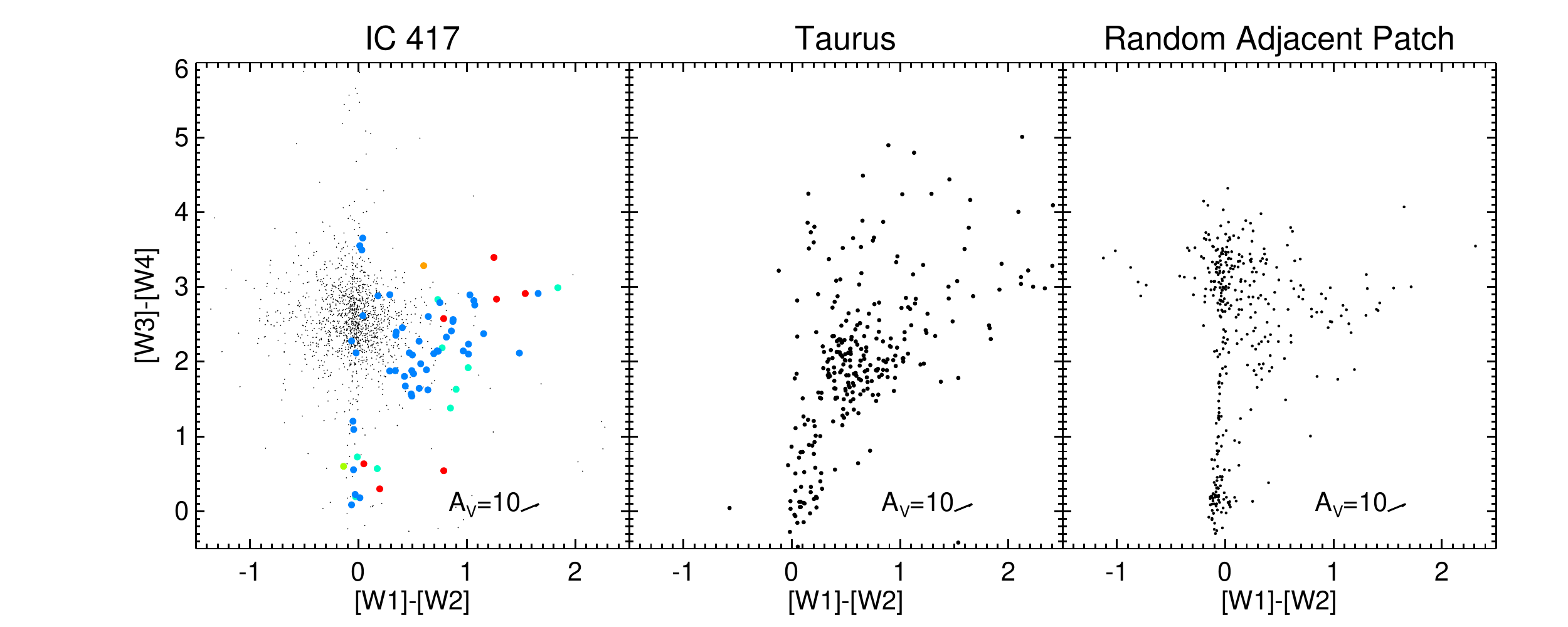}
\caption{WISE color-color diagrams.  Left: the IC~417 sample, with
notation as in  Figure~\ref{fig:appendixjhk}. Small black dots are the
ensemble catalog, and larger dots are the YSO candidates, where the
bluer the symbol color, the more reliable a YSO candidate it is. 
Center: Taurus sample of confirmed YSOs. Right: randomly selected
adjacent patch of sky near IC~417 -- e.g., there are unlikely to be
YSOs here. For all three plots, reddening vectors (following the
reddening law from Indebetouw \etal\ 2008 and Mathis 1990) are as
shown. Main sequence stars should have zero color in this plot. Young
stars should be red, up and to the right in this plot, and the Taurus
sample is found there. The IC~417 YSO candidate sample is clearly more
like the Taurus sample than the background stars found in the adjacent
patch of sky. }
\label{fig:appendixwise}
\end{figure}

\begin{figure}[htb!]
\epsscale{1.1}
\plotone{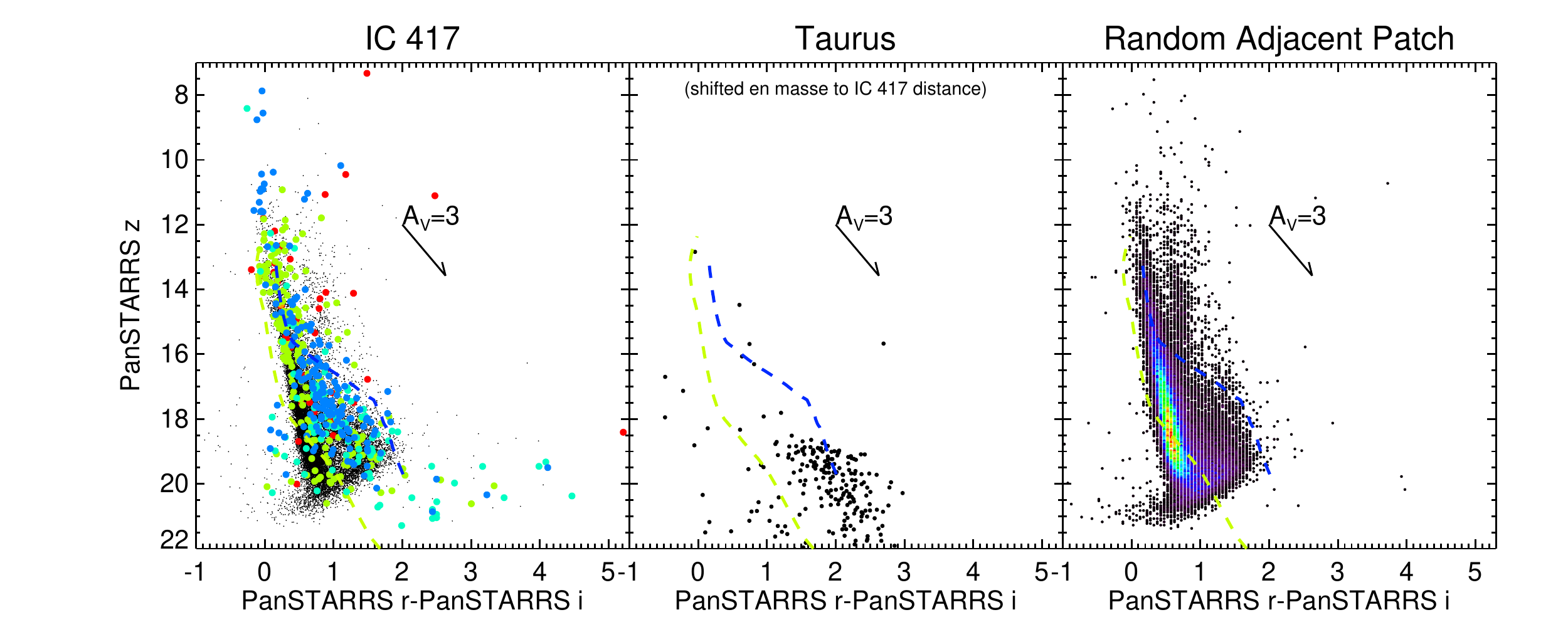}
\caption{PanSTARRS $z$ vs.\ $r-i$ color-magnitude diagrams, with
notation as in  Figure~\ref{fig:appendixjhk}.  Left: the IC~417
sample. Small black dots are the ensemble catalog, and larger dots are
the YSO candidates, where the bluer the symbol color, the more reliable
a YSO candidate it is.  Center: Taurus sample of confirmed YSOs,
shifted to be at 2 kpc, the distance of IC~417. Right: randomly
selected adjacent patch of sky near IC~417 -- e.g., there are unlikely
to be YSOs here. In this panel, colors correspond to point density
with black/purple being the lowest number of points and orange/red
being the highest. For all three plots, reddening vectors (following
the reddening law from Indebetouw \etal\ 2008 and Mathis 1990) are as
shown. The blue and yellow dashed lines are 6 Myr and 9 Myr isochrones
from PARSEC models (Bressan \etal\ 2012), respectively.  Young stars
should be clustered around the isochrone corresponding to their age,
but there is a fundamental degenerate uncertainty between age and distance,
further complicated by reddening. The distance to Taurus is well
known, but the distance to IC~417 is much less well known. The IC~417
YSO candidates hug the isochrones and are largely found  between the
two isochrones, as the Taurus stars are.}
\label{fig:appendixpanriz}
\end{figure}

\begin{figure}[htb!]
\epsscale{1.1}
\plotone{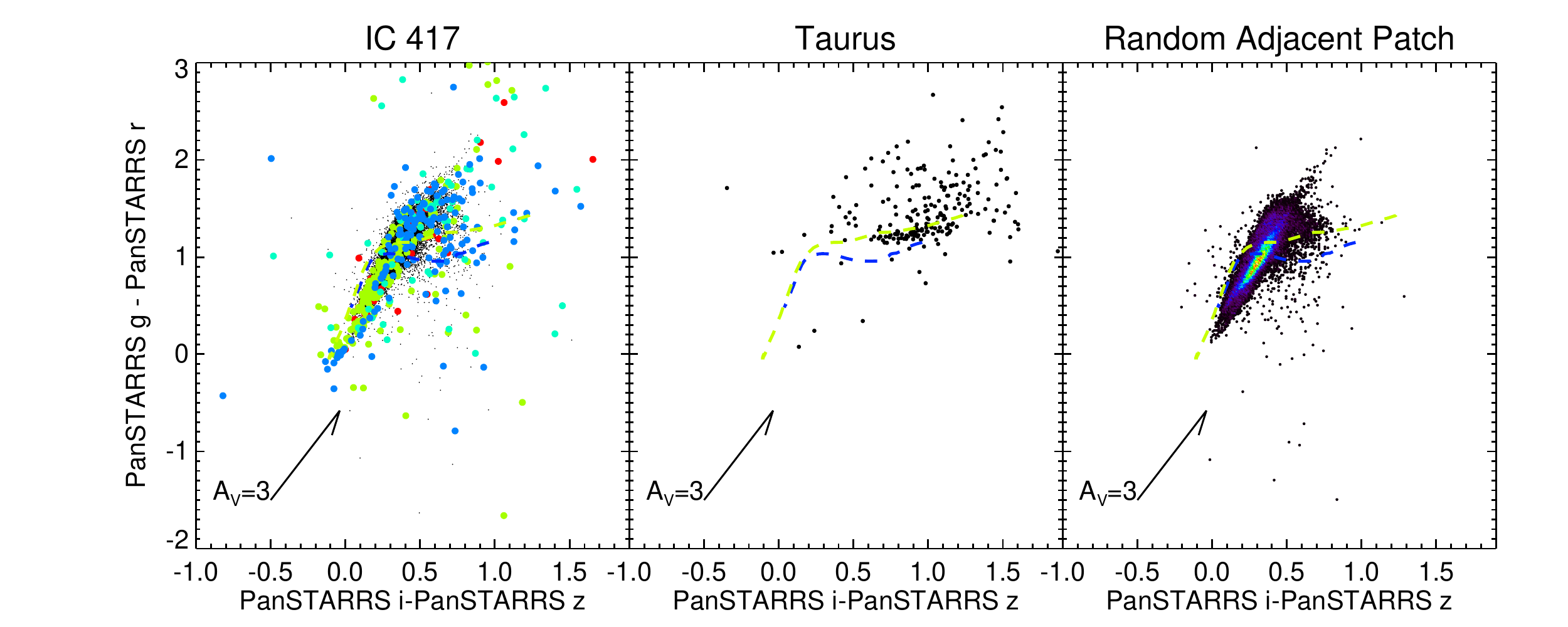}
\caption{PanSTARRS $griz$ color-color diagrams, with notation as in 
Figure~\ref{fig:appendixjhk}.  Left: the IC~417 sample. Small black
dots are the ensemble catalog, and larger dots are the YSO candidates,
where the bluer the symbol color, the more reliable a YSO candidate it
is.  Center: Taurus sample of confirmed YSOs. Right: randomly selected
adjacent patch of sky near IC~417 -- e.g., there are unlikely to be
YSOs here. In this panel, colors correspond to point density with
black/purple being the lowest number of points and orange/red being
the highest. For all three plots, reddening vectors (following the
reddening law from Indebetouw \etal\ 2008 and Mathis 1990) are as
shown. The blue and yellow dashed lines are 6 Myr and 9 Myr isochrones
from PARSEC models (Bressan \etal\ 2012), respectively.  Young stars
should be near the isochrones, but the reddening complicates things
here, as reddening pushes things up and to the right in this plot.
Stars with large blue excesses from accretion will be found below the
isochrones. Some Taurus stars are found there, as are some IC~417
stars. As noted in the text, this was not as effective a way to find
YSOs as we'd hoped.}
\label{fig:appendixpangriz}
\end{figure}

\begin{figure}[htb!]
\epsscale{1.1}
\plotone{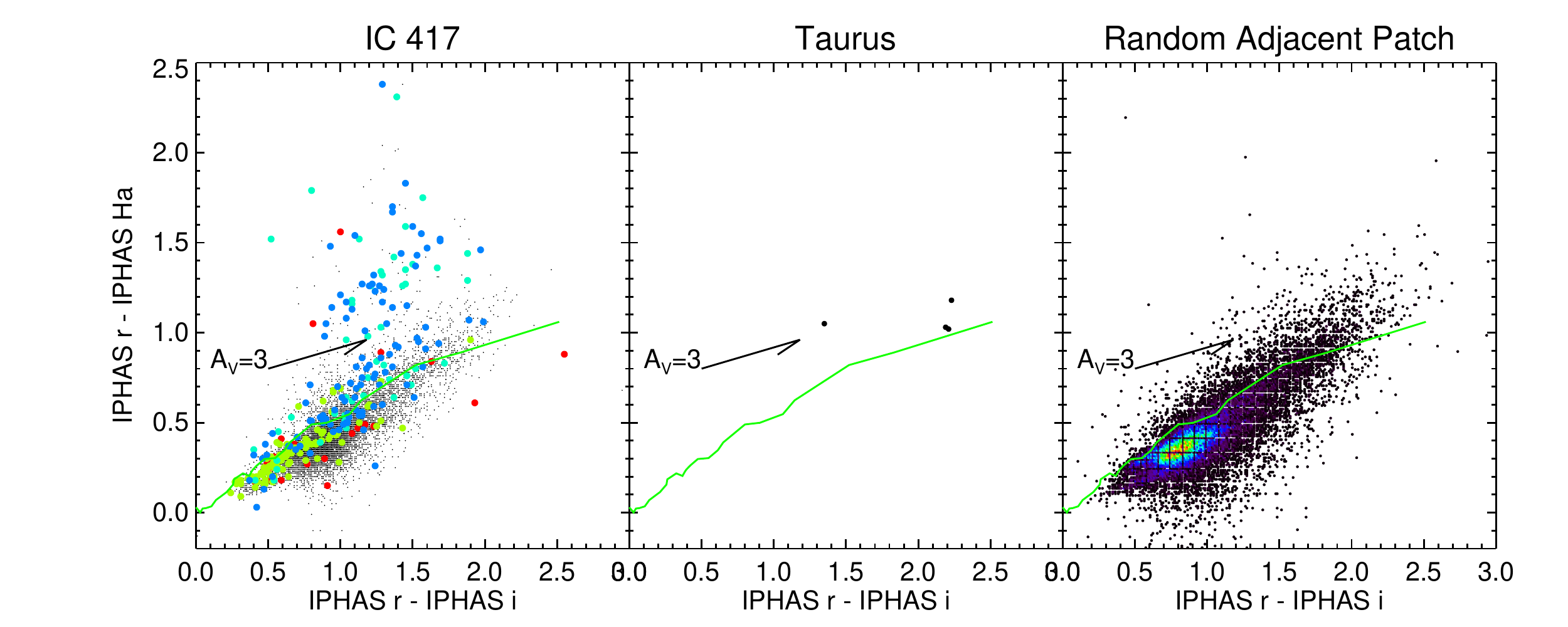}
\caption{IPHAS $riH\alpha$ color-color diagrams, with notation as in 
Figure~\ref{fig:appendixjhk}.  Left: the IC~417 sample. Small black
dots are the ensemble catalog, and larger dots are the YSO candidates,
where the bluer the symbol color, the more reliable a YSO candidate it
is.  Center: Taurus sample of confirmed YSOs (only 4 stars were
observed by IPHAS). Right: randomly selected adjacent patch of sky
near IC~417 -- e.g., there are unlikely to be YSOs here. In this
panel, colors correspond to point density with black/purple being the
lowest number of points and orange/red being the highest. For all
three plots, reddening vectors (following the reddening law from
Indebetouw \etal\ 2008 and Mathis 1990) are as shown. The IPHAS ZAMS
is shown in green and is from Drew \etal\ (2005). The IPHAS data
appear quantized due to the precision with which the magnitudes are
reported in H$\alpha$. Young stars that are bright in H$\alpha$ will
be above the ZAMS. Many more IC~417 stars are found there than stars
from the random patch of sky.}
\label{fig:appendixiphas}
\end{figure}

\begin{figure}[htb!]
\plotone{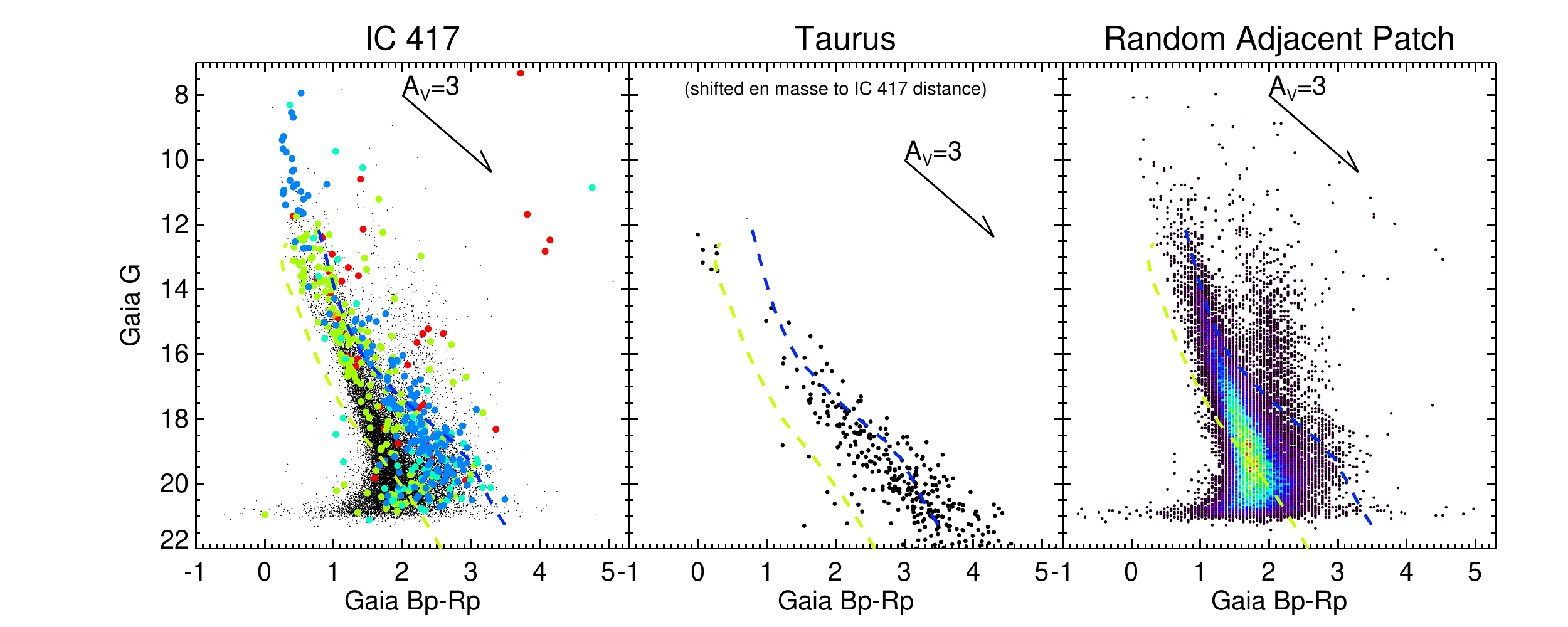}
\caption{Gaia $G$ vs.\ $G_{BP}-G_{RP}$ color-magnitude diagrams, with
notation as in  Figure~\ref{fig:appendixjhk}.  Left: the IC~417
sample. Small black dots are the ensemble catalog, and larger dots are
the YSO candidates, where the bluer the symbol color, the more reliable
a YSO candidate it is.  Center: Taurus sample of confirmed YSOs,
shifted en masse to be at 2 kpc, the distance of IC~417. Right:
randomly selected adjacent patch of sky near IC~417 -- e.g., there are
unlikely to be YSOs here. In this panel, colors correspond to point
density with black/purple being the lowest number of points and
orange/red being the highest. For all three plots, reddening vectors
(following the reddening law from Indebetouw \etal\ 2008 and Mathis
1990) are as shown. The  blue and yellow dashed lines are 6 Myr and 9
Myr isochrones from PARSEC models (Bressan \etal\ 2012),
respectively.  Young stars should be clustered around the isochrone
corresponding to their age, but there is a fundamental degenerate
uncertainty between age and distance, further complicated by
reddening. The distance to Taurus is well known, but the distance to
IC~417 is much less well known. The IC~417 YSO candidates hug the
isochrones and are largely found  between the two isochrones. The
Taurus stars do a better job of clustering around the 6 Myr
isochrone.}
\label{fig:appendixgaia}
\end{figure}

\begin{figure}[htb!]
\plotone{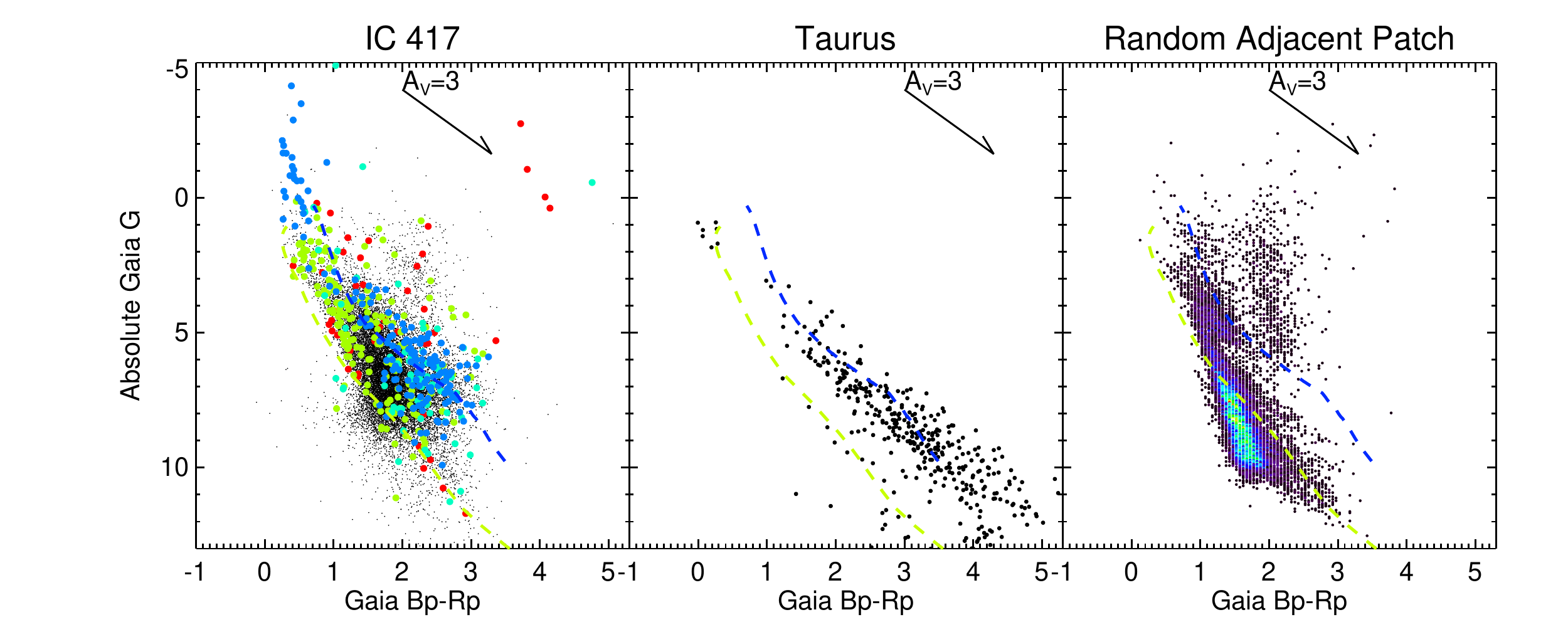}
\caption{Gaia absolute $G$ vs.\ $G_{BP}-G_{RP}$ color-magnitude
diagrams, with notation as in  Figure~\ref{fig:appendixjhk}.  Left:
the IC~417 sample, as  in Figure~\ref{fig:cmds2} above. Small black
dots are the ensemble catalog, and larger dots are the YSO candidates,
where the bluer the symbol color, the more reliable a YSO candidate it
is.  Center: Taurus sample of confirmed YSOs. Right: randomly selected
adjacent patch of sky near IC~417 -- e.g., there are unlikely to be
YSOs here. In this panel, colors correspond to point density with
black/purple being the lowest number of points and orange/red being
the highest. For all three plots, reddening vectors (following the
reddening law from Indebetouw \etal\ 2008 and Mathis 1990) are as
shown. The  blue and yellow dashed lines are 6 Myr and 9 Myr
isochrones from PARSEC models (Bressan \etal\ 2012), respectively. 
Young stars should be clustered around the isochrone corresponding to
their age, but there is a fundamental degenerate uncertainty between
age and distance, further complicated by reddening. The distance  to
Taurus is well known, but the distance to IC~417 is much less well
known. The IC~417 YSO candidates hug the isochrones and are largely
found  between the two isochrones. The Taurus stars do a better job of
clustering around the 6 Myr isochrone. The random patch of sky has no
similar clustering.}
\label{fig:appendixgaiaabs}
\end{figure}

Figures~\ref{fig:appendixjhk}-\ref{fig:appendixgaiaabs} show all eight
of the color-color and color magnitude diagrams typically used for our
source inspection in IC~417, to wit, $J-H$ vs.~$H-K_s$,  [I1]
vs.~[I1]$-$[I2], [W3]$-$[W4] vs.~[W1]$-$[W2], Pan-STARRS  $z$
vs.~$r-i$, Pan-STARRS $g-r$ vs.~$i-z$, IPHAS $r-H\alpha$ vs.~$r-i$, 
and Gaia DR3 $G$ vs.~$G_{BP}-G_{RP}$ observed and absolute.   We note
for completeness two items. (1) Individual stars in IC~417 may have
required additional exploration, e.g., if PanSTARRS $z$ was missing or
obviously inconsistent with the rest of the SED, we may have made,
e.g.,  a PanSTARRS $r$ vs.\ $r-i$ diagram in addition to the $z$ vs.\
$r-i$ diagram seen in Fig.~\ref{fig:appendixpanriz}. (2) Because
there were three Gaia releases over the course of our work (DR2, EDR3,
\& DR3), we made multiple versions of the Gaia CMDs, plus histograms
of Gaia parallaxes and distances, corresponding to each release. The
data shown here are from DR3.

In Figures~\ref{fig:appendixjhk}-\ref{fig:appendixgaiaabs}, for
comparison, the Taurus sample of known YSOs is shown, along with the
randomly selected adjacent patch of sky near IC~417.  The stars from
IC~417 are shown as they are earlier in the paper, with the targets
ranked 5 (highest) as blue, 4 as cyan, 3 as green, 2 as orange, and  1
(lowest) as red; the bluer the symbol, the more reliable a YSO
candidate it is. 

In each case, the higher-quality YSO candidates from IC~417 (rank
5,4,3) are more like the Taurus sample than the background random
patch of sky, with the higher quality YSO candidates bearing a
stronger resemblance to the Taurus sample than the lower quality
candidates. The degree to which this is the case is a function of  the
specific color space. We now discuss the distributions in each Figure,
for Figures~\ref{fig:appendixjhk}-\ref{fig:appendixgaiaabs}.

{\bf $J-H$ vs.~$H-K_s$, Fig.~\ref{fig:appendixjhk}.} In the $JHK_s$
diagram, the distribution of Taurus YSOs suggests they are nearly all
low-mass (few are located in positions close to the high-mass end of
the ZAMS as shown), some having large enough IR excesses so as to
affect the $JHK_s$ colors (some near the T~Tauri locus), and many
subject to high \av\ (smeared up and to the right of the ZAMS and the
T~Tauri locus). The Taurus stars are located where YSOs are expected
to be found in this diagram. The random patch of sky has a significant
blob of stars clumped at the ZAMS, very few (in terms of fraction
or absolute number) stars apparently reddened, and very few (in terms
of fraction or absolute number) stars near the T~Tauri locus. In
contrast, the IC~417 sample has many high-quality  YSO candidates
clustered around the ZAMS, both the high- and low-mass ends, some near
the T~Tauri locus, and many subject to high \av. The most outlying
points in this diagram tend to be the low-ranking YSO candidates. It
is not surprising that IC~417 has more high-mass stars than Taurus;
Taurus is known to have few high-mass stars.  The higher-quality YSO
candidates bear a far stronger resemblance to the Taurus sample than
the random sky sample.

{\bf [I1] vs.~[I1]$-$[I2], Fig.~\ref{fig:appendixi1i1i2}.} The IRAC
color-magnitude diagram is the best here in terms of both containing
most of the target stars and being least sensitive to reddening. The
Taurus sample in the Figure has been shifted out to 2~kpc for more
direct comparison to IC~417,  where it becomes clear, again, that
Taurus has few high mass stars. Many -- but not all -- of the Taurus
stars have significant IR excesses in this diagram; that is, they are
significantly to the right of the green line. Not all of the Taurus
stars have excesses in this diagram because some have IR excesses that
start at wavelengths longer than 4.5 \mum, and some do not have IR
excesses. The Taurus stars are located where YSOs are expected to be
found in this diagram. Relatively few of the stars in the random patch
of sky are significantly to the right of the green line.  IC~417's
highest quality YSO candidates for the most part have IR excesses;
that is, they are largely significantly to the right of the green
line. Some of the lower quality YSO candidates have smaller excesses
in this diagram. Many of the lowest quality YSO candidates are
extreme  outliers in this diagram, some in the unphysical location far
to the left of the green line.  The higher-quality YSO candidates
bear a far stronger resemblance to the Taurus sample than the random
sky sample.

{\bf [W3]$-$[W4] vs.~[W1]$-$[W2], Fig.~\ref{fig:appendixwise}.} Like
the IRAC color-magnitude diagram, this WISE color-color diagram is not
very sensitive to the effects of reddening, but a low fraction of our
stars have detections at all four WISE bands, in no small part because
IC~417 is at $\sim$2 kpc.  Since Taurus is only $\sim$140 pc away,
most of the Taurus stars have detections at all four WISE bands. The
Taurus WISE plot is well-populated. The stars without circumstellar
disks have colors near zero; the stars with large IR excesses (and
therefore likely circumstellar disks) are distributed up and to the
right from that (0,0) locus.  The Taurus stars are located where YSOs
are expected to be found in this diagram.  The targets in the random
patch of sky include a locus of dust-free stars at (0,0), and then a
clump near (0,3) that are likely background galaxies seen  through the
outer galaxy.  The distribution of targets to the right,  with both
[W1]$-$[W2] and [W3]$-$[W4] greater than $\sim$0.5 or 1 mags, 
plausibly have colors consistent with YSOs, as can be seen in
comparison to the Taurus plot. However, the overall distribution of 
Taurus YSOs looks far different than the overall distribution from the
random patch of sky. IC~417's highest quality YSO candidates are 
primarily in the right part of the diagram to be consistent with the
concentration of Taurus YSOs near ($\sim$1, $\sim$2). Some have
smaller excesses in [W1]$-$[W2] and/or [W3]$-$[W4] than most YSOs. The
background stars in IC~417 look like the background stars in the 
random patch of sky, except that they are not quite as red in
[W3]$-$[W4].  Several of the lowest quality YSO candidates are
extreme  outliers in this diagram.  The higher-quality YSO candidates
bear a stronger resemblance to the Taurus sample than to the background
or the random sky sample.

{\bf Pan-STARRS $z$ vs.~$r-i$, Fig.~\ref{fig:appendixpanriz}.} This is
the first optical diagram considered here, and the reddening is now a
bigger concern than in the IR diagrams above.  The Taurus sample is
again shifted out to 2~kpc for a fairer comparison to IC~417. Most of
the Taurus stars hug the 6 Myr isochrone, which is where YSOs are
expected to be in this diagram, at least for YSOs of about that age. 
It is apparent that there are no high-mass stars in
Taurus, and that there is considerable scatter in the data, since
so many Taurus YSOs appear below the 9 Myr isochrone in this plot.
This is likely due not only to observational uncertainties but also
reddening corrections (or lack thereof).  The random patch of sky
looks much different than the Taurus YSO sample.  The locus where
there are the most stars is closest to the 9 Myr isochrone, but is not
terribly well-aligned with it. There are two `prongs' near $z\sim14$;
the bluer one is from main sequence stars, and the redder one comes
from reddened background giants, smeared by reddening down to their
current observed location.  There is a fundamental degeneracy in the
uncertainties here -- there is likely an age spread in the star
formation regions, and probably a range of distances as well. If we
don't know the distance well to any young cluster, we can only poorly
constrain the age of said young cluster, because both age and distance
slide the isochrones up and down in this diagram, and reddening only
worsens these uncertainties.  Taurus is close, and well-studied;
IC~417 is far and not well-studied.  The IC~417 highest quality YSO
candidates hug the 6 Myr isochrone; the lower quality YSO candidates
are dispersed more broadly between the 6 and 9 Myr isochrones, as are
the Taurus stars.  The YSO candidates that are most likely to be
giants are the ones near the reddening vector annotation, and those
are most often the lowest-ranking YSO candidates.  The higher-quality
YSO candidates bear a stronger resemblance to the Taurus sample than
to the random sky sample.

{\bf Pan-STARRS $g-r$ vs.~$i-z$, Fig.~\ref{fig:appendixpangriz}.} 
This optical color-color diagram incorporates the shortest wavelengths
to which we have access, and as a result, it is most sensitive
to reddening and has the most scatter of all the diagrams we used. 
Stars with a significant blue excess (like YSOs with significant
blue excess due to accretion) should appear in the lower half 
of this diagram. What makes this hard is that reddening acts 
to push the stars up, out of the regime of obvious blue excess, and 
star forming regions have considerable, often patchy, dust creating
reddening. The Taurus sample shows this clearly; there are a few
stars clearly below the blue isochrone, but there are lot more 
scattered up and to the right as a result of reddening. The 
random patch of sky has stars largely around the isochrones, 
with some smearing up and to the right due to reddening. There are
fractionally fewer stars smeared up and to the right than in the
Taurus sample, because fractionally fewer stars are subject to 
substantial reddening in the random patch of sky. There are a
few stars that appear to have a blue excess in the random patch
of sky, and these plausibly could be active stars.  The IC~417 
YSO candidate sample in this parameter space is not as 
dramatically similar to Taurus and different from the random 
patch of sky than it was in the earlier figures, but there are
still some important characteristics to note. Some YSO candidates
are in the region of blue excesses, and some are in the region of highly
reddened stars. Most, however, are consistent with where 
the ensemble of background stars are found. As noted in the 
main body of the paper, this color-color diagram
is not as effective here as others for identifying YSO candidates.

{\bf IPHAS $r-H\alpha$ vs.~$r-i$, Fig.~\ref{fig:appendixiphas}.}  This
optical diagram is the hardest to compare across samples because we do
not have these data for Taurus. Stars that are bright in H$\alpha$ due
to accretion, e.g., YSOs, will appear in the upper half of this
diagram, above the ZAMS shown. The targets from the random patch of
sky have fractionally very few stars in the top half of this diagram;
the overwhelming majority of the stars are clustered near the ZAMS
and smeared from there along the reddening vector.  The IC~417
high-quality YSO candidate sample has many stars very bright in
$r-H\alpha$ (in the top half of the diagram, e.g., where YSOs are
expected to be). There are also many YSO candidates consistent with
ZAMS colors.  The higher-quality YSO candidates include many clearly
significantly different from the random sky sample.

{\bf Observed Gaia DR3 $G$ vs.~$G_{BP}-G_{RP}$, 
Fig.~\ref{fig:appendixgaia}.}  This optical color-magnitude diagram is
similar to the Pan-STARRS $z$ vs.~$r-i$ diagram above, but has less
scatter  overall.  The Taurus sample is again shifted out to 2~kpc for
a fairer comparison to IC~417. Most of the Taurus stars hug the 6 Myr
isochrone, which is where YSOs are expected to be in this diagram, at 
least for YSOs of about that age. Note that there are stars
scattered down between and below the isochrones, likely due to
reddening. The random patch of sky has the two `prongs' 
(as in Fig.~\ref{fig:appendixpanriz}),  and
the highest density of points is most consistent with the oldest
isochrone but is not well-matched to it. The IC~417 YSO candidate
sample is clustered around the 6 Myr isochrone, including some high
mass stars. The stars that are most likely giants (e.g., those
appearing near the reddening vector annotation) are lowest ranking
YSO candidates.  The higher-quality YSO candidates bear a stronger
resemblance to the Taurus sample than to the  random sky sample.

{\bf Absolute Gaia DR3 $G$ vs.~$G_{BP}-G_{RP}$, 
Fig.~\ref{fig:appendixgaiaabs}.}  This absolute color-magnitude
diagram is another version of the prior optical color-magnitude 
diagram, but now with everything shifted to 10 pc. As before, the 
Taurus sample is well-clustered around the 6 Myr isochrone, but with 
some stars scattered down between and below the isochrones. The 
random patch of sky still has two `prongs' (the one due to giants
is now more distinct from that due to MS stars, making the effects of
reddening far more obvious), and its densest portion is kind
of close to the 9 Myr isochrone. The IC~417 YSO candidates cluster
around the 6 Myr isochrone, with some scatter down between and below
the isochrones. The stars that are most likely to be giants are the
least confident YSO candidates.  The higher-quality YSO candidates
bear a stronger resemblance to the Taurus sample than to the random
sky sample.


\section{Brief Examples of Source Ranking}
\label{app:briefexamples}

\begin{figure}[htb!]
\epsscale{0.8}
\plotone{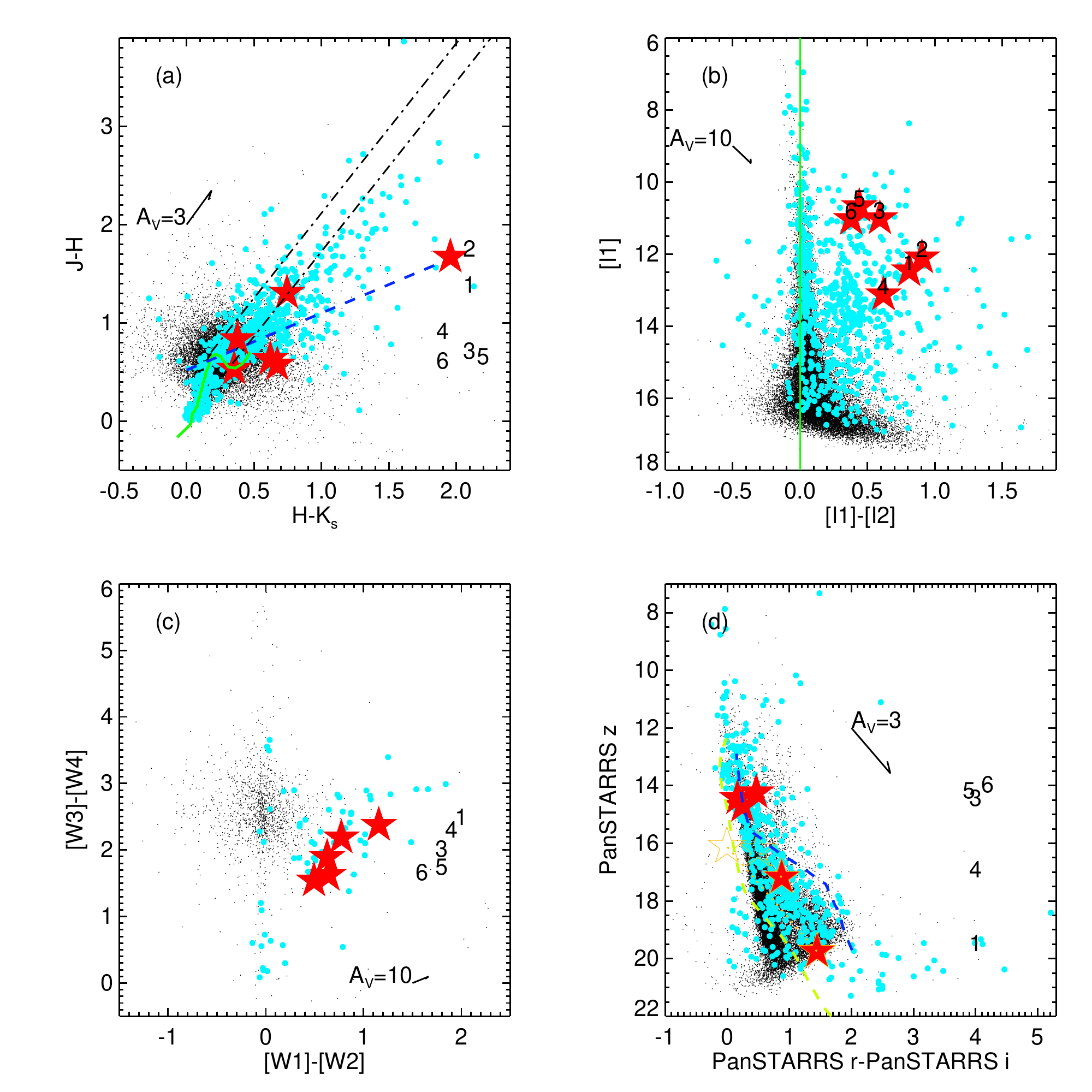}
\caption{Color-color and color-magnitude diagrams for the first six
sources in \S\ref{app:briefexamples} (red stars). Black numbers in
each panel correspond to the source numbers here in the text; in panel
b the numbers are on the corresponding symbols, but in panels a, c,
and d, they are offset to the right hand side for clarity, but are
aligned in roughly the same orientation as the symbols. Black points
are the entire IC~417 catalog, cyan points are the YSO candidates. As
in earlier similar plots, green lines (where they appear) are ZAMS,
blue/yellow dashed lines (where they appear) are PARSEC isochrones, 
and reddening vectors are computed as described in the text. The
individual reddening, where possible, was estimated from the star's 
placement in the $JHK_s$ (first panel) and applied in the optical 
color-magnitude diagram (last panel); the offset orange hollow star 
is where the star would appear if dereddened according to the assumed 
reddening law and $JHK_s$-derived magnitude. }
\label{fig:appendixfeatured1a}
\end{figure}

\begin{figure}[htb!]
\epsscale{0.8}
\plotone{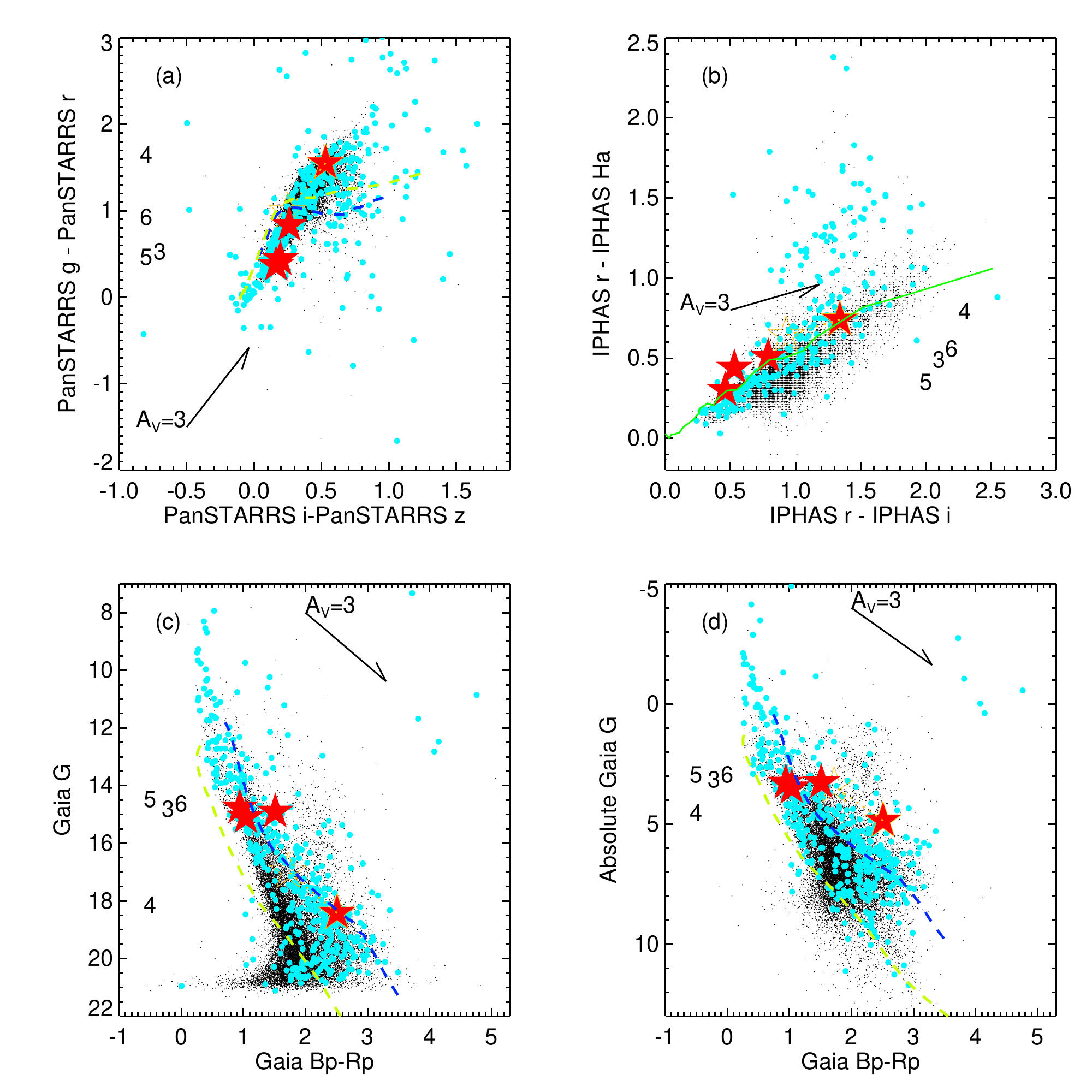}
\caption{Color-color and color-magnitude diagrams for 
the first six sources in \S\ref{app:briefexamples} (red stars). 
Notation is as in previous figure. }
\label{fig:appendixfeatured2a}
\end{figure}

\begin{figure}[htb!]
\epsscale{0.8}
\plotone{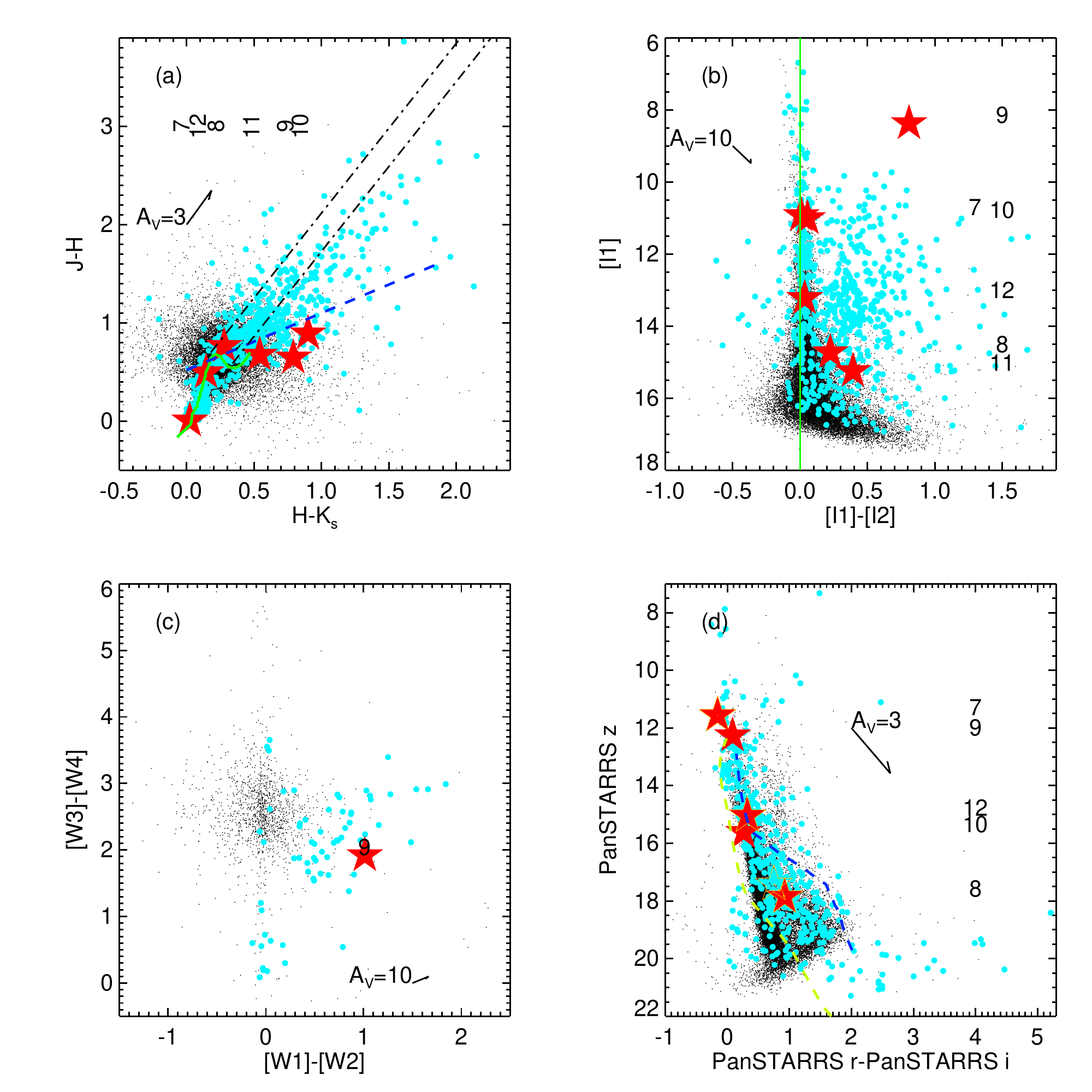}
\caption{Color-color and color-magnitude diagrams for 
the second six sources in \S\ref{app:briefexamples} (red stars). 
Notation is as in previous figure, except the source numbers 
in panel a are along the top rather than along the side. }
\label{fig:appendixfeatured1b}
\end{figure}

\begin{figure}[htb!]
\epsscale{0.8}
\plotone{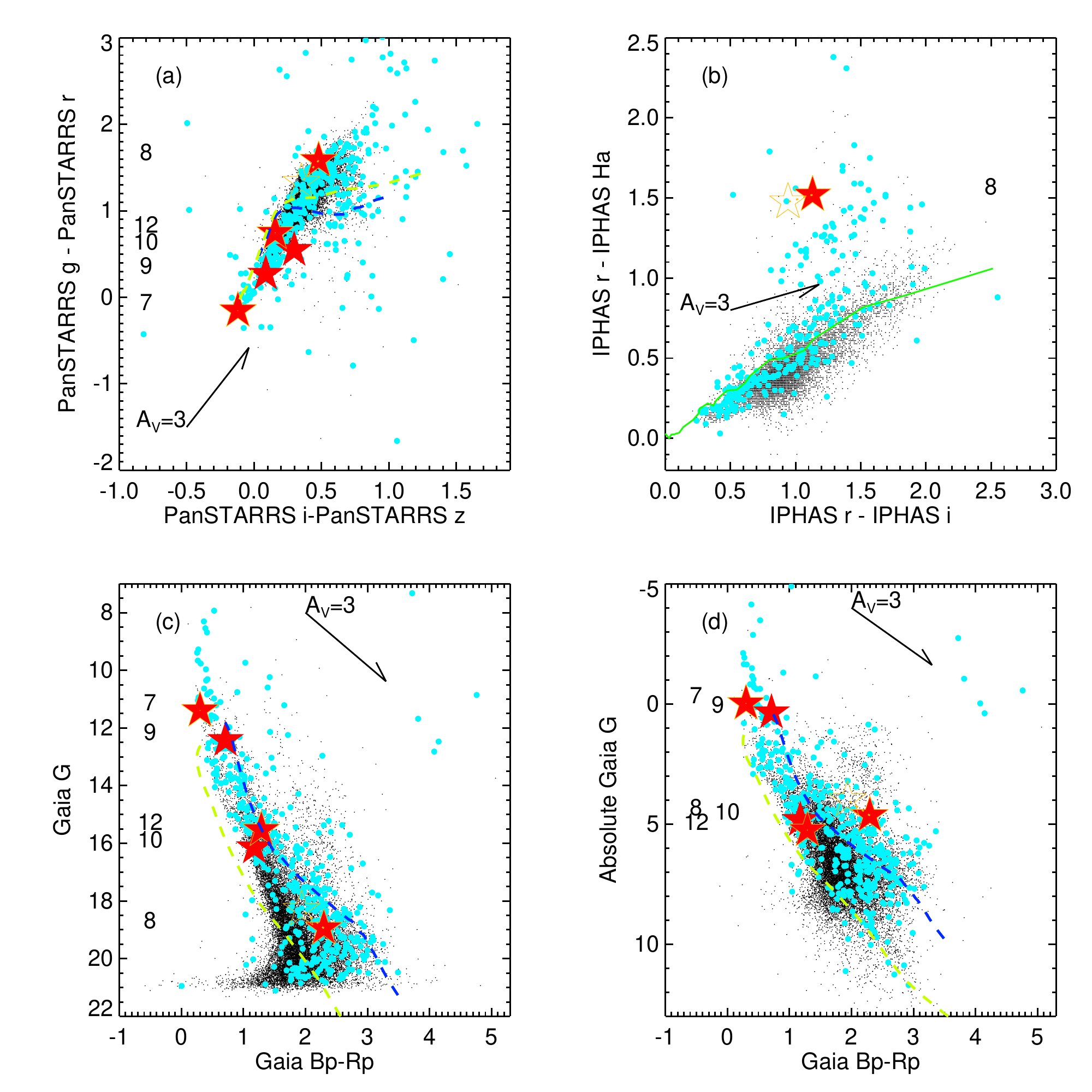}
\caption{Color-color and color-magnitude diagrams for 
the second six sources in \S\ref{app:briefexamples} (red stars). 
Notation is as in previous figure.}
\label{fig:appendixfeatured2b}
\end{figure}

Figure~\ref{fig:sampleseds} shared twelve example SEDs; we now
briefly discuss each of them in turn with an explanation of their
final rank. The color-color and color-magnitude diagrams from 
the prior section (App~\ref{app:textbook}) are included here in two
pairs of figures where the twelve stars are highlighted.
Figs.~\ref{fig:appendixfeatured1a} and \ref{fig:appendixfeatured2a} have:
1-J052807.89+341842.1;
2-J052858.77+342232.5;
3-J052718.35+344033.4;
4-J052736.37+344940.6;
5-J052705.83+343312.0; and
6-J052708.88+345031.5.
Figs.~\ref{fig:appendixfeatured1b} and \ref{fig:appendixfeatured2b} have:
7-J052717.77+342601.1;
8-J052743.82+344028.0;
9-J052919.14+341747.1;
10-J052825.85+342309.6;
11-J052807.15+342732.3; and
12-J052811.74+341625.0.



{\bf Source J052807.89+341842.1} has a final rank 5 and is SED Class I. (It
is star 1 in  Fig.~\ref{fig:appendixfeatured1a} and is too faint to
appear in Fig.~\ref{fig:appendixfeatured2a}.) It was selected as a YSO by
Pandey \etal\ (2020) and Winston \etal\ (2020), and independently by
us based on WISE IR excess. Its SED shows an unambiguous, large IR
excess. The reddening as calculated from $JHK_s$ suggests \av\ of
nearly 7 magnitudes, but the $JHK_s$ diagram indicates that the
circumstellar disk is not strongly affecting the near-IR. It appears
in the right place to be a YSO (e.g., large IR excesses) in panels b and
c of Fig.~\ref{fig:appendixfeatured1a}; it is very faint in panel d,
consistent with  \av$\sim$7 mag. It is highly likely that this is a
legitimate YSO. 

{\bf Source J052858.77+342232.5} has a final rank 4 and, like the prior
source, is SED Class I.  (It is star 2 in 
Fig.~\ref{fig:appendixfeatured1a} and, also like the prior source, is
too faint to appear in Fig.~\ref{fig:appendixfeatured2a}.) It is
within the NS polygon (in projection on the sky).  It was selected by
Winston \etal\ (2020) as a YSO, and independently by us based on WISE
IR excess. Like the prior source, it has an unambiguous, large IR
excess in the SED.  However, this source has fewer points delineating
its SED, which is the primary reason it has a lower final ranking, 4,
than the previous source. It appears in 
Fig.~\ref{fig:appendixfeatured1a}, panel a, as being in the right
place to have its $JHK_s$ affected by IR excess due to a circumstellar
disk.  It appears in the right place to be a YSO (e.g., large IR
excesses) in panels b and c. It is too faint to appear in panel d.

{\bf Source J052718.35+344033.4} is another final rank 5, but this is an
SED  Class flat. (It is star 3 in Figs.~\ref{fig:appendixfeatured1a}
and  ~\ref{fig:appendixfeatured2a}.) It was selected by Winston \etal\
(2020) as a YSO, and  independently by us based on WISE IR excess; it
has an unambiguous IR excess. Looking at the SED, it seems like this
source might have an H$\alpha$ excess, but the size of the error bars
suggests that this is not a significant excess (e.g.,
$\chi_{\rm{IPHAS}}<3$).  In Fig.~\ref{fig:appendixfeatured1a}, panel
a, it appears within the `pack' of stars clumped near the ZAMS. In
panel b, it has an excess, but it doesn't have enough WISE data to
appear in panel c. In panel d, it is consistent with the 6 Myr
isochrone. In all four of the panels in
Fig.~\ref{fig:appendixfeatured2a}, it is  consistent with the `pack'
of stars near the ZAMS. In panel b,  upon close inspection, it is
indeed slightly above the ZAMS, but given the errors, it's not
significantly above the ZAMS. The Gaia parallax and distance is
exactly right for IC~417.

{\bf Source J052736.37+344940.6} is another SED flat Class, but this is a
final rank 4. (It is star 4 in Figs.~\ref{fig:appendixfeatured1a} and
\ref{fig:appendixfeatured2a}.)   It was selected by Winston \etal\
(2020) as a YSO. It has a clear IR excess in the SED.  In
Fig.~\ref{fig:appendixfeatured1a}, panel a, it appears within the
`pack' of stars clumped near the ZAMS, but its position suggests some
reddening (\av$\sim$1.8 mag); panels b and c show that it is in the
right place to be a YSO.  Panel d also has it in the right place to be
a $\sim$6 Myr old YSO; the reddening correction moves it slightly
closer to the 6 Myr isochrone. It has enough PanSTARRS data to have it
appear in Fig.~\ref{fig:appendixfeatured2a}, panel a, but the
reddening pushes it significantly up and to the right. It is on the
ZAMS in panel b, and on the 6 Myr isochrone in panel c.  Panel d,
however, which is the absolute Gaia CMD, has this star a little high.
The Gaia data suggest it may be a little too far -- Bailer-Jones
\etal\ (2021) have a distance of 5148 pc, Gaia DR3 has 14,750 pc, but
a parallax whose error bars comfortably overlap with the expected
parallax range for IC~417. Based on the large IR excess, it is
probably young, but based on the distance, we have placed it as a 4,
not a 5. 

{\bf Source J052705.83+343312.0} is final rank 5, SED Class II. (It is star
5 in Figs.~\ref{fig:appendixfeatured1a} and
\ref{fig:appendixfeatured2a}.) It is in our list because it was
selected by Winston \etal\ (2020) as a YSO, and independently by us
based on WISE IR excess; it clearly has an IR excess in its SED. In
Fig.~\ref{fig:appendixfeatured1a}, panel a, for its $J-H$, it is fairly blue in
$H-K_s$, so as to render it impossible to estimate reddening using our
approach, since no amount of reddening will move this back to the
ZAMS.  Fig.~\ref{fig:appendixfeatured1a}, panels b and c show that it
is in the right place to be a YSO. Panel d suggests that it is on the
6 Myr isochrone, and suggests that perhaps it is a higher mass star
than many of the other  stars in the entire sample.
Fig.~\ref{fig:appendixfeatured2a}, panels a and b also  places this
star in a relatively high-mass location.  This is another one where
the SED and even its placement in panel b seems like it may also have
an H$\alpha$ excess, but the size of the error bars suggests that this
is not a significant excess.  Fig.~\ref{fig:appendixfeatured2a} panels
c and d  are the Gaia CMDs; in both, this star is in between the two
isochrones. It is at the right distance to be in IC~417. Given all of
this, we ranked it as a 5. Interestingly, Gaia DR3 reports that this
is an eclipsing binary, but the Zwicky Transient Facility (ZTF; Bellm
\etal\ 2019) light curve for it reveals that it  is not a binary, but
rather much more likely to be a dipper (e.g., Cody \etal\ 2014 and
references therein) -- these stars have large downward excursions
likely due  to eclipses by dust in the inner disk. We will explore the
ZTF light curves more for these target stars in a future paper.

{\bf Source J052708.88+345031.5} is another final rank 5, SED Class II. (It
is star 6 in Figs.~\ref{fig:appendixfeatured1a} and
\ref{fig:appendixfeatured2a}.) It was selected as a YSO by Pandey
\etal\ (2020) and Winston \etal\ (2020), and independently by us based
on WISE IR excess. It has substantial IR excess, but not significant
H$\alpha$ excess. It is essentially on the ZAMS in
Fig.~\ref{fig:appendixfeatured1a},  panel a, and in the regions with
IR excesses in panels b and c.  In panel d, it is a little above the 6
Myr isochrone. In  Fig.~\ref{fig:appendixfeatured2a}, panel a, it is
in the `pack'  of stars near the isochrones; it is marginally above
the ZAMS in panel b (consistent with not being significantly above the
ZAMS). In both panels c and d, it is a little above the 6 Myr
isochrone, as for the  prior figure. It is at the right distance to be
in IC~417.

{\bf Source J052717.77+342601.1} is also known as LS V +34 13; it is a final
rank 5, SED Class III. (It is star 7 in 
Figs.~\ref{fig:appendixfeatured1b} and \ref{fig:appendixfeatured2b}.)
It is identified as a B2IV by MN16. Gaia DR3 says that it is a white
dwarf, but the SED supports it being likely an O or B star. It is
therefore likely young. There is not much  reddening towards this
star; Fig.~\ref{fig:appendixfeatured1b}, panel a shows it at the right
place to be an unreddened early-type star.  Panel b suggests that it
does not have an IR excess in IRAC bands, and it doesn't have all four
WISE bands, so it cannot appear in panel c. 
Fig.~\ref{fig:appendixfeatured1b} panel d and 
Fig.~\ref{fig:appendixfeatured2b} panel a both again place it 
consistent with being a high-mass young star. It does not have enough
IPHAS data to appear in panel b. Panels c and d both have it high in
the diagram, beyond the end of the  isochrones. It is at the right
distance to be in IC~417.

{\bf Source J052743.82+344028.0} is a final rank 4, SED Class III. (It
is star 8 in Figs.~\ref{fig:appendixfeatured1b} and
\ref{fig:appendixfeatured2b}.) It was identified by Witham \etal\
(2008) as an H$\alpha$-bright star, and that is obvious even in the
SED.  It does not appear to have much of an IR excess; if it does have
an IR excess, it is small, and determining the size of such an excess
would require a spectral type. There is not much reddening towards
this star;  Fig.~\ref{fig:appendixfeatured1b}, panel a, shows it very
close to  the ZAMS, and the inferred reddening is \av$\sim$0.8 mag. In
panel b, this star is within the broad `pack' of stars at the faint
end of the distribution where the errors get larger. It doesn't appear
in  panel c. It is faint in panel d, and appears between the
isochrones. In Fig.~\ref{fig:appendixfeatured2b}, panel a, it is
clearly reddened to be above the isochrones. In panel b, it is
obviously and unambiguously bright in H$\alpha$. In panels c and d, it
is faint and  between the isochrones. The Gaia DR2 Bailer-Jones
distance is ok (2192 pc), the  DR3 Bailer Jones distance is slightly
too far (7249 pc), and the Gaia DR3 distance is ok (3085 pc).  Because
of this ambiguity, and since the sole indicator of youth, per se, is 
the H$\alpha$ excess, which could also just be an indicator of
elevated stellar activity, we have ranked this star a 4 rather than a
5.

{\bf Source J052919.14+341747.1} is one of the most peculiar oddballs with
final rank 4*. It is SED Class flat, and star 9 in 
Figs.~\ref{fig:appendixfeatured1b} and \ref{fig:appendixfeatured2b}.)
It was identified in J08 as having H$\alpha$ excess and was flagged in
Winston \etal\ (2020) and Pandey \etal\ (2020) as a YSO, as did we in 
our initial WISE IR excess search. Its SED is unlike most YSO SEDs; it
has MSX, AKARI, and PACS counterparts that all appear to be
consistent. In  Fig.~\ref{fig:appendixfeatured1b}, panel a it seems to
be so blue in  $H-K_s$ given its $J-H$ that guessing a reddening value
is not easy. It is very bright in panel b, with a large IR excess; it
also is the  only one of this set to appear in panel c. It is on the
isochrone and consistent with a high-mass star in
Fig.~\ref{fig:appendixfeatured1b}, panel d as well as
Fig.~\ref{fig:appendixfeatured2b}, panel a. It  does not have IPHAS
data so it does not appear in panel b. In panels c and d, it is on the
isochrone consistent with a 6 Myr star. It has a Gaia DR3 distance
completely consistent with IC~417.  A spectrum is needed to understand
this object.

{\bf Source J052825.85+342309.6} is another oddball with final rank 4*; it
is  SED Class II (and star 10 in Figs.~\ref{fig:appendixfeatured1b}
and  \ref{fig:appendixfeatured2b}). It was identified in J08 as having
H$\alpha$ excess and  in Winston \etal\ (2020) as a YSO. Its SED is
unlike most YSO SEDs; it is too broad. In
Fig.~\ref{fig:appendixfeatured1b}, panel a, it appears to be in the
T~Tauri locus. In panel b, it appears to  not have much of an IR
excess at IRAC bands, and does not have enough data to appear in panel
c. It is between the isochrones in panel d. In 
Fig.~\ref{fig:appendixfeatured2b} it is consistent with possibly 
having a small blue excess. It does not have enough data to appear in 
panel b. In panel c, it is between the isochrones, but in panel d, it
is a little above the 6 yr isochrone. Nonetheless, it has a  Gaia DR3
distance completely consistent with IC~417.   A spectrum is needed to
understand this object.

{\bf Source J052807.15+342732.3} is final rank 2, SED Class II. (It is star
11 in Fig.~\ref{fig:appendixfeatured1b} and does not have enough data
for Fig.~\ref{fig:appendixfeatured2b}.) It is on our list of candidate
YSOs because it was identified as such in J17. In
Fig.~\ref{fig:appendixfeatured1b}, panel a, it is either  on the
reddest portion of the ZAMS or the bluest portion of the T~Tauri
locus. It is faint, but has an IR excess in panel b; it does not
appear in  panel c or d. It is low-ranked primarily because it has so
few points in its SED compared to others on our list, though it does
seem to have an IR excess. 

Finally, {\bf J052811.74+341625.0} is final rank 3, SED Class III. (It is
star 12 in Figs.~\ref{fig:appendixfeatured1b} and
\ref{fig:appendixfeatured2b}.) It is identified as a variable in Lata
\etal\ (2019), but that variability is the only indication of youth.
Fig.~\ref{fig:appendixfeatured1b},  panel a has it on the ZAMS with
very little reddening. Panel b shows no IR excess; it does not appear
in panel c. It is on the 6 Myr isochrone  in panel d. In
Fig.~\ref{fig:appendixfeatured2b}, it is on the isochrones in panel a.
It does not appear in panel b. In panel c, it is on the 6 Myr
isochrone; in panel d, it is between the isochrones.  Gaia DR3 says
that it is at the right distance to be in IC~417, so it is a rank 3.

A complete set of SEDs has been delivered to IRSA, along with the
complete data table (Table~\ref{tab:bigdata}), which includes ranks,
flags, and photometry.

\end{document}